\documentclass[10pt,twocolumn]{article}
\usepackage[letterpaper,margin=0.75in,columnsep=0.25in,headheight=12pt,headsep=14pt,footskip=24pt]{geometry}
\usepackage{fontspec}
\newfontfamily\cjkfont{texgyretermes-regular.otf}
\usepackage{amsmath}
\usepackage{unicode-math}
\usepackage{placeins}
\usepackage{microtype}
\usepackage{booktabs,tabularx,array,ragged2e}
\usepackage{graphicx}
\usepackage{tikz}
\usetikzlibrary{arrows.meta,patterns,positioning}
\usepackage{caption}
\usepackage{enumitem}
\usepackage{titlesec}
\titleformat{\section}{\large\bfseries}{\thesection}{0.6em}{}
\titleformat{\subsection}{\normalsize\bfseries}{\thesubsection}{0.6em}{}
\titlespacing*{\section}{0pt}{1.4ex plus 0.6ex minus 0.3ex}{0.8ex}
\titlespacing*{\subsection}{0pt}{1.1ex plus 0.5ex minus 0.3ex}{0.5ex}
\titleformat{\paragraph}[runin]{\normalfont\normalsize\bfseries}{}{0pt}{}[.]
\titlespacing*{\paragraph}{0pt}{0.9ex plus 0.3ex minus 0.2ex}{0.55em}
\usepackage{fancyhdr}
\usepackage[hyphens]{url}
\usepackage[colorlinks=true,linkcolor=black,citecolor=black,urlcolor=blue!50!black,pdftitle={User First Intelligence: Three Measurements of Search and AI-Answer Analytics (SEO and GEO/AEO), and a Knowledge-Graph Framework for Query Discovery Beyond Search Volume},pdfauthor={Venkata Pagadala}]{hyperref}
\usepackage{balance}

\newlist{reflist}{itemize}{1}
\setlist[reflist]{label={},leftmargin=1.2em,itemindent=-1.2em,itemsep=2.5pt,parsep=0pt,topsep=3pt,font=\small}
\usepackage{etoolbox}
\AtBeginEnvironment{reflist}{\small\sloppy}
\begin{document}
\makeatletter
\twocolumn[{%
\begin{@twocolumnfalse}
\begin{center}
{\LARGE\bfseries User First Intelligence: Three Measurements of Search and AI-Answer Analytics (SEO and GEO/AEO), and a Knowledge-Graph Framework for Query Discovery Beyond Search Volume\par}
\vspace{1.3em}
{\large Venkata Pagadala\par}
\vspace{0.35em}
{\normalsize Research Scholar (Ph.D. in progress, Artificial Intelligence), KL University (Koneru Lakshmaiah Education Foundation), India \\ vdepagadala@gmail.com\par}
\vspace{0.35em}
{\small\itshape Manuscript version 5.5, 14 November 2026. Preprint, not yet peer reviewed.\par}
\end{center}
\vspace{0.6em}
\begin{center}
\begin{minipage}{0.94\textwidth}
\small
\noindent\textbf{Abstract.} People say what they want to know in many places that are not a search box, and the commercial tools that search and AI-answer optimization rely on see only part of it, through models whose figures are often reported as counts. This paper reports three measurements of what such instruments show and builds a framework on them. On the area-code section of a large US telecommunications provider’s website (381 code pages and 53 hubs), two commercial analytics platforms’ stored records from the section’s first month differ on estimated monthly visits by 7.5x to 27.1x (8.6x to 29.9x with both read worldwide). Across twelve sections of the domain the ratio runs from 0.20x to 7.24x, with no detectable relationship to section age (Spearman −0.11, a test of low power), while the platforms agree on which section is larger in 60 of 66 pairs, though in only 9 of the 15 within a factor of three. On 37,193 reviews of 378 telecom stores linked to a 42,994-node keyword taxonomy, on links two blind AI annotators judged about two-thirds correct, review demand and search volume barely agree on topics (+0.18) and agree on branches to a degree that depends on attribution and link quality (+0.47 to +0.75 on all links, up to +0.91 on multi-word labels, post hoc), while sizing them very differently. On the 372 area codes, in a comparison whose predictions were fixed before collection, a public count of requests for each code’s Wikipedia article and the vendor’s keyword volume agree roughly on ranking (+0.45, bootstrap 95\% interval +0.34 to +0.54, over 240 articles), an agreement that disappears once the number of keywords matched per code is held constant (−0.12, post hoc), and not on magnitudes (searches per request from 0.3 to 244 between the 10th and 90th percentiles). In one vendor’s index of AI answers (Ahrefs Brand Radar, US, October 2026) that name or cite the ten largest holdings of each S\&P 500 sector fund, of four predictions written before collection one held, one held in direction only and two failed; video and community sites took a median 21\% of citation incidences on Google’s AI Overviews and AI Mode against 0.6\% to 9.2\% on ChatGPT, Copilot, Perplexity and Gemini, in every sector; a later read of education answers in thirteen countries, not predicted, shows the same contrast in every country (14.0\% against 0.6\%). Traffic estimates and cited sources therefore have to be measured per instrument and per answer surface. User First Intelligence (UFI), the framework built on that premise, collects what people ask beyond the website, links it to an industry’s entities under an ontology, and scores where the organization does not answer. An open implementation is released; its scoring function is specified, not evaluated. No causal claim is made.
\vspace{0.7em}

\noindent\textbf{Keywords:} User First Intelligence; query discovery; search engine optimization (SEO); answer engine optimization (AEO); ontology; knowledge graph; competency questions; search measurement; instrument validity; AI Overviews; AI answer citations; generative engine optimization (GEO); deployment measurement study; reference content
\end{minipage}
\end{center}
\vspace{1.6em}
\end{@twocolumnfalse}
}]
\makeatother

\section{Introduction}
A content decision starts with a question someone asked. The place it was asked is often not a search engine: it is a forum thread, a question site, a comment under a video, a product review, a support ticket, a post in a group. Keyword-volume tools observe the part of that demand that reaches a search box in a countable phrasing; Section 8 shows this deployment ranking for question forms, misspellings and non-English phrasings that carry little or no reported volume in those tools, and AI answer surfaces now sit between many questions and any page at all, citing different kinds of source depending on which answer engine is asked, in one vendor’s index (Section 7). User First Intelligence is a framework built on that premise: go beyond the website to where the questions are, link them to the entities of an industry held under an ontology, keep the sources that establish structure, coverage, answerability and demand in separate roles so that a competitor’s coverage cannot pose as user demand, express user needs as competency questions that a knowledge graph can be asked, and score where the organization fails to answer, with the form of the answer meant to be measured rather than assumed (Figure 1); the released score does not yet do so, because its coverage rubric prefers per-entity pages (Section 9.1, Appendix B). The deployment reported here was built from a gap found in the framework’s graph when it held one of those surfaces, competitor keyword exports, and before the others were implemented (Section 3.2); it is a case of the graph finding the kind of gap the framework targets, not a run of the whole framework and not a test of its score. It also exposed the measurement problem the rest of this paper is about. The framework is meant to serve both traditional search engine optimization (SEO) and its counterpart for AI answer engines, known as generative engine optimization (GEO) [Aggarwal et al. 2024] or, in practitioner usage, answer engine optimization (AEO).

Much of the empirical evidence in applied search research rests on commercial analytics platforms. Papers report that a technique moved visibility, that a competitor holds a share of some keyword space, that AI answers cite one kind of source more than another. The underlying numbers come from Ahrefs, Semrush, Similarweb or a comparable vendor, and they are often reported the way measurements are reported: as quantities, usually from one vendor and without a statement of uncertainty, a practice the validity studies in Section 2 caution against [Kamerer 2013; Jansen, Jung \& Salminen 2022] or show to be unreliable [Prantl \& Prantl 2018]. The practice is growing: in OpenAlex, 456 works name Ahrefs, Semrush or Similarweb in their title or abstract, 27 of them published in 2020, 87 in 2025 and 110 in 2026 to mid-October (raw counts, not normalized for the growth of the index). Screened from title and abstract by two blind AI annotators, a random sample of 120 of them gives 82\% to 87\% that use a vendor’s figures as data (Wilson 95\% intervals 75\% to 92\%; Cohen’s kappa 0.63), and of those, 87\% to 88\% use one vendor’s figures only, while five to seven compare vendors (\texttt{vendor\_reliance.py} and \texttt{stakes\_screen\_check.py}, in the source bundle, which is posted with this preprint as its ancillary files). A search of OpenAlex and Crossref on 14 October 2026 and an independent search of five databases with citation chasing by a separate agent on 15 October, both recorded in the source bundle, the second rating its own saturation partial, found vendor tools compared with first-party analytics and with each other on established domains (Section 2), but no comparison of two vendors’ records of a single object followed from a known start date; to the extent of those two searches, this paper reports the first. A third search, on 31 October, looked for reporting guidelines for such figures and found no checklist; the nearest guidance is the set of recommendations in the discussion of one empirical comparison of the tools [Font-Julián et al. 2026], which Appendix E credits. Which sources different AI answer engines cite has been audited on public-interest queries, for news sources, and on real search queries [Li \& Sinnamon 2024; Yang 2025; Huang et al. 2026; Allaham \& Diakopoulos 2026]; this paper adds a comparison of eight vendor-collected answer surfaces across a sector frame fixed by an external rule.

The vendors do not claim otherwise in their documentation, which describes click models, of the kind studied in information retrieval [Chuklin, Markov \& de Rijke 2015], applied to estimated volumes at estimated positions (Section 4.1). The practice treats them as counters anyway. This paper puts a number on how far two such estimates of one object can lie apart, so that the practice can state its uncertainty. For a framework that reads its outcomes on such instruments (Figure 1b), that number matters: it bounds how far one platform’s magnitudes can be trusted, which is why the framework should read its outcomes on more than one.

\subsection{Why a new section is the right instrument}
Comparing two analytics vendors on an established site tells you little. Both have years of accumulated index data, the disagreements are confounded with each vendor’s historical coverage, and there is no moment at which the true value is known.

A section that did not exist last week is different. Before launch, both instruments correctly report zero. After launch, both observe the same URL prefix, in the same period, with no prior history. Their disagreement is then attributable to their models and their indexes as applied to that object, and its trajectory over time can be read from each instrument’s history. Section 5.5 shows that the disagreement is specific to the object as well: the same two instruments disagree by different factors, in different directions, on different sections of the same site.

This paper uses one such section: 381 code pages plus hubs under a single prefix on a top-authority domain, published on 26 July 2026 and observed for 77 days, with ranking positions from the instrument’s daily series and the other quantities at seven checkpoints and as monthly histories (Section 4.1 states how each was read).

\begin{figure*}[t]
\centering
\includegraphics[width=\textwidth]{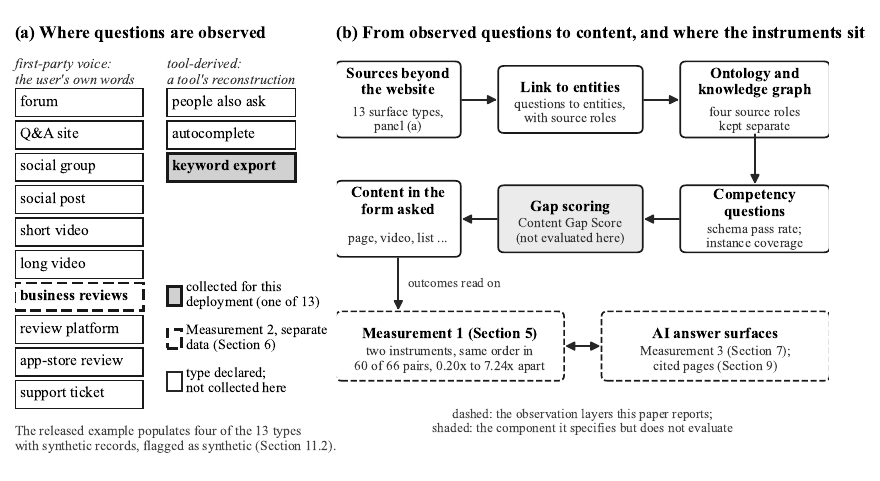}
\caption*{\textbf{Figure 1.} The User First Intelligence framework. (a) The thirteen demand-surface types the released implementation declares, grouped by whether they carry a user’s own words or a tool’s reconstruction of demand; the one surface this deployment collected is marked, and business reviews, run separately in Section 6, are marked differently. (b) The pipeline from observed questions to content, with the two observation layers this paper reports (dashed; the AI-answer layer is also measured outside the pipeline, in Section 7) and the component it specifies but does not evaluate (shaded). The deployment in this paper ran the lower layer and one surface of the upper; it did not run the pipeline end to end.}
\end{figure*}

\subsection{What this paper reports}
The paper’s original contribution is Measurement 1: two commercial analytics platforms’ estimates of the same pages differ by up to about 30 times worldwide (33 times US to US, estimated), in no fixed direction across sections, while the platforms mostly agree on how the sections rank, so no constant converts one into the other. Around it, the paper specifies, and does not evaluate, a framework for finding content gaps from demand observed beyond the search box. Each measurement bears on one part of the framework and is stated with what it does not show: Measurement 1 on the outcome layer, where a content program’s results are read on analytics platforms, Measurement 2 on the premise that what people write names different topics from what volume counts, and Measurement 3 on the answer surfaces, which cite different sources.

Three questions organize the measurements. RQ1: how far apart are two commercial platforms’ estimates of the same pages, and is the gap a constant a researcher could correct for? RQ2: do other observations of demand, people’s own words in reviews and a public count of requests for the same entities, rank topics and entities as keyword volume ranks them? RQ3: do AI answer surfaces cite the same kinds of sources, in industries chosen by an external rule? Each is answered once below, with what the answer does not show.

\begin{enumerate}[leftmargin=1.6em, itemsep=1.5pt, topsep=2pt, parsep=0pt]
\item \textbf{Measurement 1, two instruments on one prefix} (Section 5; RQ1), an instrument-validity result bearing on the framework’s outcome layer. Two commercial platforms observing one URL prefix from zero differ on estimated visits by 7.5x to 27.1x (8.6x to 29.9x worldwide against worldwide); across twelve sections of one domain the ratio spans 0.20x to 7.24x with no detectable relationship to section age, so no domain-level factor converts one platform’s magnitudes into the other’s, while the platforms agree on the ordering of sections in 60 of 66 pairs. This is the paper’s main empirical result. It does not say which platform is closer to the truth. Appendix E turns it into a reporting checklist for studies that use such figures.
\item \textbf{Measurement 2, reviews and a public count against keyword volume} (Section 6; RQ2), bearing on the demand surfaces. On 37,193 business reviews linked to a 42,994-node keyword taxonomy, the two demand signals agree on the taxonomy’s branches to a degree that depends on attribution and link quality, size them very differently, and barely agree on topics (Section 6.1). On the 372 area codes, a public count of requests for each code’s Wikipedia article and the vendor’s volume agree roughly on the ranking (+0.45 over 240 articles), an agreement that disappears given the number of keywords matched per code, and not on the magnitudes, in a comparison whose predictions were fixed before collection (Section 6.2). These are two surfaces in one industry, not the multi-surface study of Section 13.
\item \textbf{Measurement 3, what AI answers cite} (Section 7; RQ3), bearing on the answer surfaces. In one vendor’s index, across the eleven S\&P 500 sectors and eight answer surfaces, Google’s AI Overviews and AI Mode give video and community sites a larger share than every other surface in every sector, and about ten times their pooled median share (21.3\% against 2.1\%); of four predictions written before collection, one held, one held in direction only and two failed. A later read of one category in thirteen countries, with the same registered source types and no prediction, shows the same contrast in every country (Section 7.7). It motivates a per-surface term the framework does not yet have.
\item \textbf{The User First Intelligence framework} (Section 1.4, Figure 1): six components (demand surfaces, an industry ontology, a knowledge graph with separated source roles, competency questions, answer surfaces and a Content Gap Score) and two competency-question measures, with an open implementation (Section 14). The score is specified in Appendix B and not evaluated here; Section 13 specifies the studies that would test it.
\item \textbf{The deployment as context} (Sections 3.1 and 8 to 11): a 77-day series with a primary checkpoint chosen to precede three of the four identified disturbances, an observation, offered as a hypothesis because a hub pools questions about many codes, that each of the five most-cited hub pages is cited in more AI answers than any per-entity page (1.2 to 2.0 times the most-cited one), the deployment’s cost, its six enabling conditions, and six lessons stated as what it got wrong. It is a case of the kind of gap the framework targets, not a run of the framework, and no causal claim is made.
\end{enumerate}

\subsection{What this paper does not report}
A reader is owed this before the results rather than after them.

\paragraph{No causal claim}The design is observational. There is no treatment assignment, no control group, no randomization and no matched comparison. The provider’s list of sections launched in the same window is not available, so this study cannot say whether this section outperformed conventionally selected content on the same domain in the same period. Where the deployment series appears, it is a description of what was observed following publication, and “following” is the strongest word the design licenses.

\paragraph{No separation of the method from the domain}The deployment ran on a domain with an Ahrefs Domain Rating of 90 and a Semrush Authority Score of 86. Nothing here bounds how much of the outcome that explains, and Section 12 lists it among the confounds. A comparison with two carriers of similar authority does not bound it either, because it is computed within a set selected on the provider’s own presence (Section 9.2).

\paragraph{No evaluated scoring function}The framework’s Content Gap Score was not computed when the build decision was taken, nothing here tests it, and no published study has tested whether it predicts anything. It is specified in Appendix B with a reference implementation, and it is kept out of the results so that it cannot be mistaken for one.

\paragraph{No measured traffic}No figure in this paper is a counted visit. The provider’s internal analytics were not available for this study.

\paragraph{No independent verification}No person other than the author has graded, replicated or audited any part of this work. Section 14 states what has and has not been done about that, and it is the paper’s largest unremedied weakness.

\subsection{The framework, stated}
Query discovery, as the term is used in the title, is the framework’s first task: finding the questions people ask about an industry’s entities, on any surface and in their own words, before deciding which of them an organization should answer. A content gap is one outcome of discovery, not its object; Measurement 2 (Section 6) is a first run of discovery on one surface, and Section 13 specifies the run across surfaces. User First Intelligence has six components, the last of which, the Content Gap Score, is its output, and two measures of how well the ontology serves the competency questions. Appendix B gives the score’s formula and the released implementation (Section 14) gives the code; where the two differ, the code is the reference and the difference is a defect of this text.

\begin{enumerate}[leftmargin=1.6em, itemsep=1.5pt, topsep=2pt, parsep=0pt]
\item \textbf{Demand surfaces.} The places where people state a need, in their own words or as a tool’s reconstruction of it. The implementation declares thirteen surface types (Figure 1a) and ships two generic file adapters, for CSV and JSON Lines; collection from each platform is left to the deployment. Appendix F lists the one hundred concrete sources the framework is meant to collect, of which this paper has read eight. Every record carries its surface and collection date, and a license field is optional. Demand records are held in a separate demand index and joined to entities by a linker; they are not graph assertions.
\item \textbf{Industry ontology.} The classes, relations and enumerable entity sets of one industry, with competency questions written against it [Ren et al. 2014; Keet \& Ławrynowicz 2016]. The rule is that the questions are written before demand data is collected; in the released example they were elicited from a keyword export, contrary to that rule, and the example records it.
\item \textbf{A knowledge graph with separated source roles.} Every source carries one role from a closed list of four: structure (what exists, from registries, encyclopedias and standards bodies), coverage (who answers, from the organization’s and competitors’ pages), answerability (retrieved search results, judged for whether they answer a question) and demand (who asks). Coverage never counts as demand, which is the error lesson 1 of Section 11.2 records. Every graph statement carries its source, role and retrieval date, in the spirit of provenance-typed statements [Carroll et al. 2005; Lebo, Sahoo \& McGuinness 2013]. Each source has exactly one role; the implementation has no rule yet for a source that both asks and answers, such as a forum thread. Provenance is held on each statement in the implementation’s graph and reported per source; its OWL and RDF export does not yet serialize it.
\item \textbf{Competency questions.} User needs expressed as questions over the ontology and compiled to SPARQL along property paths [Wiśniewski et al. 2019]. A question the ontology cannot express is reported as a declared gap with a reason, not scored as zero.
\item \textbf{Answer surfaces.} Where answers are delivered and cited: search results and AI answer engines. The framework does not yet model them: neither its two measures nor its score takes a per-surface input, and the implementation has no answer-surface component. Section 7 is a separate measurement study that uses none of the framework’s ontology, graph, competency questions or score; it shows that the publisher types answer engines cite differ between engines, which motivates a per-surface term and does not test one.
\item \textbf{The Content Gap Score.} A ranking of entities by a prior (topic-sensitive PageRank over the structure graph, mixed with linked demand), multiplied by the organization’s own non-coverage and by a deficit that mixes coverage by a declared set of competitors with the mean, over the entity’s competency questions, of how far the best retrieved search result for any phrasing falls short of answering. Own coverage, competitor coverage and answerability are graded judgments: the package ships a rule-based grader for its tests and example and no model client, and Appendix B gives both rubrics.
\end{enumerate}

The two measures are the competency-question schema pass rate (the share of competency questions the ontology can express) and the instance coverage rate (the mean, over expressible questions, of the share of subjects for which the graph returns an answer). They keep the implementation’s names, so “coverage” names both a source role and the second rate. In this paper one demand surface is exercised on real data, business reviews in Section 6, linked to a keyword taxonomy rather than an ontology; the deployment’s competitor keyword exports count as coverage under the framework’s own rule (Section 11.2, lesson 1); and the score is specified but not evaluated.

\section{Related work}
\paragraph{Search measurement and tool validity}The validity of vendor traffic estimates has been measured mainly against first-party analytics, site by site. An agency comparison of Similarweb, Ahrefs and Semrush against Google Analytics organic visits for 25 sites found all three differing from the first-party figure in level, with the two keyword-based tools below it on total traffic by 17\% and 30\% [Langridge 2016]. A vendor study of 1,635 sites against Google Search Console reports a median deviation of 49.5\% for Ahrefs and 68.4\% for Semrush, and a correlation with the first-party figure of 0.76 and 0.74 [Soulo 2022]; it is the vendor reporting on itself and on a competitor, and is cited for its design rather than its ranking. In academic work, traffic-ranking services have been compared as proxies for organizational performance [Vaughan \& Yang 2013], and the major top-site lists were shown to disagree with one another on most of their entries and to churn by up to half of their domains a day [Scheitle et al. 2018], which motivated a ranking built to be stable and hard to manipulate [Le Pochat et al. 2019]. Tools have also been compared with one another directly: the domain-authority scores of Moz, Semrush and Ahrefs [Reyes-Lillo, Morales-Vargas \& Rovira 2023], and the organic keyword counts, keyword search volumes and URL-level traffic that Ahrefs, Semrush, Serpstat and Ubersuggest report for two academic publishers’ domains, where the keywords and URLs the tools returned overlapped so little that the choice of tool determined the findings [Font-Julián et al. 2026], with traffic compared at global and country level for the same two publishers [Font-Julián et al. 2025]; SEO tools have also been used to compare the visibility of university library sites with that of their universities [Vállez \& Ventura 2020]. Analytical results and simulation show that the processing behind third-party traffic estimates can shrink, inflate or reverse an estimated treatment effect, and caution against their use for causal inference [Khosravi \& Yoganarasimhan 2026]. Those studies compare an estimate with a ground truth across many sites, compare lists across the whole web, or compare tools on established domains at one time. This paper has no ground truth and does not compare across sites. Its cross-section is, like the last of those, a comparison of two tools at one time; what differs is the unit, twelve sections of one domain, which separates disagreement about a site from disagreement about a section of it. It adds a reading of two vendors on a section both observed from zero, made in a single retrospective read. Comparisons of estimators with first-party analytics or reference traffic data have found large and variable errors: Compete and Quantcast underestimated traffic in a panel of 18 sites (Quantcast reported on eight of them), with occasional overestimates [Kamerer 2013], and neither Alexa nor SimilarWeb ranked Czech websites accurately, although both indicated a site’s approximate significance [Prantl \& Prantl 2018]. Closest to this paper, a comparison of SimilarWeb with Google Analytics on 86 sites found significant differences in level, rank orderings that correlated, and differences systematic enough to estimate one service’s figures from the other’s for related sites [Jansen, Jung \& Salminen 2022]. The pattern recurs between two estimators: on 50 open-access repositories, SimilarWeb and Semrush differed significantly in level, by about 4\% on average for visits and for visitors, while ranking the repositories almost identically (Spearman 0.997 on total visits) [Rasool, Thakur \& Nabi 2026]; and a study of news publishers reported a decline of about 7\% in weekly visits within six weeks of blocking generative-AI crawlers, measured by SimilarWeb or by Semrush [Zhao \& Berman 2025]. Section 5.5 finds the same agreement on ordering between two estimators and the opposite on calibration: across twelve sections of one site, the ratio between them is not systematic.

\paragraph{Comparison across non-identical universes}Comparing two measurements computed over different underlying sets is an old problem in information retrieval evaluation. Pooled relevance judgments are incomplete, and systems that did not contribute to the pool are scored on a universe they did not help define [Zobel 1998]; a measure robust to incomplete judgments was proposed for exactly that case [Buckley \& Voorhees 2004], and the bias a pooled universe imposes on a new system can be estimated and adjusted [Webber \& Park 2009]. Section 5.6 makes the same move on commercial instruments: it decomposes the visits divergence into a keyword-universe term and a per-keyword term, so that a disagreement about how much surface each vendor indexes is not read as a disagreement about traffic. Unlike a test collection, neither vendor’s universe is published, so the decomposition can be computed only at the aggregate level reported here.

\paragraph{AI answer surfaces}An audit of 55,393 queries found that nearly 30\% of the domains cited by Google’s AI Overviews did not appear on the first page of results shown with them [Xu, Iqbal \& Montgomery 2026]; on 11,500 queries, the sources retrieved by Google Search, AI Overviews and Gemini overlapped little (Jaccard similarity below 0.2 on average), with the generative systems favoring Google-owned content [Grossman et al. 2026]; five generative search systems from three providers reached topical coverage comparable to Google’s organic results through markedly different retrieval footprints [Kirsten et al. 2026]; vendor measurement across 15,000 prompts reports that on average only 12\% of the URLs cited by ChatGPT, Gemini, Copilot and Perplexity rank in Google’s top ten for the same prompt, 28.6\% for Perplexity and about 8\% for the others [Linehan 2025]. Generative engine optimization was introduced as a way to raise a source’s visibility in generated answers [Aggarwal et al. 2024]; the citations of generative search engines have been audited for whether they support the statements they accompany [Liu, Zhang \& Liang 2023]; and the evaluation of retrieval has been reframed for systems that generate rather than rank [Gienapp et al. 2024]. Users click less when a summary appears and rarely click its sources: a click on a link inside an AI summary occurred in 1\% of visits to pages showing one [Chapekis \& Lieb 2025]. A critical survey of 45 generative-engine-optimization studies argues that the foundational paper’s gains establish neither organic discoverability nor durable traffic effects, and identifies topical relevance and context position as the most reproducible levers [Martinez 2026]. Audits have compared which sources answer engines cite: ChatGPT, Bing Chat and Perplexity on public-interest queries, with sources classified [Li \& Sinnamon 2024]; OpenAI, Perplexity and Google models on over 366,000 citations from a public comparison arena, which cite distinct news sources [Yang 2025]; five systems including Google AI Overviews on 11,000 real queries, with Wikipedia the most-cited domain and over-represented in the generated summaries, and cited social media under-represented in the summaries of Google’s AI Overviews [Huang et al. 2026]; ChatGPT, Copilot, Gemini and Perplexity on 712 queries, about 16\% of whose unique cited sources an AI-text detector classed as AI-generated [Allaham \& Diakopoulos 2026]; and Google’s AI Overviews on “your money or your life” queries drawn from MS MARCO Web Search, which cited AI-generated documents more often than human-authored ones even after controlling for retrieval rank [Kakimov et al. 2026]. Comparisons across industries exist, in frames their authors chose: a fixed panel of 480 questions across eight industries sent to Google’s AI Overviews, ChatGPT, Perplexity and Gemini [Hendricks 2026]; controlled experiments across verticals that found AI search engines favoring earned, third-party sources over brand-owned and social content, unlike Google [Chen et al. 2025]; query intent and Google rank as predictors of citation, intent across four chat assistants and eight verticals and rank on two assistants and ten verticals [Lee 2026]; and a vendor index of 6,095,520 citations from seven engines in 17 undefined industries [Tousseyn 2026]. Other vendor reports, built on their own tools, publish source shares by platform for categories they define: social sources made up 31\% of Perplexity’s citations against about 7\% of ChatGPT’s in nine categories it chose in January 2026 [Tinuiti 2026], 19.4\% against 5.3\% in a report that also breaks citations down for 15 industries [Punturo, Torres Llorca \& Fishkin 2026], and brand-managed sources 86\% of the citations of three engines in four industries [Yext 2025]. Vendor-tracked brand visibility has also been measured: in more than 100,000 prompt responses for over 100 brands, about 78\% of the sources engines cited were corporate websites, with YouTube leading the non-corporate sources, ahead of Reddit, editorial media and Wikipedia [Kumar 2026], and a scoping review of 23 studies, mostly preprints, maps how such presence and citation are measured [Romero Ramos et al. 2026]. One working paper a search found by title, on 80 global brands in LLM answers, could bear on the frame and could not be read [Cook 2025]. Section 7 differs in its frame, eleven sectors fixed by an external rule, in using eight vendor-collected surfaces, and in its universe, answers that name a sector’s largest companies. Section 5.3 adds a measurement observation those studies need: the two vendor definitions of “cited in an AI Overview” differ by a factor that is itself time-varying, so a GEO result reported on one vendor’s definition is not directly comparable to one reported on the other’s.

\paragraph{Personalization and the visibility of demand}The framework’s premise that a search box shows only part of what people ask has one adjacent documented finding, one observation and one open question. Long-tail studies show demand spread widely across rare items, in web search among other settings [Goel et al. 2010], though they do not measure demand that never reaches a search engine; Section 8 adds the untested observation that the section ranks for question forms, misspellings and non-English phrasings carrying little or no reported volume. Personalization of classic web search has been measured directly: about 11.7\% of Google results differed between users in a 2013 audit, with signed-in status and location the drivers and demographic attributes such as gender and age producing no substantial effect [Hannak et al. 2013]. On AI answer surfaces the answer is composed rather than ranked and the user’s stated situation enters the prompt. Chatbot responses have been shown to vary with the gender and race associated with a user’s name [Eloundou et al. 2025]; the behavioral evidence on AI summaries concerns clicks [Chapekis \& Lieb 2025]; and whether the sources an answer cites vary with the persona of the asker has not, to the author’s knowledge, been measured. It is an open question this paper does not answer and Section 13 specifies.

\paragraph{Information needs and query fragmentation}People take questions to their social networks [Morris, Teevan \& Panovich 2010] and to question-answering sites when web search fails [Liu et al. 2012], and phrase questions differently from queries [White, Richardson \& Yih 2015]. The gap between the need a user has in mind and the request a retrieval system can match is old ground [Belkin, Oddy \& Brooks 1982; Taylor 1968], as is the observation that aggregate demand concentrates differently from individual demand [Goel et al. 2010]. The taxonomy of web search [Broder 2002] distinguishes query intent rather than query granularity. Section 8 describes a deployment in which the operative distinction was granularity: one need expressed as many distinct strings (the 381 code pages rank for 29,000 keywords on one instrument).

\paragraph{Demand in users’ own words}Customer reviews have long been mined for the product aspects people raise and the opinions they attach to them [Hu \& Liu 2004], and search volume has been used as a demand signal in its own right, for example to nowcast economic indicators [Choi \& Varian 2012]. Some studies of search sample their queries, and the queries’ volumes, from keyword research tools, and a systematic method for doing so has been proposed [Schultheiß et al. 2023]. Measurement 2 compares the two signals on one taxonomy: which topics each names, rather than how well either predicts an outcome.

\paragraph{Completeness in knowledge bases}Appendix B’s scoring function draws on the study of completeness in open-world knowledge bases [Razniewski, Suchanek \& Nutt 2016; Galárraga et al. 2017; Balaraman, Razniewski \& Nutt 2018; Luggen et al. 2019; Razniewski et al. 2024], on demand-weighted completeness [Hopkinson et al. 2018], on the discovery and classification of gaps in encyclopedias [Wulczyn et al. 2016; Piccardi et al. 2018; Redi et al. 2020] and on the question-driven drafting of encyclopedia articles from scratch [Shao et al. 2024], with Ripoll et al. [2025] giving a recent typology of content gaps for a collaborative knowledge base. Since the function is not evaluated here, Appendix B positions it against the closest of these methods.

\paragraph{Closest prior work, by measurement}For Measurement 1: comparisons of estimators with first-party analytics or with one another, on established sites at one time [Jansen, Jung \& Salminen 2022; Rasool, Thakur \& Nabi 2026; Font-Julián et al. 2026]; this paper holds one site fixed, reads back two instruments’ stored histories of one section to its first month, in a single retrospective read, and compares twelve sections of the site on one date. For Measurement 2: aspect mining of reviews [Hu \& Liu 2004] and keyword tools as a source of queries and volumes [Schultheiß et al. 2023]; this paper places the two signals on one taxonomy and compares the topics each ranks; its second comparison (Section 6.2) follows the work that treats search-engine hit counts as an instrument whose consistency must be checked before use [Thelwall 2008], and Google Trends [Hölzl, Keusch \& Sajons 2025] and Wikipedia page views [Alibudbud 2023] as instruments whose settings and access dates, and for page views the handling of crawlers and redirects, should be reported, and checks one such count against a vendor’s volume on a fixed class of entities. For Measurement 3: comparisons of cited sources across engines in industries or query sets their authors chose [Hendricks 2026; Lee 2026; Huang et al. 2026]; this paper fixes the industries by an external rule, the eleven GICS sectors, and compares eight vendor-collected surfaces.

\section{Setting}
A large US telecommunications provider publishes a reference section covering the North American Numbering Plan. \textbf{As verified on 12 October 2026 by fetching every code page linked from the section’s 52 state hubs, the section holds 381 code pages}: 374 for geographic codes assigned to US states and the District of Columbia (all 372 in service and two assigned but not yet in service), and 7 for toll-free codes (the released list of the pages gives each code’s registry location, which takes 52 values: the 50 states, the District, and the plan-wide toll-free codes); a re-fetch on 14 October 2026 found the same 381 and matched them to the numbering plan administrator’s registry of that date. All 381 return HTTP 200, are canonical, and are full pages. The only geographic codes in the plan with no page are the six assigned to US territories and 451, assigned to North Carolina on 15 September 2026, after launch, and not yet in service. The section is therefore complete over the class it set out to cover, and \textbf{there is no set of uncovered codes within the class to serve as a control}.

All code pages, the state hubs and the national hub were published together at launch; no pages were added to the section during the observation window. This is the author’s statement of the deployment’s publishing history. Table 5’s series of pages holding a position (210, 365, 391, 414, 440) is therefore the indexing of a fixed set of 434 URLs by one instrument, and not growth in the section itself.

Two page counts remain and are easy to conflate. \textbf{381} is the number of code pages the section holds. \textbf{441} is the number of URLs under the prefix holding at least one ranking position on 11 October, which is selected on the outcome; it exceeds 381 plus the 53 hub pages by seven, which is within what the instrument attributes to a prefix beyond its canonical pages. Brand Radar’s count of 455 distinct cited pages, read on 11 October and listed with the paper’s other single dashboard reads in the released data, exceeds 434 by 21; that instrument does not document how it normalizes URL variants, and the difference is reported rather than reconciled. Every fact on the pages is templated from two public-domain US government sources, the numbering plan administrator’s registry and census data. The section went live on 26 July 2026, which is day 0 throughout this paper, dated from the first third-party detection of the pages.

\subsection{Six conditions that make this deployment unrepresentative}
These are stated here, before the results, as the scope of everything that follows. All six held, and most classes in most industries satisfy none of them.

\begin{enumerate}[leftmargin=1.6em, itemsep=1.5pt, topsep=2pt, parsep=0pt]
\item \textbf{The class is finite and enumerable.} There is a fixed, published list of area codes. Coverage is decidable.
\item \textbf{A free authoritative registry supplies every fact.} No research, no writer, no subject-matter review per page.
\item \textbf{The facts are uncontested.} Two sources would not disagree about which state a code serves.
\item \textbf{The domain is top-decile authority.} Ahrefs Domain Rating 90, Semrush Authority Score 86.
\item \textbf{No comparable-authority competitor covered the class page-per-entity.} Two carriers at Domain Rating 87 and 91 had, between them, overlay notices and a single lookup page.
\item \textbf{No defender responded.} The incumbent pages were specialist directories and voice vendors at Domain Rating 13 to 83, and none mounted a visible response during the window; this is an observation, with no archived record.
\end{enumerate}

Conditions 1 to 3 make production nearly free. Conditions 4 to 6 describe an undefended position. A reader considering transfer should check all six, and should note that condition 6 in particular reads as a criterion for selecting undefended targets. That is the shape of a data void, a query space with thin or absent authoritative content that whoever arrives first fills [Golebiewski \& boyd 2018], and Section 14 treats it as an ethics question rather than a tactic.

\subsection{How the class was found}
This is the author’s account, with the dated records that bear on it. The records belong to the author’s own platform and are not released, and nothing in this subsection tests the framework’s score. The author gathered the industry’s competitors, direct and indirect, collected the topics they are found on, and connected those topics and their queries in one graph on that platform; the taxonomy of Section 6.1 was exported from the same platform. By the author’s account the class had not been visible in keyword lists read by hand and became visible in the graph: the area-code queries gathered in one node, the carriers ranked for them only with pages about other things, such as store pages, and the pages written for the topic belonged to third parties, the directories and voice vendors of condition 6. By the same account the topic had been on the content backlog for some time without being taken up, and it was taken up after the graph showed it; the author judged it evergreen and likely to add to the site’s authority on the industry’s topics. The graph also showed what the topic connects to; in the snapshot named below its sub-topics include location and international calling. The platform’s records are consistent with the account. On 4 July 2026 its build log records 1,196 area-code keywords carrying 6.7 million monthly searches being gathered into one node, and a note of the same day records a competitor’s store pages ranking for area-code lookups; a snapshot committed on 18 July shows that node with 14,164 keywords. The section launched on 26 July. This is discovery from one kind of source, keyword data joined to a list of competitors. Section 11.2 states what the deployment did not do, and Section 13 describes the test that would show whether the score ranks such opportunities better than search volume does.

\section{Measurement design}
Everything in Sections 5 to 11 is an observation. The language is chosen to keep that visible.

\subsection{Instruments, definitions and denominators}
\begin{table*}[tp]
\centering\footnotesize
\caption*{\textbf{Table 1.} Instruments, definitions, denominators and read dates for the measured quantities in Sections 5 to 10. Vendor authority scores (Domain Rating, Authority Score, URL Rating) are given with their instrument where they are used.}
\begin{tabularx}{\textwidth}{>{\hsize=0.856\hsize\RaggedRight\arraybackslash}X>{\hsize=1.305\hsize\RaggedRight\arraybackslash}X>{\hsize=0.980\hsize\RaggedRight\arraybackslash}X>{\hsize=0.859\hsize\RaggedRight\arraybackslash}X}
\toprule
\textbf{Quantity} & \textbf{Definition} & \textbf{Instrument} & \textbf{Read} \\
\midrule
Ranking positions & URL under the prefix in positions 1–100, counted once per location & Ahrefs Site Explorer, prefix mode, all locations & daily series, 11 Oct 2026 \\
Ranking keywords & distinct keywords on which any URL under the prefix ranks 1–100 & Ahrefs (all locations); Semrush (US database, and worldwide) & 11 Oct; 10 Oct (worldwide 12 Oct) 2026 \\
Pages holding a position & distinct URLs with at least one position & Ahrefs & 11 Oct 2026 \\
Estimated visits & vendor click model applied to estimated position and estimated volume & Ahrefs (worldwide); Semrush (US) & 11 Oct; 10 Oct 2026 \\
AI Overview citation (a) & keywords whose AI Overview lists a URL under the prefix as a source, over all keywords the prefix ranks on whose results show an AI Overview & Ahrefs owned-AI-Overview & 11 Oct 2026 \\
AI Overview citation (b) & US keywords with an AI Overview position & Semrush & 10 Oct 2026 \\
AI answers & stored prompts whose latest answer cites a URL under the prefix, per platform & Ahrefs Brand Radar & 11 Oct 2026 \\
AI visibility (Semrush) & mentions and cited pages in the AI-visibility panel; scoped to the domain whatever subfolder is queried (Section 5.5) & Semrush & 10 Oct; 12 Oct 2026 \\
Referring domains & live linking domains to the prefix & Ahrefs; Semrush Backlink Analytics & 11 Oct 2026 \\
Section age (Table 4) & months since the first month with organic traffic in the all-time series; left-censored at September 2015 & Ahrefs Site Explorer, prefix mode, chart interval all & 12 Oct 2026 \\
US share (Table 4) & United States share of the prefix’s estimated visits, from the traffic-by-location table & Ahrefs Site Explorer & 12 Oct 2026 \\
Monthly search volume & vendor-estimated monthly searches for a phrasing, summed over a set of phrasings & DataForSEO ranked-keyword exports, three carrier domains, US database & exported 12 to 19 Oct 2025 \\
Review mentions (Section 6) & reviews whose text matches a taxonomy label, per label; volume is the vendor’s monthly searches summed over the nodes carrying the label & DataForSEO Google reviews and keyword volume; released exact-match linker & exported 8 Sep 2026 \\
Citation incidence (Section 7) & per domain, the number of stored answers citing it (“Cited in”), summed over the hosts of a registrable domain, among the 1,000 domains found in the most answers & Ahrefs Brand Radar, US, per answer surface & 12–13 Oct 2026 \\
\bottomrule
\end{tabularx}
\end{table*}

Two properties of these instruments bear on everything that follows, and the first is the subject of Section 5.

\paragraph{They are models, not counters, and not of the same unit}Estimated visits are a click model over an estimated volume at an estimated position. Ahrefs describes its metric as an estimate of the clicks from Google that a target gets each month, summed over the keywords it ranks for from position, monthly search volume and an estimated click-through rate [Ahrefs 2026b]. Semrush describes its figure as the estimated number of visitors a domain receives from unpaid search results each month [Semrush n.d.-b], computed for its domain rankings from the click-through rate at the domain’s positions and each keyword’s monthly volume [Semrush n.d.-c]. Clicks and visitors are different units, and this paper cannot say how much of the ratios below that difference explains. No figure in this paper is a measured visit. Neither vendor states an uncertainty for its estimates, so the ratios between them are ratios of two point estimates and carry none; what the paper can show is how they range across months and sections.

\paragraph{How the figures were read}Every vendor figure but one (Semrush’s worldwide keyword count in Section 5.2, read by the author) was read from the vendors’ web interfaces by an AI browser assistant operating the author’s logged-in browser, under the author’s direction, with no API (Section 14). The monthly figures in Tables 2 and 3 are each instrument’s own history as read on 10 and 11 October, not values captured month by month; so are the checkpoint values in Table 5, read on 11 October from the Ahrefs daily series and from Brand Radar’s stored answers, as its caption states. A vendor may revise its history, and the 13 October re-read of Semrush found its completed months unrevised.

\paragraph{The Ahrefs and Semrush visit figures are not the same quantity}Ahrefs’ is worldwide; Semrush’s is the US database. The section ranks in 105 countries with 18.5\% of Ahrefs’ estimated visits from outside the US (18.6\% in the 12 October read of Table 4), so the Semrush figure covers a strict subset of the Ahrefs figure’s territory and is nonetheless between 7.5 and 27.1 times larger. Table 2 therefore reports three ratios: Semrush US over Ahrefs worldwide, which understates the gap because the larger figure covers less territory; worldwide over worldwide (8.6x to 29.9x); and US over US (9.2x to 33.3x), estimated by applying Ahrefs’ US share of its own visits, 81.5\%, read once in October, to every month.

\subsection{The query universe, and what cannot be computed on it}
The evaluation query set was not frozen before launch; the section was live before this study was designed. The outcome quantities in Sections 8 to 10 are therefore measured on universes defined after the fact: every keyword the section ranks for, and every prompt whose answer cites it.

Counts and positions on such a universe are meaningful, because the universe is the vendor’s index rather than this paper’s selection. \textbf{Shares and percentages computed on it are not}, because the denominator was selected on the outcome. Every share in this paper is one of the following, and is labeled at the point of use: a share whose denominator is outcome-selected, flagged as such and drawing no comparative conclusion; or a ratio between two instruments on the same object, which is this paper’s primary result and is not affected, because both denominators are the same prefix. The rule governs the deployment universes of Sections 5, 8 and 9. Section 7’s shares are computed on a different universe, answers that name a sector’s largest companies, which is selected separately for each answer surface but not on this deployment’s outcome; its comparisons are stated as conditional on that universe, are made within one instrument across differently selected universes, and are descriptive. The share going to the sector’s own company domains is partly fixed by that selection, and no comparison is drawn from it.

\subsection{The primary checkpoint}
Four disturbances fall in the observation window. Google’s August 2026 spam update ran from 18 to 21 August, days 23 to 26 [Google 2026]. The September 2026 spam update began on 24 September, day 60, which is also the day of the estimated-traffic peak. Ahrefs posted a notice on 29 September, day 65, that Google had changed how it serves results and that SERP data might be inconsistent from that date, which covers days 65 to 75. And an automated link network began arriving in the day-37 to day-45 interval.

\textbf{Day 37 (1 September 2026) is the primary checkpoint}: the last observation preceding the last three. It does not precede the August update, which falls in the month of the largest visits ratio (27.1x), and whose effect on this section is not estimated. Day 37 carries a cost that is stated rather than worked around: it is not one of the fifteen-day checkpoints, and the 11 October read gives four of the nine quantities for it, not all nine. The claims this paper makes at day 37 concern ranking positions, top-three positions, estimated visits and the link profile, and nothing else.

\section{Measurement 1: two instruments on one prefix}
This section is the paper’s primary result. It is reported first because its figures do not rest on the author’s judgment, though the choice of the twelve sections of Section 5.5 does: every figure in it is one of two vendors’ readings of a public URL prefix, and a reader holding both subscriptions can re-read them.

\subsection{Estimated visits: a divergence of 7.5x to 27.1x on one prefix}
\begin{table*}[tp]
\centering\footnotesize
\caption*{\textbf{Table 2.} Estimated monthly visits for the prefix from two instruments. Semrush worldwide in parentheses. The June to September rows are each dashboard’s figure for the completed month, labelled “total” in the rows; how a month’s figure is formed was not recorded (fourth note below). The October row is each instrument’s current monthly run-rate as read on 10–11 October, not eleven days of accumulated traffic, and is marked accordingly. Every row is the instrument’s history as read on that date. The first ratio divides Semrush US by Ahrefs worldwide. The second divides worldwide by worldwide. The third divides Semrush US by Ahrefs worldwide times 81.5\%, Ahrefs’ US share read on 11 October; it assumes that share held in every month and is an estimate until Ahrefs’ US-only monthly figures are read.}
\begin{tabularx}{\textwidth}{>{\hsize=1.043\hsize\RaggedRight\arraybackslash}X>{\hsize=1.043\hsize\RaggedRight\arraybackslash}X>{\hsize=0.828\hsize\RaggedRight\arraybackslash}X>{\hsize=1.043\hsize\RaggedRight\arraybackslash}X>{\hsize=1.043\hsize\RaggedRight\arraybackslash}X>{\hsize=1.043\hsize\RaggedRight\arraybackslash}X>{\hsize=0.957\hsize\RaggedRight\arraybackslash}X}
\toprule
\textbf{Month} & \textbf{Days live} & \textbf{Ahrefs, worldwide} & \textbf{Semrush, US (worldwide)} & \textbf{Semrush US / Ahrefs worldwide} & \textbf{Worldwide vs worldwide} & \textbf{US vs US (estimated)} \\
\midrule
June 2026 & 0 & 0 & 0 (0) & n/a & n/a & n/a \\
July 2026 (total) & 6 & 3,862 & 100,012 (103,045) & 25.9x & 26.7x & 31.8x \\
August 2026 (total) & 31 & 32,281 & 876,153 (963,937) & 27.1x & 29.9x & 33.3x \\
September 2026 (total) & 30 & 150,113 & 1,534,362 (1,765,077) & 10.2x & 11.8x & 12.5x \\
October 2026 (run-rate) & 11 (78 in all, days 0 to 77) & 215,577 & 1,610,284 (1,858,297) & 7.5x & 8.6x & 9.2x \\
\bottomrule
\end{tabularx}
\end{table*}

\begin{figure}[t]
\centering
\includegraphics[width=\columnwidth]{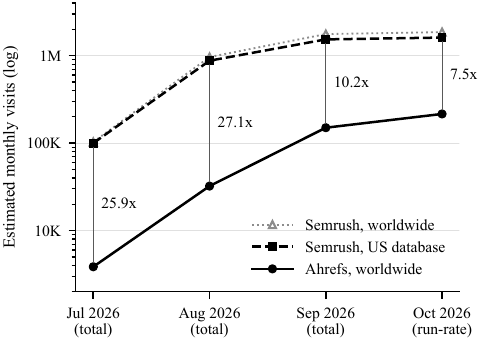}
\caption*{\textbf{Figure 2.} Estimated monthly visits for the area-code prefix from two instruments, on a logarithmic axis so that the vertical gap is the ratio. Ahrefs Site Explorer (prefix mode, all locations) and Semrush Domain Overview (subfolder mode, US database and worldwide), read 10 and 11 October 2026. June to September are each dashboard’s figure for the completed month; October is each instrument’s current monthly run-rate. The ratio labels are Semrush US over Ahrefs worldwide, as in Table 2’s first ratio column.}
\end{figure}

Four things about this table and Figure 2.

\paragraph{The instruments disagree about change as well as level}A constant difference in level would cancel in a comparison of growth within one instrument, so the question for an applied study is whether the two agree on how much a section grew. From July to August they did (8.4x on Ahrefs, 8.8x on Semrush US); from August to September Ahrefs’ estimate grew 4.65x and Semrush’s 1.75x, and into October’s run-rate 1.44x against 1.05x. A month’s lift measured on one instrument can therefore differ from the same lift on the other by a factor of more than two.

\paragraph{It is not a territory artifact}Semrush’s US-only figure exceeds Ahrefs’ worldwide figure throughout, and the section draws 18.5\% of its Ahrefs-estimated visits from outside the US. Matching territory widens the ratio rather than closing it: worldwide against worldwide it runs from 8.6x to 29.9x, and US against US, estimated, from 9.2x to 33.3x (Table 2).

\textbf{It narrows after the second month on this prefix}, from 27.1x to 10.2x to 7.5x. Section 5.5 tests whether that narrowing is a property of section age across twelve prefixes and finds that it is not; Section 5.6 takes up what the narrowing can and cannot be attributed to.

\textbf{What a month’s figure is was not recorded, and in a section’s first months it matters.} The reads took each dashboard’s monthly view as displayed and did not record how either vendor forms a month’s point. The first instrument’s own daily series (Table 5) reads 22,016 on day 15 and 51,517 on day 30 against 32,281 for August, and 92,897 on day 45 and 244,041 on day 60 against 150,113 for September, so its monthly point sits inside a month in which its daily estimate more than doubled, as an average over the month would. If the second instrument’s monthly point describes a date late in the month, part of the August and September ratios is a difference of dates and not of instruments: against the first instrument’s values on day 30 (25 August) and day 60 (24 September) they are 17.0x and 6.3x, not 27.1x and 10.2x (18.7x and 7.2x worldwide against worldwide). The growth comparison above carries the same caution. The October row and the cross-section of Section 5.5 are read on one date for both instruments and do not. The range 7.5x to 27.1x is therefore the range of the two monthly views as displayed, and the bound of Section 5.6 holds at 17.0x with a factor of 4.1 in place of 5.2.

\subsection{Ranking keywords: a factor of two, and three like for like}
On the count of distinct keywords for which any URL under the prefix ranks in positions 1–100, Ahrefs reports 29,000 for all locations and Semrush 58,586 for the US database, read a day apart: a factor of 2.0, again with the larger figure from the smaller territory. Like for like in territory (Section 5.6 notes that how Semrush aggregates its worldwide count is not documented), Semrush’s worldwide count, read by the author in its interface on 12 October, is 92.3K (the interface rounds to the hundred), 3.2 times Ahrefs’ 28,900 for all locations on the same day (Table 4); matching territory again widens the gap. Pages holding at least one position are 441 in Ahrefs.

\subsection{AI answers and AI Overviews: two definitions, converging}
\paragraph{AI-answer counts are not comparable at the prefix level}Brand Radar’s 17,206 stored answers citing the section and Semrush’s 158,100 mentions read on 10 October differ by a factor of 9.2, but that ratio is a scope mismatch, not a divergence: Semrush’s companion count of 29,600 cited pages is 68 times the 434 URLs the section holds, so its figures are scoped to the domain rather than the prefix, as Section 5.5 confirms by observation. No comparison is drawn from those counts, nor about the ordering of AI answer platforms, whose per-platform breakdown did not sum to its total (Appendix A).

What can be compared on this prefix is reported above and below: visits, ranking keywords, referring domains and AI Overview counts, each with an instrument, a definition, a territory and a read date in Section 4.1. The Ahrefs figure of 17,206 stored answers is reported in Table 5 as a single-instrument count and is not paired with a second reading.

For AI Overview citation the vendors use different operational definitions, given in Section 4.1. Definition (a) is a ratio over AI Overviews the prefix’s keywords actually appear on; definition (b) is a count of US keywords with an AI Overview position. Both are reported as “AI Overview visibility” in their respective products.

\begin{table*}[tp]
\centering\footnotesize
\caption*{\textbf{Table 3.} AI Overview citation under two vendor definitions, same prefix, same months.}
\begin{tabularx}{\textwidth}{>{\hsize=1.075\hsize\RaggedRight\arraybackslash}X>{\hsize=1.109\hsize\RaggedRight\arraybackslash}X>{\hsize=1.080\hsize\RaggedRight\arraybackslash}X>{\hsize=0.824\hsize\RaggedRight\arraybackslash}X>{\hsize=0.911\hsize\RaggedRight\arraybackslash}X}
\toprule
\textbf{Month} & \textbf{(a) Ahrefs: cited / all AIO keywords} & \textbf{(a) as \%, 95\% Wilson CI} & \textbf{(b) Semrush count} & \textbf{(b) / (a) numerator} \\
\midrule
July 2026 & 47 of 284 & 16.5\% [12.7, 21.3] & 138 & 2.94x \\
August 2026 & 851 of 6,749 & 12.6\% [11.8, 13.4] & 2,221 & 2.61x \\
September 2026 & 3,380 of 22,665 & 14.9\% [14.5, 15.4] & 8,187 & 2.42x \\
October 2026 (run-rate) & 7,455 of 37,238 & 20.0\% [19.6, 20.4] & 10,018 & 1.34x \\
\bottomrule
\end{tabularx}
\end{table*}

\begin{figure*}[t]
\centering
\includegraphics[width=\textwidth]{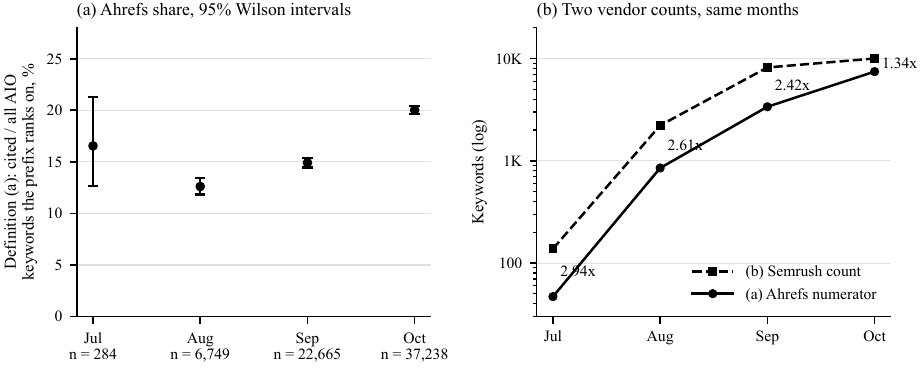}
\caption*{\textbf{Figure 3.} AI Overview citation under two vendor definitions, same prefix, same months. (a) Ahrefs definition (a) as a share of all AI Overview keywords the prefix ranks on, with 95\% Wilson intervals; the denominators grow from 284 to 37,238. (b) The two vendor counts, Semrush keywords with an AI Overview position against the Ahrefs numerator, on a logarithmic axis, with their ratio labeled. Sources as in Table 3.}
\end{figure*}

The ratio between the two counts falls from 2.94 to 1.34 over four months (Figure 3b). Two cautions on reading it. The comparison is between a count (b) and the numerator of a ratio (a), and the two vendors’ territories differ, so what the column measures is partly a territory difference and not purely a definitional one. And the series has three intervals, so “converging” describes the observed sequence and is not a trend estimate. The July row rests on 47 events over six days live, and its interval is correspondingly wide. A study reporting an AI Overview visibility figure on one vendor’s definition, for a section of a given age, is not reporting a quantity comparable to the same figure from the other vendor, and on this prefix the size of the incomparability changed over its first four months.

The (a) column’s percentage is the one proportion in this paper whose denominator grows by two orders of magnitude across the series, from 284 to 37,238, and it is not monotone. Its denominator is outcome-selected: it counts only AI Overviews on keywords the prefix already ranks on. It is reported because numerator and denominator come from one instrument on one date, and no trend is drawn from it. Its Wilson interval [Wilson 1927] treats the month’s AI Overview keywords as draws from a stable process: it shows binomial noise in the ratio, not the instrument’s error.

\subsection{Referring domains: 11.4x, and disagreement about composition}
Both instruments report live referring domains to the same prefix on the same day, and both are in the Section 4.1 table with a definition and a read date, which makes this the cleanest paired comparison in the study after estimated visits.

Ahrefs reports 913 referring domains to the prefix; Semrush Backlink Analytics indexes 80 for the subfolder: a factor of 11.4. They also disagree about what those domains are. Ahrefs puts 61.9\% on .info, 26.4\% on .asia and 10.2\% on .xyz; Semrush puts 75\% on .xyz and 88\% at Authority Score 0 to 10. Both descriptions agree the domains are low-quality automated networks and disagree about their composition by a wide margin, which matters for any study classifying link profiles from one vendor’s TLD distribution.

\begin{figure}[t]
\centering
\includegraphics[width=\columnwidth]{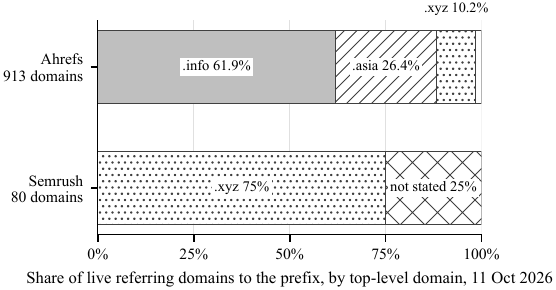}
\caption*{\textbf{Figure 4.} Composition of live referring domains to the prefix by top-level domain, as reported by each instrument on 11 October 2026. Ahrefs: 913 domains; Semrush Backlink Analytics: 80 domains, of which the instrument reports 75\% on .xyz and does not break down the remainder.}
\end{figure}

Figure 4 shows the two compositions side by side. A third figure from the first instrument complicates even its own count: its daily series gives 909 at day 75 against 913 on 11 October, and its new-and-lost report for September gives 917 new against 95 lost, which with nineteen at day 37 implies about 841 at the end of September, an earlier date; the three figures are consistent with a count still rising. Section 10 sets out why no claim here turns on the exact count.

\subsection{Twelve prefixes on one domain: the divergence is prefix-specific, with no detectable relationship to age}
Sections 5.1 to 5.4 are one prefix. That the narrowing of the visits ratio there reflects section age, each vendor’s index filling in at its own rate, was stated as a hypothesis, with a cross-prefix test, before the test was run (Appendix A). The test was run on 12 October 2026, on both instruments, for twelve sections of the same domain: the area-code section and eleven others chosen by the author to span ages from under two years to the instrument’s horizon; the selection was purposive, not random, and the results describe these twelve. Every figure below is a dashboard read on one date. The raw values, the read method, the age proxy and the United States share of each prefix’s estimated visits are in the released data files, and \textbf{every statistic quoted in this section is printed by a script}: \texttt{analyze.py}, released beside them, or \texttt{instrument\_extra.py}, in the source bundle, which prints the ordering correlation’s significance and intervals, the spread, the size-banded agreement, the ratio columns of Table 4 and the remaining per-prefix figures.

\begin{table*}[tp]
\centering\footnotesize
\caption*{\textbf{Table 4.} Twelve URL prefixes on one domain, both instruments, read 12 October 2026. Age is months since the first month with organic traffic in the Ahrefs all-time series, read to the nearest month or quarter; the series begins in September 2015, so the oldest age (/es-us/local/) is left-censored at that horizon and the next two lie within four months of it. Ahrefs visits are worldwide and Semrush visits are the US database; the “US share” column is the United States share of Ahrefs’ estimate from its traffic-by-location table, and the matched ratio divides Semrush US by Ahrefs US-only. The keyword ratio is Semrush over Ahrefs, like the visits ratios; the referring-domain ratio is Ahrefs over Semrush. Referring-domain and keyword counts are the 12 October reads and differ slightly from Sections 5.2 and 5.4’s 10–11 October reads (29,000 and 58,586 keywords; 80 Semrush domains, 11.4x).}
\begin{tabularx}{\textwidth}{>{\hsize=1.534\hsize\RaggedRight\arraybackslash}X>{\hsize=0.739\hsize\RaggedRight\arraybackslash}X>{\hsize=1.100\hsize\RaggedRight\arraybackslash}X>{\hsize=1.534\hsize\RaggedRight\arraybackslash}X>{\hsize=1.317\hsize\RaggedRight\arraybackslash}X>{\hsize=0.739\hsize\RaggedRight\arraybackslash}X>{\hsize=1.172\hsize\RaggedRight\arraybackslash}X>{\hsize=0.811\hsize\RaggedRight\arraybackslash}X>{\hsize=0.883\hsize\RaggedRight\arraybackslash}X>{\hsize=0.739\hsize\RaggedRight\arraybackslash}X>{\hsize=0.811\hsize\RaggedRight\arraybackslash}X>{\hsize=0.883\hsize\RaggedRight\arraybackslash}X>{\hsize=0.739\hsize\RaggedRight\arraybackslash}X}
\toprule
\textbf{Prefix} & \textbf{Age (mo)} & \textbf{Ahrefs visits} & \textbf{Semrush visits (US)} & \textbf{Ratio, unmatched} & \textbf{US share} & \textbf{Ratio, matched} & \textbf{Ahrefs kw} & \textbf{Semrush kw} & \textbf{Kw ratio} & \textbf{Ahrefs RD} & \textbf{Semrush RD} & \textbf{RD ratio} \\
\midrule
/local/area-codes/ & 3 & 221,000 & 1,600,000 & \textbf{7.24} & 81.4\% & \textbf{8.89} & 28,900 & 59,000 & 2.0 & 913 & 81 & \textbf{11.3} \\
/product-compare/ & 21 & 5,500 & 4,100 & 0.75 & 69.6\% & 1.07 & 716 & 2,300 & 3.2 & 187 & 41 & 4.6 \\
/local/bundles/ & 36 & 25,600 & 5,200 & 0.20 & 99.2\% & 0.20 & 2,100 & 33,800 & 16.1 & 115 & 194 & 0.6 \\
/device-compare/ & 39 & 26,700 & 26,400 & 0.99 & 82.6\% & 1.20 & 2,400 & 12,600 & 5.2 & 47 & 45 & 1.0 \\
/local/cell-phones/ & 57 & 37,300 & 135,100 & 3.62 & 98.8\% & 3.67 & 3,600 & 37,500 & 10.4 & 144 & 225 & 0.6 \\
/support/ & 93 & 1,900,000 & 1,957,880 & 1.03 & 98.7\% & 1.04 & 38,700 & 351,854 & 9.1 & 10,600 & 7,056 & 1.5 \\
/local/fiber/ & 95 & 102,000 & 47,300 & 0.46 & 99.8\% & 0.46 & 5,400 & 31,900 & 5.9 & 258 & 245 & 1.1 \\
/buy/ & 95 & 644,000 & 1,014,124 & 1.57 & 98.5\% & 1.60 & 19,800 & 206,365 & 10.4 & 7,400 & 3,246 & 2.3 \\
/local/internet/ & 114 & 243,000 & 220,600 & 0.91 & 93.6\% & 0.97 & 13,300 & 79,700 & 6.0 & 1,700 & 516 & 3.3 \\
/stores/ & 129 & 312,000 & 2,051,269 & 6.57 & 99.4\% & 6.61 & 14,700 & 306,952 & 20.9 & 4,100 & 3,527 & 1.2 \\
/deals/ & 132 & 214,000 & 219,600 & 1.03 & 99.3\% & 1.03 & 5,300 & 32,000 & 6.0 & 1,600 & 2,042 & 0.8 \\
/es-us/local/ & 133 & 8,900 & 2,200 & 0.25 & 42.6\% & 0.58 & 1,300 & 4,800 & 3.7 & 284 & 62 & 4.6 \\
\bottomrule
\end{tabularx}
\end{table*}

\begin{figure*}[t]
\centering
\includegraphics[width=\textwidth]{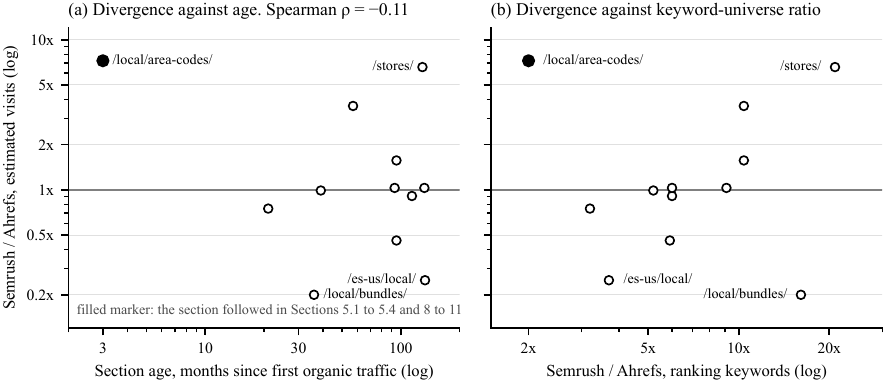}
\caption*{\textbf{Figure 5.} Twelve URL prefixes on one domain, both instruments, 12 October 2026. Each marker is one prefix; the filled marker is the area-code section. (a) The Semrush over Ahrefs ratio on estimated visits, unmatched for territory, against section age in months since the first month with organic traffic in the Ahrefs all-time series, both axes logarithmic; the horizontal line is a ratio of one. (b) The same ratio against the Semrush over Ahrefs ratio on ranking keywords. Data in Table 4 and in the released file.}
\end{figure*}

Six things follow (Figure 5); the first is that the age hypothesis fails.

\textbf{The cross-section shows no detectable relationship between the ratio and section age.} Spearman’s rank correlation between age and the unmatched visits ratio is −0.11 over twelve prefixes, with a Fisher 95\% interval of −0.64 to +0.49 and a permutation p of 0.73; territory-matched it is −0.28; excluding the area-code section it is +0.15. Twelve prefixes give about 38\% power to detect a correlation of 0.5 (37.8\% by the Fisher approximation), so this is not a demonstration that no age effect exists. What it does establish is narrower and sufficient: the ratio does not converge on a common value as sections mature, because the seven sections aged between seven and eleven years sit anywhere from 0.25x to 6.57x unmatched (0.46x to 6.61x matched). The age hypothesis, that each vendor’s index fills in at its own rate and the ratio settles as a section ages, is therefore withdrawn, and the narrowing of the area-code ratio from 27.1x to 7.5x over three months is reported as an observation on one section and not attributed. Only one prefix in the cross-section is under twenty-one months old, so the first months of a section’s life are not characterized here beyond that one case.

\paragraph{The ratio is prefix-specific, and not one-directional}Unmatched for territory, Semrush’s estimate runs from one fifth of Ahrefs’ (/local/bundles/, 0.20x) to seven times it (/local/area-codes/, 7.24x), a spread of 1.55 orders of magnitude, with a median of 1.01 and six of twelve prefixes below one. Matched for territory the spread is wider (0.20x to 8.89x), the median is 1.06, and four of twelve are below one, because two of the six sub-one prefixes cross one once their non-US traffic is removed (/product-compare/, 69.6\% US, from 0.75x to 1.07x; /device-compare/, 82.6\% US, from 0.99x to 1.20x); /es-us/local/, at 42.6\% US, moves from 0.25x to 0.58x and stays below one. The territory objection therefore changes the count and not the conclusion: on the same domain, on the same day, the vendors’ estimates differ by factors from well below one to well above it, and no single correction factor maps one onto the other. Taken across the twelve, the unmatched ratio has a geometric mean of 1.13 (bootstrap 95\% interval 0.62 to 2.12) and a geometric standard deviation of 3.1 (1.9 to 4.1), a normal range on the log scale of about 0.12x to 10.6x; with twelve purposively chosen sections this describes them and is not a prediction for another section.

\paragraph{The vendors agree on the ordering of sections and disagree on their magnitudes}Across the twelve prefixes, the rank correlation between the two instruments’ visit estimates is +0.94 (bootstrap 95\% interval over prefixes +0.70 to +1.00; p < 0.001; Kendall’s tau +0.82, interval +0.57 to +1.00; 60 of 66 pairs concordant). The vendors agree on which section is larger in 60 of 66 pairs; what has unknown sign and magnitude is the \textbf{correction factor} between the two instruments, which no domain-level constant captures. The agreement depends on how different two sections are in size: of the pairs whose sizes, taken as the geometric mean of the two estimates, differ by more than a factor of three, all 51 are ordered alike; of the 15 pairs within a factor of three, 9 are. Read by the factor on each instrument instead, all 40 pairs more than three times apart on both are ordered alike, as are 16 of the 18 more than three times apart on exactly one (the other two are six to seven times apart on one instrument and about equal on the other) and 4 of the 8 within a factor of three on both (\texttt{instrument\_extra.py}, section 10). The consequence for practice is restated in Section 5.6 in that form.

\paragraph{Keyword universes diverge consistently; visit estimates do not follow them}Semrush reports more ranking keywords than Ahrefs on every prefix, by between 2.0x and 20.9x. The rank correlation between the keyword ratio and the visits ratio is +0.17 over all twelve, and +0.52 with the area-code section excluded; the second figure is reported because the first is the one the paper is entitled to use and a reader should see both. /local/bundles/ has the second-largest keyword ratio (16.1x) and the smallest visits ratio (0.20x). Across sections, then, the keyword-universe ratio has no detectable relationship with the visits ratio (two-sided p = 0.60 over all twelve and 0.10 without the area-code section; \texttt{instrument\_extra.py}), so the index-coverage mechanism behind the age hypothesis finds no support here; how much of the area-code section’s own divergence the keyword universe accounts for depends on the territory pairing (Section 5.6).

\paragraph{The section this paper studied is where the instruments disagree most}It has the highest visits ratio (7.24x unmatched, 8.89x matched; /stores/ is next at 6.57x) and the highest referring-domain ratio (11.3x) of the twelve. The referring-domain outlier has a visible cause: Ahrefs indexed the automated link network described in Section 10 and Semrush largely did not. The visits outlier does not have a visible cause here, and two candidates are stated as hypotheses only: the section’s query distribution is unusually long-tailed (Section 8), which click models over estimated volume may treat very differently; and the section was three months old, which the cross-section cannot characterize.

\paragraph{A measurement result about one of the instruments, established in passing}Semrush’s AI-visibility panel reported 158.3K mentions and 29.5K cited pages for every one of the twelve subfolders, identically, on 12 October (Section 5.3 quotes the 10 October read of 158,100 and 29,600). Those figures are domain-scoped whatever prefix is queried. This confirms by observation what Section 5.3 infers from the page count: the two instruments’ AI-answer counts cannot be compared at the prefix level.

\paragraph{Limits}Twelve prefixes, one domain, one read date; the age proxy is a chart read, left-censored at September 2015 for one prefix and close to that horizon for two more; one prefix under twenty-one months; the territory match uses each prefix’s US share as the instrument reports it. A second domain and a second read date are the next step, and neither needs this provider.

\subsection{What follows for applied search research}
\paragraph{What divergence licenses, and what it does not}There is no ground truth here. Neither vendor’s figure can be checked against a counted visit, because the provider’s analytics were not available for this study. Divergence is therefore not error. What two readings A and B of one quantity with true value T license is a bound on the larger error and nothing more:

max(|A − T|, |B − T|) ≥ |A − B| / 2.

At the August ratio of 27.1x, the two readings cannot both lie within a factor of 5.2, the square root of 27.1, of the true value, so at least one of the two estimates lies a factor of five or more from the quantity both are estimating, and this study cannot say which. This treats the two figures as readings of one quantity, which is an approximation: they differ in territory, which only strengthens the bound, and in unit (Section 4.1), which it does not carry. In absolute terms the bound is larger: the larger of the two errors is at least 13 times the smaller reading. That is the whole of the inference, and it is a property of the pair, not a finding about either vendor: a reader who compares such estimates across instruments should know that on a two-month-old section they can stand a factor of five or more apart, without the comparison itself showing which is closer.

\paragraph{What the narrowing does and does not support}The ratio falls 27.1 → 10.2 → 7.5 over three months on this prefix. Section age, with index coverage as the mechanism, was the hypothesis; Section 5.5 tested it across twelve prefixes and found no detectable relationship between age and the ratio. Within this one prefix the narrowing is also confounded: section age is collinear with calendar time, and the window contains the September 2026 spam update, the arrival of several hundred automated referring domains, and a vendor data-reliability notice covering two of the four months. The narrowing is reported as an observation on one section and is not attributed.

Where it can be checked, the index-coverage term is not negligible, and its size depends on the territory pairing. With Semrush’s US figures against Ahrefs’ worldwide ones, read on 10 and 11 October, the divergence decomposes as

7.47x = 2.02x (keyword universe) × 3.70x (estimated visits per ranking keyword),

so the keyword universe accounts for about 35\% of the divergence in log terms. With both read worldwide (keywords read on 12 October), it is 8.62x = 3.19x (keyword universe) × 2.70x (per keyword), and the keyword universe accounts for about 54\%. How Semrush aggregates its worldwide keyword count across national databases is not documented in the report, so the decomposition is reported both ways and no claim is made about which component is larger.

\paragraph{Why the headline ratios understate the disagreement}The natural objection to Table 2 is territory: Semrush’s figure is US and Ahrefs’ is worldwide, so the comparison is unfair to Ahrefs. It is, and in the direction that makes the reported ratios conservative. Matching worldwide to worldwide gives 8.62x for October and 26.7x to 29.9x in the first two months (Table 2). Matching US to US, deflating the Ahrefs figure by the 18.5\% non-US share of its own estimated visits, gives \textbf{9.17x} for October against the 7.5x unmatched. Applying the same share to July, August and September gives 31.8x, 33.3x and 12.5x (Table 2); the share was read once, so those three are estimates that assume it held. The headline range keeps the unmatched pairing because it is the conservative one; every matched pairing is larger.

\paragraph{Consequence for practice}The design this bears on is the standard one in applied content work: publish something, read one dashboard, compare it against the rest of the site or against a competitor. Section 5.5 is the finding that bears on it most directly. Across twelve sections of one domain the vendor ratio on estimated visits spans 0.20x to 7.24x (0.20x to 8.89x territory-matched), so no domain-level calibration factor can convert one vendor’s figure into the other’s. The vendors agree on which of two sections is larger in 60 of 66 pairs, so a single-vendor ranking of sections agrees in direction with the other instrument’s when sections differ several-fold on both instruments (all 40 pairs more than three times apart on both) and less often when they are close (9 of the 15 pairs within a factor of three on the geometric mean of the two estimates); with no ground truth, agreement between the two is all that can be shown; measured on one instrument alone, two pairs six to seven times apart were ordered the other way on the second; the magnitude of any difference, and any ratio between a section’s figure on one vendor and another section’s on the other, can differ by a factor of five below to nearly nine above. Growth measured on the two instruments also differed, by a factor of 2.7 in one month (Section 5.1). Both instruments remain the appropriate tools for ranking sections and for directional comparison; this result concerns the comparability of magnitudes across instruments, not the usefulness of either. It bears on the framework’s outcome layer (Figure 1b): a content program that reads its results on one platform cannot take the magnitudes at face value. Whether demand that never reaches a search box differs in topic and form is the framework’s separate hypothesis, which Measurement 2 begins to test and Section 13 specifies. Reporting both makes the range of estimates visible; reporting one does not. Appendix E states these consequences as a reporting checklist, with an accounting identity for the gap between two instruments.

\paragraph{The limits of this result}One domain, one vendor pair, no ground truth; the longitudinal series is one prefix over 77 days and the cross-section is twelve prefixes on one date. What is established is that disagreements of up to an order of magnitude occur in both directions between sections of one site, from a factor of five below to nearly nine above, that no relationship with section age is detectable across twelve sections, only one of them under twenty-one months old, and that no relationship with keyword-universe size is detectable across them. What is not established is which instrument is closer to the truth for any section, or what determines the ratio. Section 13 sets out the next step: a second domain and a second read date.

\section{Measurement 2: reviews and a counted public signal against keyword volume}

\subsection{Reviews against keyword volume on one taxonomy}
The framework’s premise can be tested wherever a first-party demand surface and a tool-derived one describe the same industry. A separate dataset in the same industry allows a first run. It holds 47,485 Google reviews of 379 retail locations of major US telecom carriers, obtained through DataForSEO under the author’s own license and exported on 8 September 2026, of which 37,193 carry text, at 378 of the locations. The rule by which the locations were chosen was not recorded. The export holds between 1 and 201 reviews per location (median 145) and dates reviews only relatively (for example “a year ago”), so the period they cover is not known; where owner responses name a carrier, at 26 locations, they name five different carriers. It also holds a 42,994-node telecom taxonomy built by the author’s clustering engine, on the author’s own platform, from 1,898,584 keywords (the export labels 1,895,189 of them “corpus”, 2,781 ranked keywords of three carrier domains and 614 authored), each node carrying DataForSEO monthly search volume. The licensed export contains reviewer display names; they were excluded from the analysis and are held only in the author’s private copy of the export, and only per-label counts are reported. Reviews were linked to taxonomy labels with the released exact-match linker (\texttt{TokenLinker}), so every attribution is auditable from the label. A label is linkable if it has at least two words or at least six characters and is not one of ten generic terms the script lists (among them “phone”, “store”, “service” and “internet”). The taxonomy repeats generic labels under many parents, so a label is one linking unit and its volume is the sum over the nodes that carry it. Of the 37,193 reviews, 20,663 (55.6\%) linked to at least one of 27,817 linkable labels, at 2.25 labels per linked review, and 1,008 labels were mentioned at least once.

At the level of the taxonomy’s 17 top-level branches the two signals agree only roughly, and size the branches very differently. The taxonomy’s volumes are hierarchical, so each branch’s volume is taken from its own top node, and each label’s mentions are split across the branches where the label occurs; the rank correlation between the share of review mentions and the share of search volume is then +0.75 (bootstrap 95\% interval over branches +0.44 to +0.90; p < 0.001, 17 branches; Figure 6b), and +0.47 (+0.01 to +0.78; p = 0.06) if each label’s mentions go to the branch holding most of its nodes, with ties broken by the script’s fixed order (59 of the 1,008 labels tie; under random tie-breaks the figure has a median of +0.50, and 95\% of draws lie between +0.42 and +0.55; \texttt{m2\_tiebreak\_check.py}); the branch-level agreement therefore depends on how a label that occurs in several branches is attributed. An earlier computation gave +0.78; it summed nested volumes over only the labels that reviews mention, which selects the volume measure on the outcome, and it is superseded. The shares differ widely: carriers draw 66.8\% of review mentions and 14.9\% of search volume, devices 11.0\% and 32.7\%, broadband 4.0\% and 18.1\%; the reviews are of carrier stores, which may itself raise the carriers’ share of mentions. At the level of topics the two barely agree. Across the 1,008 labels that at least one review mentions, the rank correlation between review mentions and monthly volume is +0.18 (bootstrap 95\% interval over labels +0.12 to +0.25; Kendall’s tau +0.13); counting the 26,809 linkable labels that no review mentions as zero, it is +0.27 (+0.26 to +0.28). It is +0.28 among the 213 labels with at least twenty mentions, and +0.16 among the 106 labels mentioned at least twenty times in one- and two-star reviews (Figure 6a). Labels share reviews and nest in one taxonomy, so they are not independent draws: these intervals, and the p-values below 0.001 the scripts print, are too narrow, and the label-level figures are descriptive. A bootstrap over reviews does not repair this here, because label counts are small integers and resampling or reweighting reviews shifts their ties, and the rank correlation with them (\texttt{supplement\_r10.py}, section 12). Of the 1,008 labels, 367 (36.4\%) carry under 1,000 monthly searches and 106 (10.5\%) carry none. A label’s volume is summed over the nodes that carry it, which can include nested nodes; computed instead on each node’s own volume (its volume less its children’s), the two label-level correlations are +0.16 and +0.26. Several variants are reported (attribution rule, nested or own volumes, mentioned or all labels, and, after the link check, multi-word labels only); none is adjusted for multiple comparisons, and the multi-word variants are post hoc. The per-label counts behind every correlation in this section are in the source bundle in anonymized form, without label names and with volumes as ranks, with a script that recomputes them (Section 14).

\paragraph{How good the links are}Two blind annotators, separate AI agents on different models given only a random sample and a fixed protocol and none of the hypotheses, judged 200 links drawn uniformly from the 46,517: 65\% to 67\% are correct (Wilson 95\% intervals about 58\% to 73\%), and the two agree on 95.5\% of the items (Cohen’s kappa 0.90). The errors lie almost entirely in single-word labels, which are right about half the time (for example “Thanks”, “People”, “Customers” and place names that are also first names), while multi-word labels are 95\% to 96\% right. Restricted, after this check, to multi-word labels, the branch-level correlation is +0.91 (bootstrap 95\% interval +0.74 to +0.98) with split attribution and +0.65 with modal attribution (at the top of its range under random tie-breaks, median +0.53, 95\% of draws +0.44 to +0.645), the label-level correlation over mentioned labels is +0.24 (+0.16 to +0.33), and carriers still draw 56\% of review mentions against 15\% of search volume. The conclusions about topics and about shares hold; the branch-level rank agreement is much stronger when the links are cleaner, so what survives at that level is the difference in shares. The annotators are models, not people; a check of the same sample by a person is outstanding.

The complaint the author’s practice had identified is the cleanest case. Of 6,620 one- and two-star reviews, 924 (14.0\%, Wilson 95\% interval 13.1\% to 14.8\%) match a pattern for waiting (wait, waited, waiting, long line, hours in line), against 1.9\% (1.8\% to 2.1\%) of five-star reviews. The same annotators judged 100 of the 924 matches: 83\% to 85\% are complaints about waiting (Cohen’s kappa 0.93), and about nine in ten of those concern waiting in the store, so roughly 12\% of one- and two-star reviews complain about a wait, most of them in a store. The whole taxonomy, built from 1.9 million keywords, holds 11 labels containing “wait”, with 622 searches a month between them, and none concerns waiting in a store (at the level of keywords, 532 of the 1,898,584 match the waiting pattern, with 114,329 searches, 0.01\% of the file’s volume, and nine name a store, shop or location, with 205 searches, some of them about an app store; \texttt{waiting\_keywords\_check.py}): they cover call waiting, waitlists, activation and unlock delays, call-center wait times and a campaign name; the one that reviews mention carries 12. Eight labels name a remedy, appointments, queues or callbacks, with 13,170 searches a month between them, most of it for appointment reminders; the label for scheduling a store appointment carries none, and no label concerns check-in. A complaint named in about one in eight one- and two-star reviews, though no script ranks it against other complaints, is close to invisible in this taxonomy’s volumes, and what it calls for is a product or service change, such as appointment booking or digital check-in, rather than a page.

\begin{figure*}[t]
\centering
\includegraphics[width=\textwidth]{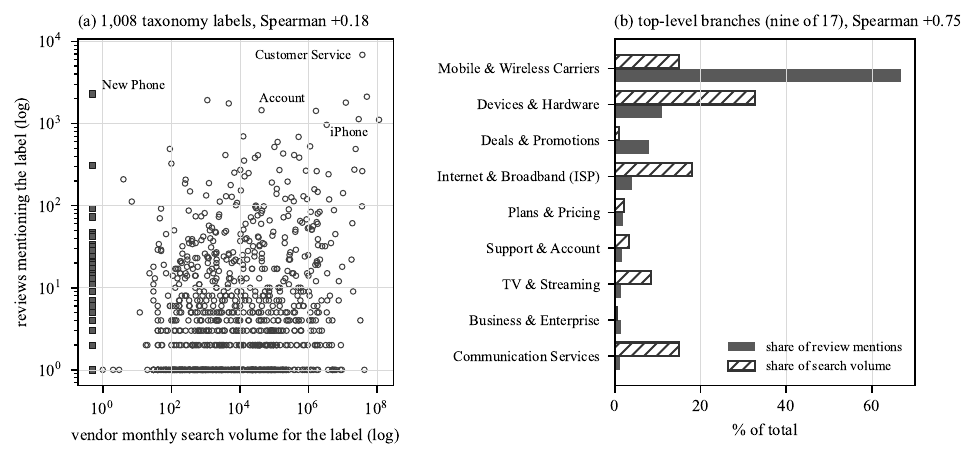}
\caption*{\textbf{Figure 6.} First-party demand against tool-derived demand on one industry taxonomy: 37,193 Google reviews of 378 retail locations of major US telecom carriers, linked to a 42,994-node taxonomy carrying vendor monthly search volume (both sourced through DataForSEO). (a) Each marker is one of the 1,008 taxonomy labels mentioned in at least one review; filled squares carry zero reported volume. (b) Share of review mentions against share of search volume for the nine largest of the 17 top-level branches. Only aggregate counts are shown.}
\end{figure*}

This is a first run, not the study specified in Section 13. It uses one first-party surface on one platform in one industry, exact-match links of which about two in three are correct, relative review dates, and a taxonomy whose labels are themselves search phrasings, which biases the comparison toward agreement. The reviews describe store locations rather than the area-code section of Sections 5 and 8 to 11, so this result tests the framework’s demand premise and not the deployment. Within those limits it is consistent with the framework’s narrowest claim: on one industry, what people write in their own words and what a volume tool counts agree on the broad areas of demand to a degree that depends on attribution and link quality, size them very differently and barely agree on topics; agreement among complaint labels (+0.16) is lower than among all frequent labels (+0.28), a difference these data do not test.

\subsection{A public request count against the vendor’s volume on the 372 area codes}
The reviews above are one first-party surface. A second comparison uses a demand signal that is tallied from request logs rather than estimated from a model, and that anyone can re-read to the digit. Every one of the 372 geographic area codes in service at the section’s launch has an English Wikipedia article, alone or shared with the codes of its overlay complex (240 articles: 125 about one code, 102 about two, 13 about three to five), and the Wikimedia Foundation publishes for each article the monthly number of requests it classifies as human. The vendor side is the table frozen for Study R1 on 14 October 2026 (Section 13): for each code, the sum of DataForSEO monthly search volume over the pre-launch keywords that name it. The plan, four predictions and the analysis script were reviewed in draft by four AI agents, committed, and submitted for an OpenTimestamps stamp at 05:10 UTC on 31 October 2026, before any request count was collected apart from a few probes of the endpoints that the plan lists (\texttt{wikipedia\_demand/wikipedia\_demand\_372.md}). The collection followed at once, 1,497 requests to two public APIs from 05:11 to 05:19 UTC, before the stamp reached a Bitcoin block (the earliest is block 973,934, mined at 05:44 UTC): the proof shows that the plan existed within 25 minutes of the end of the read and does not by itself order the two; the order rests on the commit history and the collection log, which the author’s agents wrote. It read the article title, every redirect to it and every title on each code’s resolution path, so that a request by any name for an article is counted once; no title was missing and none lacked data. The items a review of page-view studies asks authors to state, the pages, the language, the collection dates and the period covered, the type of data (monthly here), the agent type and whether crawlers and redirects were included [Alibudbud 2023], are fixed in the plan. The window is October 2024 to September 2025, the twelve months before the vendor exports.

\textbf{The two instruments agree roughly on the ranking and not at all on the magnitudes.} Across the 240 articles, the rank correlation between the vendor’s summed volume and the year’s requests is +0.45 (bootstrap 95\% interval +0.34 to +0.54; Kendall’s tau +0.30; Figure 7), and +0.54 (+0.45 to +0.62) between the number of matched keywords and requests. Prediction P1 (positive, with an interval excluding zero) held; P2 (a point estimate between +0.40 and +0.80, the band inside which Measurement 2’s branch-level estimates fall) held at its boundary, because the interval crosses +0.40; P4 (every code covered by an article that names it) held. The magnitudes are another matter. Requests vary little across articles, a geometric standard deviation of 1.5 (7,989 to 120,065 in the year, 87.5\% of articles between 15,000 and 50,000), while the vendor’s summed volumes have one of 16.7 (50 to 3,704,650 searches a month), so monthly searches per monthly request run from 0.3 at the 10th percentile to 244 at the 90th (median 35), a geometric standard deviation of 14.5 against the 3.1 that Measurement 1 reports for the vendor pair across twelve prefixes (P3, held; the plan calls this bar lax, since at comparable spreads it fails only if log requests track log volume at a Pearson correlation above about 0.9, and at the spreads observed it could not have failed: the ratio’s geometric standard deviation cannot fall below 11.1 at any correlation). Whatever the public count measures, it is not a scaled copy of the vendor’s volume: the vendor’s figure spreads the class across five orders of magnitude where the count spreads it across one. Much of that spread is not the volume model alone: summed volume and the number of matched keywords rank the articles nearly alike (+0.90), and the logarithm of the keyword count accounts for 77\% of the variance in the logarithm of volume, so the spread reflects how many keywords the carriers’ exports matched for each code as well as the volume per keyword, whose geometric standard deviation is 7.0. The same goes for the ranking: holding the keyword count constant, the rank correlation between volume and requests is −0.12 (bootstrap 95\% interval −0.24 to 0.00), so the agreement above may reflect how many keywords the carriers ranked for rather than the volume model (post hoc checks; \texttt{wiki\_demand\_supplement.py}). A vendor side built from the same phrasings for all 372 codes, not conditioned on the carriers’ rankings, would separate the two; it needs vendor data this study did not buy.

\begin{figure}[t]
\centering
\includegraphics[width=\columnwidth]{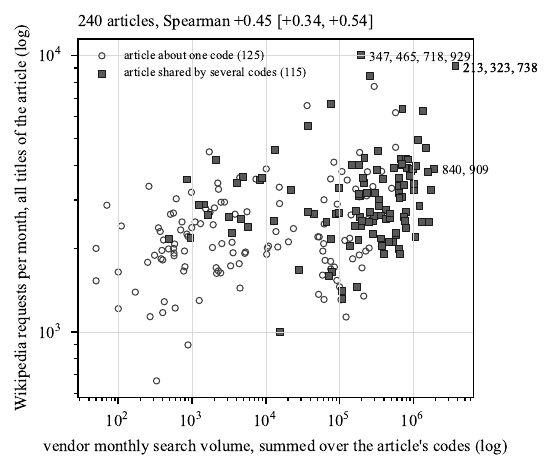}
\caption*{\textbf{Figure 7.} A public request count against the vendor’s volume on the 372 area codes. Each marker is one English Wikipedia article (240): its requests per month, Wikimedia pageviews API, agent type user, October 2024 to September 2025, summed over the article’s title and every redirect to it, against the DataForSEO monthly search volume summed over the codes the article covers, from the table frozen for Study R1. Filled squares are articles shared by several codes; both axes logarithmic. Data and scripts in the source bundle (\texttt{wikipedia\_demand/}).}
\end{figure}

\textbf{The ranking agreement holds under every pre-specified variation and is weaker among single-code articles.} Shifting the window a month earlier gives +0.45; the six months outside the Foundation’s reclassification of automated traffic, +0.45; twelve times the median month, which removes single-month spikes (one article, for codes 327 and 870, had a month 4.6 times its median), +0.44; dropping the six articles that hold a code entering service after October 2025 or a code with no matched keyword, +0.46; the first two full months after the section’s launch, +0.39 (+0.29 to +0.49), with the public count itself stable between the windows (+0.85). Counting an article’s current title alone gives +0.31, partly because ten articles carry their current title only since January 2025 or later, apparently renamed as overlay codes were added, with 44\% to 100\% of their requests arriving through former titles that are now redirects (without those ten, +0.37 on the title alone against +0.43 on all titles; \texttt{wiki\_demand\_supplement.py}, written after the data were read, prints these and the other descriptive figures the stamped script does not). Pooled, 9.3\% of requests arrive through redirect titles; for the median article, 0.7\%. Among the 125 articles about one code the correlation is +0.30 (+0.12 to +0.46): a shared article pools several codes’ demand on both sides, which raises agreement.

\paragraph{What this does and does not show}A Wikipedia request and a keyword search are different acts joined by the search result page: how many searches become article visits depends on whether the engine answers the query in place and on which sources it cites, which varies by code, so the two readings are coupled rather than independent, and the dispersion cannot be attributed to either alone. Requests are worldwide where the volume is the US database; the Foundation’s human and automated split uses rules it revised inside the window; and the query universe behind the vendor side inherits Study R1’s selection on the carriers’ rankings. Within those limits the result has the same shape as Section 6.1’s: on the same 372 entities, a free public count and a commercial volume model agree only roughly on which carry more demand (+0.45, against +0.47 to +0.75 on all links for the taxonomy’s branches), an agreement carried by the keyword count, and size them very differently. The exploratory correlation between the Content Gap Score and the count, +0.41, is not a test of the score, which contains the vendor’s volume; it needs the per-code scores held back with Study R1, so a run on the released files prints every other figure and not this one. The plan was written, stamped and run by AI agents (Section 14). Everything here is public data: the per-title monthly counts, the API responses they came from, the per-article and per-code tables and the scripts are in the source bundle, and the analysis reruns from them.

\section{Measurement 3: what AI answers cite, across eleven sectors and eight answer surfaces}
This measurement uses none of the framework’s machinery (its ontology, graph, competency questions or score); it bears on the framework’s answer-surface component. It asks one question the framework’s design raises, whether the kinds of publisher an answer engine cites differ between engines, and it asks it in one vendor’s index.

\subsection{Pilot}
The pilot is a per-surface export of the Brand Radar cited-domains table for six online-learning brands, made on 12 October between 22:21 and 22:26, under an hour before the first sector export: youtube.com was the most-cited non-brand domain on all four Google views and ranked 28th on ChatGPT, 4th on Perplexity, 5th on Gemini and 9th on Copilot, and reddit.com was in the top three on every surface except Copilot (240th); these figures are printed by the project’s copy of \texttt{cross\_sector\_sensitivity.py}, whose pilot section is not released because the pilot’s brand set and exports are withheld. The predictions below were written after seeing it, so the cross-sector study is a replication across sectors of a pattern first seen in one market. Like the cross-sector study, the pilot covers answers that name or cite a listed brand. An earlier pilot, computed by the author’s own analytics platform, has been withdrawn (Appendix A).

\subsection{Design}
The design was written in a dated design file on 12 October 2026, before the first sector export, and amended once before that export (A1). The file was not deposited with any registry and was first committed to version control after collection, so its date rests on the author’s attestation; the word used here is “written before collection”, not “pre-registered”. Amendment A2, made after collection and before any manuscript text, set aside eight exports that repeated a view or mixed two platforms, recorded the RTX re-run described under Limits, fixed how P1 is scored (YouTube first among non-company domains on at least one query set of each Google product), defined “non-company” as not one of the ten domains entered, and added the exploratory overlap measure. A third amendment, A3, made after the review of this section, records corrections: how the exports were made, the stopped prompt-level sample, and the company-property list used in one sensitivity check. A fourth, A4, records the author’s decision not to wait for the vendor’s written confirmation before using the data (Section 14). Two steps the design named were not completed: the per-surface answer counts were not read, only each sector’s all-surface total, and the prompt-level sample was stopped after 14 of 88 exports (Section 13). The frame is the eleven GICS sectors of the S\&P 500, an exhaustive partition maintained by S\&P Dow Jones Indices and MSCI, so no industry was chosen for its result. The companies are the ten largest holdings by index weight in each sector’s Select Sector SPDR fund, read from State Street’s public fund pages (holdings as of 9 October 2026): 109 companies, because Communication Services counts one issuer, Alphabet, through two share classes. The organization studied in Sections 5 and 8 to 11 is one of them, by that rule. The instrument is Ahrefs Brand Radar, Explore view, United States, read in one collection session from 23:05 on 12 October to 01:15 on 13 October 2026 (exports until 01:05, answer counts from 01:10) through its web interface, with no API; each export was triggered by an AI browser assistant operating the author’s logged-in session (Section 14). For each sector, the cited-domains table was exported for eight vendor views of six answer products (AI Overviews and AI Mode, each on Ahrefs’ prompt set and on its search-query set; ChatGPT; Perplexity; Gemini; Copilot). Each export holds the 1,000 domains found in the most answers, the most this table exports, which reduce to 760 to 975 registrable domains; the 88 exports are analyzed, and an all-surfaces export per sector was made and is used only for the company-dominance figures below. The universe is the set of answers that name or cite at least one of the sector’s ten companies. Summed over sectors it holds 20,389,967 answer records, from 18,763 (Utilities) to 7,766,466 (Consumer Discretionary), not distinct across sectors because an answer naming companies from two sectors is counted in both; the totals were transcribed from the vendor’s overview page rather than exported. By the vendor’s own description, which is vendor documentation [Bek 2026], its queries come from Google’s “People Also Ask” corpus and its keyword database; it tracks about 282 million queries a month on AI Overviews, 24 million on AI Mode and 13 to 15 million on each chat assistant; chat answers are refreshed monthly over a 90-day reporting window and Google’s continuously. Whether the surfaces share one query set is not stated, and model versions are not documented; of the search settings, the vendor states that prompts run through the free, public web interfaces. Source types were fixed before collection: video, forums and question sites, social networks and user-publishing platforms (together, “video and community”), reference, news, government, the sector’s own companies, and other web. The quantity is a domain’s citation incidence, the number of stored answers citing it (“Cited in”), summed over the hosts of a registrable domain; shares are of the summed incidence over the exported domains, within each surface, so an answer citing two domains counts once for each.

\begin{figure*}[t]
\centering
\includegraphics[width=\textwidth]{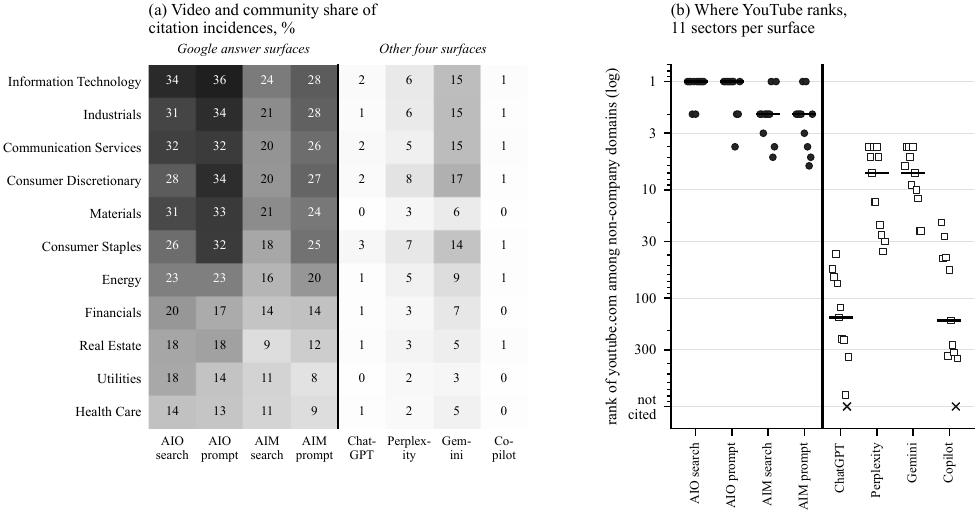}
\caption*{\textbf{Figure 8.} What AI answers cite, by sector and answer surface: Ahrefs Brand Radar, United States, answers naming the ten largest holdings of each S\&P 500 sector fund (nine companies in Communication Services), one collection session on 12 and 13 October 2026. (a) Share of citation incidences going to video and community domains (YouTube, TikTok, Vimeo, Reddit, Quora, Stack Overflow and Stack Exchange, Facebook, Instagram, LinkedIn, X, Pinterest, Threads, Medium, Substack, DEV, GitHub), among the 1,000 domains found in the most answers; sectors ordered by the Google mean. AIO and AIM are Google’s AI Overviews and AI Mode, each on Ahrefs’ search-query and prompt sets. (b) The rank of youtube.com among domains that are not one of the sector’s companies, one marker per sector; bars are medians; a cross marks a sector where YouTube is not cited among the exported domains.}
\end{figure*}

\subsection{Results}
\paragraph{Descriptive}The share going to video and community platforms differs most between answer surfaces. In every one of the eleven sectors, the lowest of the four Google views gave video and community domains a larger share of citation incidences than the highest of ChatGPT, Copilot, Perplexity and Gemini (Figure 8a). Across sectors and views, the median share was 21.3\% on the Google views (range 8.1\% to 35.8\%; 95\% bootstrap interval over sectors 17.2\% to 26.7\%) against 2.1\% on the other four (range 0.3\% to 16.7\%; interval 1.7\% to 2.7\%), and by surface 0.6\% on Copilot, 0.9\% on ChatGPT, 4.6\% on Perplexity and 9.2\% on Gemini. The bootstrap resamples sectors: it shows how much the median depends on which sectors are in the frame, not sampling error in the vendor’s index. Restricted to the two Google prompt views, which are the closest match to the prompt-based surfaces, the Google median is 24.6\%. youtube.com was among the three most-cited domains other than the sector’s own companies in 38 of the 44 sector-by-Google-view cells and in none of the 44 others; its median rank was 1 on AI Overviews, 2 on AI Mode, 7 on Perplexity and Gemini, 151 on ChatGPT and 161 on Copilot (Figure 8b), where a rank in a sparse cell can be a tie among domains cited little or not at all. The contrast is not YouTube alone: without it, the medians are 12.7\% and 1.6\%, and the lowest Google view exceeds the highest other surface in 7 of 11 sectors. The design cannot say whether the contrast belongs to search-integrated answer products or to Google: every high-share surface is Google’s, Gemini, also Google’s, is the highest of the other four, and Perplexity, which searches the web for every answer, is low.

\paragraph{Source types fixed before collection}The design’s question also expected reference, job and forum sources to lead on the chat products. That half is not supported: the median reference share was 2.7\% on the other four surfaces against 3.2\% on the Google views, and forum sites 1.1\% against 5.7\%. On those four, the largest classified type was the sector’s own company domains (median 14.1\%), a share the selection rule partly fixes, so it is not compared across surfaces. “Other web”, which the fixed list does not classify, held a median 61\% of incidences on the Google views and 72\% on the other four (90\% on Copilot), so the classified types describe a minority of the mass, and the claim here is about the share going to seventeen named video and community platforms rather than about every kind of source. Across the 88 cells, answer surface explains 77\% of the variance of that share (bootstrap 95\% interval over sectors 70\% to 91\%) and sector 14\% (5\% to 20\%); for news (12\% against 73\%), government (26\% against 49\%) and the companies’ own domains (40\% against 45\%), sector explains more (the share of the total sum of squares of the 88 cell shares explained by the surface means and by the sector means, unweighted; the remainder is interaction). Appendix C gives every type’s median and range by surface.

\paragraph{By sector (exploratory)}The cited sources also differ between sectors, most visibly on ChatGPT, whose most-cited domain other than the sector’s own companies was a search engine (Information Technology), the Food and Drug Administration (Health Care), a personal-finance publisher (Financials), a streaming guide (Consumer Discretionary), a Disney property (Communication Services), a graphics-card maker (Industrials, the ambiguous name described under Limits), reddit.com (Consumer Staples), the Securities and Exchange Commission (Energy, Utilities and Real Estate) and a paint maker (Materials). The ten most-cited non-company domains overlap between sectors more on the Google views (mean pairwise Jaccard 0.36 to 0.40 across the 55 sector pairs) than on the other four (0.08 to 0.23), a measure not planned before collection. Appendix C (Tables C3 and C4) lists the most-cited sources for every sector on every surface. Three things limit this reading. The one prediction about a sector difference, P4, did not exceed chance; with the company-property list its direction reversed on search queries (22.5\% against 24.2\%), and in no check did it approach significance (p = 0.23 to 0.63). In several sectors one company dominates: by the largest company domain’s share of the ten companies’ citation incidences in the all-surface export, Sherwin-Williams holds 94\% of Materials, Amazon 75\% of Consumer Discretionary, Walmart 65\% of Consumer Staples, Apple 62\% of Information Technology and Uber 55\% of Industrials, so a sector difference here is a difference between company sets. And some of it is mechanical: properties of a sector’s own companies that were not entered as company domains, such as go.com, chase.com and mobil.com among ChatGPT’s top three, are sector-specific by construction.

\subsection{The predictions written before collection}
Of the four, one held, one held in direction only and two failed. P1, that youtube.com is the most-cited non-company domain on AI Overviews and AI Mode in a majority of sectors, failed as written: it held on AI Overviews (9 of 11 sectors on search queries, 8 of 11 on prompts) and not on AI Mode (2 of 11 on each), where google.com, almost all of it www.google.com (median 93\% and 98\% of its incidences on the two views), usually ranked first and YouTube second. It fails under every domain reduction tried; it would hold (9 of 11) if google.com, Google’s answer surfaces citing Google’s own pages, were set aside, which the design did not provide for. P2, that YouTube ranks outside the five most-cited non-company domains on ChatGPT, Perplexity and Copilot in a majority of sectors, held on all three (11, 6 and 11 of 11). P3, that Reddit is among the ten most-cited non-company domains on every surface except Copilot in a majority of sectors, failed: it held in 3 of 11, and ChatGPT was a failing surface in all eight sectors where it did not hold. P4, that the video and community share on AI Overviews is higher in five consumer-facing sectors than in five others named in the design file (Financials is in neither group), held in direction (26.9\% against 24.2\% on search queries, 29.6\% against 24.4\% on prompts) and not beyond chance (exact one-sided permutation p = 0.27 and 0.21).

\subsection{Sensitivity}
None of the following was planned before collection. The descriptive contrast was recomputed with “Found in” in place of “Cited in” (Google median 21.5\%, other four 3.3\%; lowest Google view above the highest other surface in 11 of 11 sectors), with the largest single host in place of summed hosts (22.0\%, 2.4\%, 11 of 11), and with 64 properties of the sector’s companies, from a hand-built list in the source bundle beside the design file (Section 14), counted as company domains (19.8\%, 2.1\%, 10 of 11). The exception is Communication Services, where YouTube, Facebook and Instagram belong to two of the sector’s companies: counted as theirs, the Google views fall from 20.2\% to 32.4\% to 5.9\% to 12.7\%, and Gemini’s 9.2\% exceeds AI Mode on search queries. Restricting each export to its first 100, 250 or 500 rows gives Google medians of 30.6\%, 26.3\% and 23.3\% against 3.0\%, 2.6\% and 2.3\%. Shares therefore fall as rows are added: truncation at 1,000 rows biases them upward, the fixed type list biases them downward, and the net direction is unknown, while the ratio between the two groups stays near ten. Ranks 501 to 1,000 carry a median 7.6\% to 13.8\% of each export’s incidences, depending on the surface.

\subsection{Limits}
The universe is conditional on the sector’s companies appearing in the answer, so answers built only from other sources are absent, and surfaces differ in how often they name companies, so each surface’s universe is selected differently. The exports are ordered by “Found in” while the shares use “Cited in”. Company domains are the ten entered; the property list behind the sensitivity was built from the exports and is not drawn from filings. Two entered names matched other things. “RTX” also matched answers about graphics cards; a re-run with “RTX Corporation” left the Google results essentially unchanged (video and community share 31.3\% to 30.1\% on AI Overviews search queries; YouTube’s rank stayed 1 on AI Overviews and moved from 2 to 1 on AI Mode prompts), moved Perplexity from 6.2\% to 5.5\% and Gemini from 15.0\% to 13.4\%, and changed Copilot’s top non-company domain from amazon.com to a flight-route site, while its ChatGPT export failed twice on the vendor’s side, so the original ChatGPT file, led by graphics-hardware sites, stands; without Industrials the ChatGPT median is 0.8\%, P2 holds on ChatGPT and Copilot in 10 of 10 sectors and on Perplexity in 6 of 10, and P3 holds in 3 of 10. “Visa” also matched answers about travel visas: four immigration and border domains of the US government hold up to 2.2\% of a Financials surface’s incidences, as government sources outside the video and community share. The other names were screened only at the domain level, by reading each sector’s 25 most-cited non-company domains on every surface; the screen found two possible further cases, a nutrition site among Information Technology’s Copilot sources and a laboratory-supplies site of a different company named Merck, and it cannot detect off-topic answers that cite on-topic domains. The prompt and search-query sets are the vendor’s, not observed user demand. The surfaces are tracked on query sets of very different size (Section 7.2), and whether they share one set is not established, so a difference between surfaces may partly reflect the questions each is asked as well as how it answers. The vendor’s index refreshes, so a re-export reproduces the procedure and not the figures. The read is one collection session, one country and one vendor’s answer index. Citations per answer differ by surface, so shares are compared within surfaces; medians over cells weight every sector equally, although cells range from 860 to 18,381,848 incidences (Appendix C gives each cell’s size). Vendor reports on other universes and tools put Perplexity’s share of social sources far higher (19.4\% to 31\%; Section 2) than the 4.6\% median video-and-community share found here, so how a surface ranks on source mix depends on the instrument and the universe of answers, which is Measurement 1’s lesson applied to answer surfaces.

\paragraph{What Measurement 3 establishes}The result is descriptive and narrow: in one vendor’s index, in one collection session, among answers that name a sector’s largest companies, the share of citations going to video and community platforms differs more between answer surfaces than between sectors, while news, government and company sources vary more by sector; and Google’s AI Overviews and AI Mode give video and community domains about ten times the pooled median share of the other four surfaces, and more than twice Gemini’s. If that holds outside this index, a coverage or answerability judgment made against one answer surface may not transfer to another; the framework has no term for that yet.

\subsection{The same contrast in one category across thirteen countries}
Section 7’s contrast was read in one country, in a universe conditional on companies. A later collection relaxes both limits for one category. On 13 and 14 November 2026 the author exported from the same vendor’s index every answer whose prompt the vendor files under Jobs and Education > Education, with no site or company filter, for thirteen countries: 3,724,208 answers dated 17 August to 14 November 2026, in 44 export files, each checked against the row count the interface showed: 34 match it exactly and ten fall short by one to seven rows, 19 in all (one further slice, 1,492 ChatGPT answers for France, failed to export and is missing; the collection’s ledger of files, row counts and checksums is in the source bundle). These are answer-level rows with each answer’s cited pages, not the cited-domain tables of Section 7.2. The author analysed and published the collection, with different source categories, on 14 November 2026 [Pagadala 2026b], before it was read for this paper, so nothing here was predicted and this subsection is a replication read after the fact. What was fixed in advance is the measure: the source types registered on 12 October 2026 and Section 7’s statistic, applied unchanged by a script that stops if the lists differ (\texttt{intl\_surface\_cells.py}).

\paragraph{The contrast holds in every country}Over the 104 cells (thirteen countries by eight vendor views; the smallest holds 1,335 answers), video and community sources take a median 14.0\% of citation incidences on Google’s four views (bootstrap 95\% interval over countries 10.5\% to 17.0\%) against 0.6\% (0.4\% to 1.0\%) on ChatGPT, Copilot, Perplexity and Gemini. The lowest Google view is above the highest of the other four in 13 of 13 countries, by the narrowest margin in Japan (1.2\% against 0.6\%; Table C5). The view accounts for 76\% of the sum of squares of the cell shares and the country for 13\%, against 77\% for surface and 14\% for sector in Section 7.3. Counted by answers, a median 36.8\% of the answers on Google’s views cite at least one such source, against 1.3\% on the others.

\paragraph{Limits}This is one vendor’s index, one category and the vendor’s own filing of prompts: by a labelled sample the author reports, roughly one answer in ten is not about education [Pagadala 2026b]. The registered lists name platforms prominent in the United States, so community sources local to a country, a question site in Japan for example, count as other websites, and the shares outside the United States are floors. Google’s views hold 91.6\% of the answers. The surfaces are again tracked on query sets of different size, so a difference between surfaces may partly reflect the questions each is asked. Shares here are over every cited domain of every answer, not over a cell’s 1,000 most-cited domains, so their levels are not comparable with Section 7.3’s; the ordering of surfaces is. How the exports were made, and under which permission and terms, is stated in Section 14; the rows are not redistributed, and the per-cell counts are in the source bundle (\texttt{edu\_ai\_citations\_intl/}).

\section{The deployment series}
This section is context for Section 5, not a claim. It describes what two instruments recorded following the publication of 381 code pages and their hubs, on a domain and a topic that satisfy all six conditions in Section 3.1.

\begin{table*}[tp]
\centering\footnotesize
\caption*{\textbf{Table 5.} Outcomes at each checkpoint. Ahrefs daily series for the prefix, all locations; AI answers from Brand Radar at the nearest stored date; read 11 October 2026. Day 37 is the primary checkpoint. “n/r” means not recorded: the 11 October read gives no value for that date, and none is interpolated here. “≈” marks a value read from the daily chart rather than from a table.}
\begin{tabularx}{\textwidth}{>{\hsize=1.380\hsize\RaggedRight\arraybackslash}X>{\hsize=0.924\hsize\RaggedRight\arraybackslash}X>{\hsize=0.981\hsize\RaggedRight\arraybackslash}X>{\hsize=0.981\hsize\RaggedRight\arraybackslash}X>{\hsize=0.924\hsize\RaggedRight\arraybackslash}X>{\hsize=0.924\hsize\RaggedRight\arraybackslash}X>{\hsize=0.981\hsize\RaggedRight\arraybackslash}X>{\hsize=0.924\hsize\RaggedRight\arraybackslash}X>{\hsize=0.981\hsize\RaggedRight\arraybackslash}X}
\toprule
\textbf{Metric} & \textbf{Day 0 (26 Jul)} & \textbf{Day 15 (10 Aug)} & \textbf{Day 30 (25 Aug)} & \textbf{\textbf{Day 37 (1 Sep)}} & \textbf{Day 45 (9 Sep)} & \textbf{Day 60 (24 Sep)} & \textbf{Day 75 (9 Oct)} & \textbf{Day 77 (11 Oct)} \\
\midrule
Pages holding a position & 2 & 210 & 365 & \textbf{n/r} & 391 & 414 & 440 & 441 \\
Ranking positions, all & 2 & 3,385 & 14,905 & \textbf{15,830} & 20,326 & 37,141 & 43,302 & 43,980 \\
Positions 1–3 & 0 & 727 & 1,801 & \textbf{1,874} & 2,649 & 5,822 & 8,753 & 9,239 \\
Positions 4–10 & 1 & 1,494 & 10,151 & \textbf{n/r} & 15,104 & 28,733 & 33,203 & 33,560 \\
Positions 11–20 & 0 & 909 & 2,639 & \textbf{n/r} & 2,331 & 2,336 & 1,276 & n/r \\
Positions 21–100 (derived) & 1 & 255 & 314 & \textbf{n/r} & 242 & 250 & 70 & n/r \\
Estimated visits (Ahrefs) & 2 & 22,016 & 51,517 & \textbf{≈51,800} & 92,897 & 244,041 & 221,933 & n/r \\
Live referring domains & 0 & 1 & 19 & \textbf{19} & 415 & 828 & 909 & 913 \\
AI answers citing the section & 0 & 1,603 & 3,725 & \textbf{n/r} & 6,390 & 10,266 & 14,823 & 17,206 \\
\bottomrule
\end{tabularx}
\end{table*}

The 21–100 row is derived as the total less the three reported bands, because Section 4.1 defines a ranking position as positions 1–100 while the instrument reports only three bands and the total. It is arithmetic on vendor estimates and not a fourth measurement. It closes every full column. The row is small, and the instrument does not document its band definitions in the report, so the row is not interpreted. Day 77 is the last read of the series and the one the abstract and this section quote; positions 11–20 were not captured that day, so the 21–100 row cannot be derived for it, and the daily visits series was read only to day 75 (the October run-rate in Table 2 is a different construct). Its referring-domain figure is the instrument’s overview count, which differs from its daily series (Section 10).

Day 75 falls inside the window the Ahrefs data notice covers, and day 60 is the first day of the spam update. The estimated-visits peak at day 60 coincides with that day, and the subsequent fall is consistent with either a data artifact or a ranking change; this study does not distinguish them.

\textbf{At the primary checkpoint}, 37 days after publication, the section held 15,830 ranking positions, 1,874 of them in positions 1 to 3, with an estimated 51,800 monthly visits and nineteen referring domains. By day 77 it held 43,980 positions across 441 URLs holding at least one position, ranked for 29,000 keywords in one instrument and 58,586 in the other, was recorded as a cited source in Google AI Overviews on 7,455 keywords, and appeared in 17,206 stored AI answers.

\paragraph{Concentration}Of the 441 URLs holding a position on 11 October, the ten highest-traffic carry 22.8\% of estimated visits and the top 50 carry 64.9\%, both from one instrument and both on a universe selected by the outcome. Fifty of 441 is 11\% of the pages carrying about two thirds of the estimated traffic, which is concentration, not breadth; an earlier claim of breadth is withdrawn (Appendix A).

\paragraph{Demand the volume tools do not report}Among the 58,586 US keywords, the lowest-traffic rows are misspellings, non-English forms, numeric fragments and question forms with 40 to 170 reported searches a month or none at all. The section ranks in 105 countries, with 18.5\% of Ahrefs-estimated visits from outside the US. This is stated as a distributional observation and not as a test; the per-page threshold test that would make it one requires the full per-page keyword export matched to pre-launch reported volume, which has not been run.

\section{What the answers cite, and what they do not}

\subsection{Enumerating pages are cited more often than any single code page}
The pages most cited in AI answers are state hubs that enumerate codes, cities and maps: New York 474 answers, California 337, Texas 322, Florida 304, and the national hub 294. Each exceeds any individual code page: 214 Dallas 238, 786 Miami 228, 404 Atlanta 218 (Figure 9). A hub lists many codes, so it pools the questions about all of them: per code it lists, a state hub is cited less often than the most-cited code pages. The observation is about which page an answer engine cites, not about which page answers a given question better.

\begin{figure}[t]
\centering
\includegraphics[width=\columnwidth]{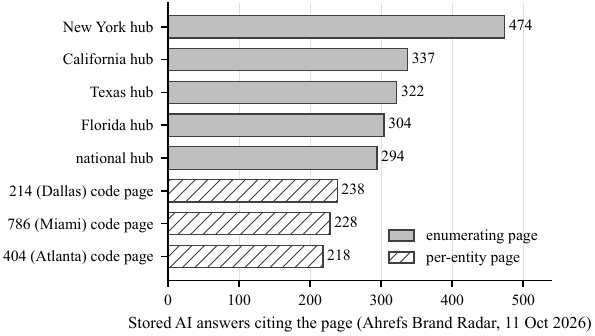}
\caption*{\textbf{Figure 9.} The eight pages under the prefix most often cited in stored AI answers, Ahrefs Brand Radar, 11 October 2026. Enumerating pages (state hubs and the national hub) are solid; per-entity code pages are hatched.}
\end{figure}

A per-entity content strategy predicts that per-entity pages are cited most. The question “where is area code 214” is answered most directly by the 214 page, and that page is cited less than the page listing every Texas code. The hypothesis this raises, to be tested per question rather than per page, is that an answer engine prefers a page that answers a family of questions to pages that each answer one; if it holds, the unit a content program should publish is not the unit the question names. This is the framework’s form question in miniature: the form in which a question is best answered is an empirical property to be measured. The released coverage rubric does not leave it open: it grades a dedicated page about an entity 1.0 and a list or section page carrying a fact about it 0.5, a preference for the per-entity form that this observation questions (Appendix B).

It is a hypothesis from one section. It is the observation in this paper most worth testing elsewhere, per question, and testing it needs no access to this deployment: any site with both hub and per-entity pages on an enumerable class can check it.

\subsection{Who else is cited, and what that does not establish}
The queries are held, and the AI answers supplied, by encyclopedic and directory sites. The five domains with the largest keyword overlap with the section are two area-code directories, a time-zone reference and two voice vendors, at Domain Ratings between 13 and 83. Wikipedia, at 97, is co-cited in roughly 11,000 of the answers that cite the section. Two other national carriers of comparable authority are co-cited 9 times (Domain Rating 91) and 240 times (Domain Rating 87), against the section’s 17,206. Neither published a page per code.

\paragraph{This establishes nothing about necessity}The counts are co-citations \emph{within the set of answers that cite the provider}, a set selected because the provider is in it. A carrier absent from that set may be present across answers the provider is absent from, and no instrument here is pointed at those answers. The counts bound the carriers’ co-presence from below within a selected set and support no statement about what authority alone produces, nor about whether content on the topic was necessary.

The comparison that would establish it is outcome-independent and has not been run: the registry’s codes crossed with a fixed set of verbalizations, queried without reference to who is cited, so that each domain’s presence is measured on a denominator none of them selected. Section 13 specifies it.

\section{The link profile}
The section had zero live referring domains through 31 July, one on 10 August, and nineteen on 25 August and still nineteen on 1 September, by which date it already held 15,830 ranking positions.

The referring-domain count on 11 October is reported by the instrument as 913, while the same instrument’s daily series gives 909 at day 75 and its new-and-lost report for September gives 917 new against 95 lost, which with nineteen at day 37 implies about 841 at the end of September. The three figures are for different dates (the end of September, day 75 and day 77), so they need not agree, and their order is consistent with the network still arriving. No claim here turns on the exact count; the shape is nineteen through day 37, several hundred within three weeks, roughly nine hundred by day 75.

The character of the arrivals is not in doubt. They are automated networks: 61.9\% .info, 26.4\% .asia, 10.2\% .xyz, every linking page at URL Rating below 10, top anchors advertising link sellers. The second instrument indexes 80 referring domains for the subfolder, 88\% from domains at Authority Score 0–10 and 75\% on .xyz.

\subsection{What the link profile can and cannot explain}
\begin{table*}[tp]
\centering\footnotesize
\caption*{\textbf{Table 6.} Position growth by interval, with the link status of each.}
\begin{tabularx}{\textwidth}{>{\hsize=0.852\hsize\RaggedRight\arraybackslash}X>{\hsize=1.062\hsize\RaggedRight\arraybackslash}X>{\hsize=0.893\hsize\RaggedRight\arraybackslash}X>{\hsize=1.006\hsize\RaggedRight\arraybackslash}X>{\hsize=0.774\hsize\RaggedRight\arraybackslash}X>{\hsize=1.413\hsize\RaggedRight\arraybackslash}X}
\toprule
\textbf{Interval} & \textbf{Referring domains} & \textbf{Domains gained} & \textbf{Positions gained} & \textbf{Rate} & \textbf{Link status} \\
\midrule
Day 0 to 15 & 0 to 1 & 1 & 3,383 & 226 / day & clean \\
Day 15 to 30 & 1 to 19 & 18 & 11,520 & 768 / day & clean \\
Day 30 to 37 & 19 to 19 & 0 & 925 & 132 / day & clean \\
Day 37 to 45 & 19 to 415 & 396 & 4,496 & 562 / day & network arriving \\
Day 45 to 60 & 415 to 828 & 413 & 16,815 & 1,121 / day & network present \\
Day 60 to 75 & 828 to 909 & 81 & 6,161 & 411 / day & network present, data notice \\
\bottomrule
\end{tabularx}
\end{table*}

\begin{figure*}[t]
\centering
\includegraphics[width=\textwidth]{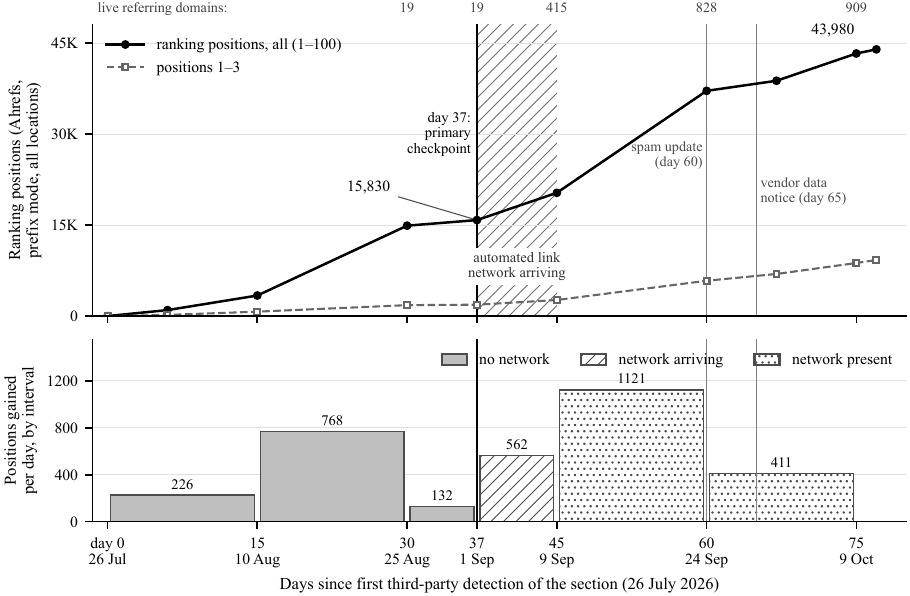}
\caption*{\textbf{Figure 10.} Ranking positions and the link profile over 77 days. Upper panel: ranking positions in positions 1 to 100 and in positions 1 to 3, Ahrefs daily series for the prefix, all locations, read 11 October 2026; the hatched band is the interval in which the automated link network arrived; the vertical rules mark the primary checkpoint (day 37), the September 2026 spam update (day 60) and the vendor data-reliability notice (day 65), and the August 2026 spam update (days 23 to 26) is not drawn; live referring domains at each checkpoint are printed along the top. Lower panel: positions gained per day in each interval of Table 6, shaded by link status. The upper panel also plots two reads from the same daily series that are not checkpoints in Table 5: day 6 (984 positions, 186 in 1–3) and day 67 (38,801 and 6,942).}
\end{figure*}

Of the 43,302 positions held at day 75, 15,830 (36.6\%) were held at day 37 and 27,472 (63.4\%) were acquired after the network began arriving (Figure 10). Because the arrival is dated only to an eight-day interval, that share is itself bounded rather than exact: 53.1\% to 63.4\%. For AI answers, day 37 is not recorded, so the corresponding share is bounded 56.9\% to 74.9\%.

\paragraph{What survives}A result of 15,830 ranking positions and 1,874 top-three positions was established with nineteen referring domains, a count unchanged for the preceding week. It was reached before the automated network arrived, so the early positions did not need that network; it does not show that links played no part, because nineteen referring domains are not none and their quality is not assessed here.

\paragraph{What does not}Any claim about the full series, including that link acquisition plays no part in the outcome: most of the series postdates the network.

\paragraph{What the table does not show, against the temptation to read it as showing it}The fastest clean interval, 768 a day across days 15 to 30, which holds the August spam update, is 37\% faster than the 562 a day of the arrival interval. Reading the 132-a-day trough immediately before the arrival as a baseline would suggest an acceleration; that baseline sits at the series minimum, and the inference does not survive showing all six intervals. A section of 434 pages entering an index does not grow linearly. The network is therefore \textbf{unbounded as a confound on the continuation}, and nothing here shows that it accelerated anything.

\paragraph{Disclosure}The provider has an exposure this paper discloses rather than analyzes: a spam link profile pointed at a live section, which the provider has been asked to review. A manual action before a later checkpoint would confound any continuation of this series. The author did not commission, purchase or solicit any link to the section.

\section{What the deployment cost, and what it got wrong}

\subsection{Cost}
These figures are the author’s reconstruction from project records, not an accounting extract, and are stated to the nearest useful unit rather than implied to be precise. They are included because a deployment paper without them gives a reader no basis for deciding whether to attempt the same thing (Table 7).

\begin{table*}[tp]
\centering\small
\caption*{\textbf{Table 7.} What the deployment cost, as reconstructed by the author from project records.}
\begin{tabularx}{\textwidth}{>{\hsize=0.825\hsize\RaggedRight\arraybackslash}X>{\hsize=0.760\hsize\RaggedRight\arraybackslash}X>{\hsize=1.415\hsize\RaggedRight\arraybackslash}X}
\toprule
\textbf{Item} & \textbf{Figure} & \textbf{Basis} \\
\midrule
Analysis and class selection & ≈ 3 person-weeks & taxonomy construction, classifier runs, the re-homing pass and a structured review, Nov 2025 to Jul 2026, part-time alongside other duties \\
Template and generation build & ≈ 2 person-weeks & one page template, the registry-to-page generator, internal linking and structured markup \\
Content production, marginal & ≈ 0 & every fact templated from the public registry; no writer, editor or per-page review \\
Editorial and link acquisition & 0 & no outreach, no placement, no paid promotion \\
Third-party data & ≈ USD 6,000 / year & the author’s own Ahrefs and Semrush subscriptions, already held for other work; no incremental spend attributable to this deployment \\
Engineering to publish the section & within existing platform capacity & generated into an existing CMS and publishing pipeline; no new infrastructure \\
\bottomrule
\end{tabularx}
\end{table*}

Two things follow for a reader considering transfer. The marginal cost per page was near zero \emph{because} conditions 1 to 3 of Section 3.1 held, and they do not hold for most classes. And the dominant cost was analysis rather than production, which inverts the usual content-program cost structure and is the part a reader should expect to pay.

\subsection{Six lessons, stated as this deployment’s failures}
Each is something this deployment got wrong, which is why its cost can be stated.

\begin{enumerate}[leftmargin=1.6em, itemsep=1.5pt, topsep=2pt, parsep=0pt]
\item \textbf{Separate structure sources from coverage sources, or do not claim a method.} This deployment did not, so its demand layer derives from what competitors already rank for. Cost: an evaluated method claim; the score is specified in Appendix B, and this deployment cannot serve as its validation.
\item \textbf{Freeze the evaluation query set before launch.} This deployment did not, so every share in this paper would otherwise be computed on a denominator selected by the outcome. Cost: all comparative claims. The released implementation makes freezing a command so the failure is harder to repeat.
\item \textbf{Collect at least one first-party demand surface before claiming anything about users.} This deployment collected no first-party surface: of the thirteen demand-surface types the released implementation declares (Figure 1a), only competitor keyword exports, a tool-derived surface, were linked to any entity; no forum, Q\&A, video, social, review or support data was. Cost: the deployment cannot show that the framework’s user-question layer did any work; that layer is described in Section 1.4, exercised in the released example on synthetic records, flagged as synthetic, on four of the thirteen surface types, and untested on this deployment’s data; Section 6 runs the comparison on a separate review dataset.
\item \textbf{Record the recommendation and its date.} The document that turned analysis into a build decision is not in retrievable form. Cost: the causal chain has no anchor at its most important link.
\item \textbf{Get a second grader early.} Grading judgments with AI agents the author built (Section 14), and planning independent annotation for later, means the reliability statistic arrives after the claims that depend on it. Cost: Appendix B’s inputs are unvalidated.
\item \textbf{Expect the class that scores well to have no commercial page to attach to.} Area codes are a reference topic on a commercial domain. The score has no feasibility term and someone must decide how business relevance enters.
\end{enumerate}

\section{Threats to validity}
\paragraph{Measurement 1 (Section 5)}Instrumentation is its subject rather than a caveat on it, and no outcome is a measured visit. The July to September ratios of Table 2 compare monthly points whose construction was not recorded, and part of them may be a difference of dates (Section 5.1). Trade press reports that Google has served altered result pages to suspected rank trackers since September 2026, with social and forum sites replacing tracked ones: 8.0\% of one tracker’s pages since 13 September [Rijo 2026a] and 5\% to 20\% by Ahrefs’ account, depending on the provider, with Ahrefs cleaning up past results and Semrush dating the problem to late summer [Rijo 2026b]. These are secondary reports, and the effect here is not measured. Every read in this paper falls after that start; the day-37 figures describe earlier dates but were read from histories a vendor may revise, and a swap favoring social and forum sites would bear on Sections 5, 7 and 8 to 10 alike. One instrument posted a data-reliability notice covering days 65 to 75. “AI Overview citation” has two operational definitions here, which differ by a factor that falls from 2.94 to 1.34; both are reported, and a comparison of AI-answer counts is withdrawn as a scope mismatch (Section 5.3). The estimated-visit figures are vendor click models that do not distinguish automated from human requests, and a section attracting an automated link network may also attract automated requests; no instrument available here separates them. Ordering agreement is weaker for sections within a factor of three of each other, and the share of the divergence the keyword universe explains depends on the territory pairing. External validity: twelve purposively chosen prefixes on one domain, one of them followed for 77 days; one vendor pair; one read date for the cross-section; no ground truth; and dashboard reads that were not archived, so they can be checked only by reading the dashboards again. Internal analytics would resolve much of this and are not available.

\paragraph{Measurement 2 (Section 6)}One first-party surface on one platform in one industry; exact-match links of measured but modest precision (about two in three correct, half for single-word labels; judged by blind AI annotators, not by people); relative review dates; and a taxonomy whose labels are themselves search phrasings, which biases the comparison toward agreement. The reviews describe store locations rather than the area-code section. No person has checked the links or the waiting matches, and a review is a different construct from an information need. Section 6.2’s public count is coupled to search demand through the result page, worldwide where the vendor’s volume is US-only, and split from automated traffic by rules the Foundation revised inside the window; its plan was written, stamped and run by AI agents, and the author had not read it when it was stamped.

\paragraph{Measurement 3 (Section 7)}One vendor’s answer index, one country, one collection session, a universe conditional on the sector’s companies appearing in the answer, and company sets that one company dominates in several sectors. Section 7.6 lists the rest. Section 7.7 relaxes two of these for one category, after the fact and with no prediction: thirteen countries, and no company filter.

\paragraph{The deployment (Sections 8 to 11)}Confounding, with directions. Domain authority: direction positive, magnitude unknown, \textbf{no bound available}. The carrier comparison in Section 9.2 is not a bound: it is computed on a set selected on the provider’s presence. Content quality and execution: templating, registry facts, structured markup and internal linking are perfectly confounded with selection; direction positive; no estimate. Topic contest: no defender responded. The automated link network: between 53.1\% and 63.4\% of day-75 positions postdate its arrival, the range reflecting that the arrival is dated only to an eight-day interval; direction unknown, magnitude unestimated, and confounded with index maturation and with the September spam update. Algorithm updates: two named, dated updates fall in the window, the August spam update (days 23 to 26), before the primary checkpoint and inside its fastest interval, and the September one, on the day of the traffic peak; whether unannounced ones did is not knowable from here. Selection: The query universes were defined after launch. Counts are reported on them; shares are reported only where the denominator is stated and flagged, or where the quantity is a ratio between two instruments on the same prefix, which selection does not affect. “Visibility” is not value: no revenue, conversion or impression figure appears in this paper and none should be inferred. External validity: one organization, one industry, one class, a top-authority domain, 77 days, six stated conditions.

\paragraph{Author independence}The author designed the approach, deployed it as an employee of the organization studied and remains employed there, directed every measurement, wrote this paper with the AI tools Section 14 names, owns an AI-visibility analytics platform, and may develop commercial tools and a book based on the approach. No person other than the author has performed any part of the evaluation; the reproduction and annotation checks of Sections 6 and 14 were made by AI agents the author ran. A reader should treat every judgment here as made by an interested party, and should weight the instrument readings of Section 5, which are independently reproducible from two commercial subscriptions, far above anything else in the paper. That section is the primary result partly for that reason.

\section{What would test what this paper declines to claim}
\paragraph{No within-class control exists}A control of geographic codes with no page, queried on the same instruments as the published ones, would test whether publishing a page did anything; Section 3 records that no such set exists (Appendix A records the control that was planned and withdrawn). The section holds a page for every geographic code in service in a US state or the District of Columbia; the six territory codes without pages are too few and too different in kind to serve. The paper therefore claims no test of whether publishing a page did anything. The closest substitute is external: a comparable-authority domain that covers the class partially, which this study has not identified.

\paragraph{The enumerating-pages test, which needs no access to this deployment}Section 9.1 reports that each of the five most-cited hub pages is cited in more AI answers than any of the individual code pages they enumerate (1.2 to 2.0 times the most-cited code page). Any site holding both hub and per-entity pages over an enumerable class can test whether that ordering holds per question, on either instrument, with no reference to this provider.

\paragraph{An outcome-independent query universe}The registry’s geographic codes crossed with a fixed set of question verbalizations yields a query set that is complete with respect to the class, defined without reference to any outcome, and hashable. The released implementation writes it: \texttt{ufi freeze} produces the set with its digest. The example ships a 924-query demonstration over the 66 codes in its registry extract, at digest \texttt{f05ce279942be5f8}; the set for the full class would be built the same way.

\paragraph{A prospective study on an ordinary domain}A second industry, a domain outside the top decile of authority, the ontology and parameters registered before collection, the query set frozen and hashed before launch. Three classes with matched page counts: top class by mean score, top class by volume, and a class in the bottom quartile of the score. Primary outcome: top-ten presence on the frozen set at day 37, read on both instruments. Five scoring parameters are locked: \textbf{α = 0.5, β = 0.5, damping = 0.85, at least three phrasings per question and a mean class aggregate, lock digest \texttt{7290a2235d8de05f}, locked 12 October 2026}. The lock covers those five values, two of which, the phrasing minimum and the class aggregate, the released code records but does not consult; it does not cover the scoring code, which changed later that day (Appendix B), or the ontology, questions, rubrics and grader. A test rather than a fit needs all of them pinned to a tagged commit and the design deposited with a third-party timestamp before collection.

\textbf{A retrospective test of the score on this deployment (Study R1), fixed before its outcomes were read.} Every one of the 372 geographic codes in service received a page, so the score cannot be tested here as a guide to which pages to build, but it can be tested as a ranking: whether, from data dated nine months before launch, it orders the codes’ later outcomes better than search volume alone, on each instrument. The plan, the score of every code under the locked parameters and the analysis script were committed on 14 October 2026 and stamped with OpenTimestamps before any of its registered outcomes (each code page’s ranking keywords, visits and query coverage) was read for any code (the plan names the website and is held back; the manifest of its files’ digests and its proof are in \texttt{framework\_test\_r1/} in the source bundle); the stamp is anchored in Bitcoin block 971,506, mined at 04:05 UTC on 15 October 2026. Earlier reads for this paper had shown some per-page information, the pages most cited in AI answers (Figure 9) and the share of visits held by the highest-traffic URLs (Section 8); a dated amendment, released with the plan’s manifest and anchored separately in Bitcoin block 971,522 (06:10 UTC on 15 October 2026), records this. The test waits on the vendors’ per-page exports, which have not been made; its result will be reported whichever way it falls.

\textbf{An annotation study}, two graders independent of the author, on a sample of the coverage and answerability items, reporting weighted kappa with bootstrap intervals. The released implementation caches judgments keyed on the rubric, so a second grader costs one pass.

\paragraph{The multi-platform demand study, which is the framework’s own test}Measurement 2 (Section 6) is a first run on one first-party surface, with a public count as a second (Section 6.2); the study specified here extends it. For one enumerable class, collect user questions from the first-party surfaces in Figure 1a (forum, question site, video comments and transcripts, reviews, social groups, support tickets), link them to entities with the released linker, and compare the resulting demand distribution with the keyword-volume distribution for the same entities. The prediction is that the two distributions differ in both which entities carry demand and in what form the demand takes (question, comparison, how-to, video), and that the difference is largest where reported volume is smallest. The released implementation tags each imported record with its surface type and prints which surface types returned rows, so a run that collected nothing cannot be reported as one that did.

\paragraph{A prompt-level replication of Measurement 3 (Section 7)}The same eleven sectors and eight surfaces at the level of the individual answer, with the cited pages of each, so that the source mix can be split by question type (how-to, definition, comparison, purchase) and tested within each sector rather than across them. Amendment A1 fixed a sampling rule of the 10,000 highest-volume prompts per sector and surface; because the table keeps no volume sort, the exports took the first 10,000 rows in the vendor’s default order, a change recorded after the design file’s first commit (A2, A3). Collection was stopped after two small sectors, Energy and Utilities (14 of 88 exports, 44,463 answers), when the vendor’s terms on automated use were found (Section 14), before any row was examined; those rows are stored and unanalyzed, and the study’s predictions are not yet written. Section 7.7 is a partial step of this kind in one category: answer-level rows for thirteen countries, read after the fact.

\paragraph{A persona study on AI answer surfaces}The same frozen query set issued to AI answer surfaces under a small set of stated personas (for example parent, business owner, traveler), with the cited sources and the answer text compared across personas. The question is whether the answer and its citations vary with who is asking, which a classic web search audit did not find for demographic attributes such as gender and age [Hannak et al. 2013], which has been measured for chatbot responses, for the gender and race a user’s name suggests [Eloundou et al. 2025], and which, to the author’s knowledge, has not been measured for the sources AI answer surfaces cite.

\paragraph{A second domain for Measurement 1 (Section 5.5)}The cross-section was run on one domain. The result that the vendor ratio is prefix-specific, with no detectable relationship to age, needs a second site and a second read date before it is more than an observation about one domain. Anyone with both subscriptions and any multi-section site can do it in a day; the data file released with this paper gives the read method. A protocol that chooses the second domain and its prefixes by rules fixed in advance, and re-reads the original twelve on a later date, was stamped with OpenTimestamps before any reading under it, in the same Bitcoin block; the protocol and every file its stamped manifest lists are in the source bundle.

\subsection{Further directions}
Three directions follow from the measurements’ finding that demand and answers sit on several surfaces and move between them; none tests a claim made here.

\paragraph{Surfaces that move}The thirteen surface types the implementation declares, and the hundred sources of Appendix F, are a snapshot. People ask on video, forum, newsletter, professional and short-video platforms (for example YouTube, Reddit, Substack, LinkedIn, X, Snapchat and TikTok), and which of them people use changes: among US adults, use of TikTok rose from 21\% in 2021 to 37\% in 2025 and use of Reddit from 18\% to 26\% over the same four years, while use of YouTube and Facebook held steady [Gottfried \& Park 2025]. A demand layer therefore has to be re-measured rather than fixed, keeping the relations between people, questions, entities and answers in the graph so that a shift between surfaces shows up as a change in the graph rather than as missing data. Tracking one industry’s questions by surface over time, with the same linker and ontology, would show how quickly a demand layer goes stale.

\paragraph{Agents as askers}AI agents that research, compare, plan, book and buy for a person are both a demand surface, through the questions they ask, and an answer surface, through the sources they consult. Whether they ask what the people they serve would ask, and rely on the sources Measurement 3 counts, is measurable and not addressed here.

\paragraph{Business disclosures}Some companies, Costco among them, now report how AI search affects them, on the record and with a date: on its fourth-quarter fiscal 2026 earnings call, Costco’s chief financial officer said that traffic to its site from AI search had grown by triple digits for the second consecutive quarter, from a low base [Motley Fool 2026], in a quarter with net sales of \$93.9 billion, up 11.2\% [Costco Wholesale Corporation 2026]. The author’s reading of the call, as evidence that AI answers have become a sales channel a large retailer measures, is set out in a separate note [Pagadala 2026a]. Costco is one of the 109 companies of Section 7. Logging what those companies say about AI search each quarter, beside how often AI answers cite their domains on a prompt set fixed in advance (Section 7’s shares are partly fixed by its selection rule), would test whether disclosures move with citations and whether a content program could anticipate a shift rather than read it afterwards.

\section{Reproducibility, ethics and conflicts of interest}
\paragraph{What is released}The \texttt{ufi} reference implementation at \textbf{https:/\allowbreak{}/\allowbreak{}github.com/\allowbreak{}VenkataPagadalaGIT/\allowbreak{}ufi} under Apache 2.0, version 0.2.0, cited here as commit \texttt{149994a} of 12 October 2026, because the repository has no tags and later commits kept the version number: test suite, command line interface, and a worked example that runs offline on public data. Its OWL reasoning step uses HermiT through owlready2, which needs a Java runtime; the package declares Python 3.10 or later and minimum, not pinned, dependency versions. Most of its code was written by an AI coding assistant under the author’s direction (all 22 commits at that point are AI-assisted: 21 carry the assistant as author and one as co-author). The corrected numbering-plan ontology in YAML with OWL and SHACL export, and the area-code dataset the example uses. The demand layer declares thirteen surface types, ships two generic file adapters (CSV and JSON Lines) and prints which surface types returned rows; collection from each platform is left to the deployment. The frozen 924-query set, the parameter lock and the run manifest are committed in the repository at \texttt{examples/nanp/out/}, so the digests quoted in Section 13 and Appendix B resolve to released files. It has \textbf{no DOI}; a Zenodo deposit is outstanding. The exported OWL and SHACL mint terms under \texttt{urn:\allowbreak{}ufi:\allowbreak{}<name>} by default, which makes no resolution claim; no persistent namespace has been registered.

\paragraph{Data released with this version}The twelve-prefix cross-section (Table 4) is released as two CSV files, one per instrument, each row carrying the prefix, the figures read, the read date and the read method, at \texttt{data/instrument\_divergence/} in the repository named above. Everything else this paragraph names is in the source bundle, which is posted with this preprint as its ancillary files and carries no licensed rows, review text or reviewer names; it is not in the repository at the cited commit, and it has no DOI. For Section 7, the bundle holds the design file with its amendments, the 109-company list with its source, the company-property list, the report-URL file, the analysis, sensitivity and figure scripts, the per-cell sizes, the per-sector and per-surface aggregates, the ranked top-ten non-company domain names and the checksums of the raw exports. The raw vendor exports are not redistributed, because the vendor’s terms restrict use of its data; a Brand Radar subscriber can re-export them with the recorded settings, which reproduces the procedure and not the figures, because the index refreshes. For Section 6, the bundle holds the per-label counts behind its correlations, anonymized (no label names; volumes as ranks, which is all a rank correlation uses), the link-check judgments and the annotators’ protocol, and \texttt{m2\_from\_counts.py} recomputes from them every Section 6 correlation, branch share and interval that does not need the review text; the licensed review export itself is not released. For Section 6.2, everything is public data and is in the bundle: the plan with its stamped manifest, the per-title monthly request counts and the API responses they came from, the per-article and per-code tables and the scripts; the per-article sums of vendor volume are derived aggregates, and no keyword, per-keyword volume or per-code score is released. For Section 7.7, the bundle holds the counts per country and view, the scripts and the collection’s ledger of export files with their row counts and checksums; the answer-level rows are not redistributed.

\paragraph{What a reader can verify independently}Every figure in Sections 5 and 8 to 10 can be re-read from the same two dashboards for the same prefix, subject to index drift. The exception is the Brand Radar answer count, which moved from 14,823 (9 October) to 17,206 (11 October) and 17,400 (12 October) as stored answers were refreshed; it is reported with its read date and is not claimed to be re-readable. A re-read of Semrush on 13 October reproduced the July to September figures of Table 2 exactly, in both territories; the October run-rate had moved by about 3\% (worldwide 1,910,867, US 1,659,636). The Ahrefs daily series should reproduce the checkpoint rows until it ages out of the six-month window; it was not re-read for this version. The series in Tables 2, 3, 5 and 6 are in the source bundle as CSV files; they are transcribed from those tables, and the dashboard reads behind them are not archived.

\paragraph{What is not released}The provider’s internal graph database, the crawl of its site, and its analytics. No figure in this paper is drawn from them; the cost figures in Section 11.1 are the author’s reconstruction from project records. The build log and snapshot of the author’s platform cited in Section 3.2 are not released either.

\paragraph{Ethics and data collection}No experiment was run on people, and no data about visitors to the section was collected. The deployment published reference pages and ran no experiment on visitors. Page facts derive from the numbering plan administrator’s public registry and US Census data, both public-domain US government sources. This study performed no crawl of any third-party site; the provider’s own site crawl is an internal operation on the organization’s own property. The review dataset in Section 6 was obtained through DataForSEO, a licensed data provider, under the author’s own license and its terms; the export contains reviewer display names, which were excluded from the analysis and are held only in the author’s private copy of the licensed export; no review text is reproduced, and only per-label aggregate counts are reported. No institutional ethics determination is reported for that analysis of public reviews; such analyses are commonly exempt, but the determination belongs to the institution.

The vendor figures in Sections 3, 5 and 8 to 10 were read from the vendors’ web dashboards under the author’s own paid subscriptions, held for other work (Section 11.1), and the Section 7 exports were made through the Ahrefs Brand Radar web interface, using its own export function, under a subscription held by Previsible, a digital marketing agency, with its written permission of 11 October 2026 to use the agency’s accounts and tools. In both cases the browser was operated by an AI browser assistant (Claude in Chrome, from Anthropic) under the author’s direction: it read the displayed figures and, for Section 7, clicked the interface’s export control for each export. No API was used, and raw vendor exports are not redistributed. Ahrefs’ terms of service prohibit using its services “through an automated means” except through its API (clause 4.4(j), read 13 October 2026), and searching, scraping or harvesting its data through any “engine, software, tool, agent, device or mechanism” other than its own (clause 4.4(i), read 3 November 2026); the Ahrefs reads and exports were made through its web interface by the browser assistant, no confirmation from the vendor has been obtained, and the prompt-level collection of Section 13 was stopped when the clause was found. The Section 7 design had made using and publishing the collected data conditional on the vendor’s written confirmation (amendment A3); amendment A4 records the author’s decision of 13 October 2026 not to treat that confirmation as a precondition, and a reader should weigh that decision. Semrush’s terms prohibit using or launching “any automated system” that sends more requests than a person could reasonably produce, and to “harvest or scrape any content of the Services” (section 3.3, items (k) and (p), read 15 October 2026); the Semrush figures were read from its dashboards by the same browser assistant, and Semrush has not been asked whether that falls within those terms. The author made the exports of Section 7.7 by hand on 13 and 14 November 2026 in the same vendor’s web interface; a script, run by an AI coding assistant, filed each downloaded file and checked its row count, and no script or assistant operated the vendor’s interface. By the collection’s own notes they were made in the agency’s workspace; whether the agency’s permission of 11 October covers them and their publication is not established here. An earlier set of education exports for 33 named sites, made on 1 October 2026 before that permission, is not used in this paper.

\emph{Competitive effects.} The deployment displaced incumbent pages, mostly specialist directories at Domain Rating 13 to 83, in a public search index. This is a commercial outcome with distributional consequences for smaller publishers, and the informational value added is modest, because the registry had already published the facts. Condition 6 of Section 3.1 can be read as a criterion for selecting undefended targets. That reading is available and this paper does not endorse it; a method whose transferable content is “find a finite factual class that an authoritative free registry already answers and that no one is defending” deserves to be stated in those terms so that the question can be asked.

\paragraph{Corrections to earlier versions}Appendix A summarizes the corrections that change how a figure or claim must be read; the full version history is in the changelog in the source bundle.

\paragraph{Organizational permission}The organization that operates the section has not been asked to approve this paper and has not approved it. The author’s determination is that its approval is not needed, on two grounds a reader can check against the paper. First, the paper uses none of the organization’s own data: none of its analytics, logs, crawl output or internal databases (\emph{What is not released}, above); every figure about the section is a commercial vendor’s estimate for public URL prefixes, which any subscriber can re-read, or a count of the section’s public pages. The accounts behind the figures are not the organization’s either: the Section 5 reads were made under the author’s own subscriptions, the Section 6 export and taxonomy under the author’s own license and on the author’s own platform, and the Section 7 exports under the agency’s subscription. Second, the organization is not named. This determination is the author’s alone: the organization has not confirmed it, and no legal review of it is reported. What does come from the author’s employment is marked where it appears: the publishing history in Section 3, the account of the deployment in Section 11, and the cost figures of Table 7, which the author reconstructed from project records. The twelve-prefix reads of Table 4 have been in the public repository since 12 October 2026. Not naming the organization does not make it unidentifiable: the prefixes in Table 4 and the author-independence statement in Section 12 allow a reader with access to the same instruments to identify it, and that statement is kept because the disclosure is required. The name is withheld; the identity is not claimed to be concealed.

\paragraph{Matters open at posting}This preprint is posted with four matters unsettled, each stated where it arises: no institutional ethics determination is reported for the analysis of public reviews in Section 6; neither vendor has confirmed that reads and exports made through a browser assistant fall within its terms, and the agency’s permission is not established to cover the Section 7.7 exports; earlier manuscript copies in the public repository that assert the organization’s approval or name it had not been removed when this version was built; and no person other than the author has re-read a dashboard figure (below). Versions 4.9 to 5.4 made the first and third of these, with the organization’s approval and a public release of the Section 7 artifacts, conditions of submission; the author has since decided to post with them stated instead (Appendix A).

\paragraph{Conflicts of interest}The author is employed by the organization that operates the section studied (Section 12); that organization is also one of the 109 companies in Section 7, included by the selection rule. The author owns an AI-visibility analytics platform, which computed a pilot for Section 7 that has been withdrawn (Appendix A), and may develop commercial tools and a book based on the framework. The Section 7 data were collected under the subscription of Previsible, a digital marketing agency, through a personal contact who gave the author access; the author has no business relationship with Previsible. To limit the effect of these interests, the primary result rests on two vendors’ readings of public URL prefixes that any reader with the same subscriptions can re-read, the implementation is public, and the studies in Section 13 are designed before their data exist.

\paragraph{Use of AI tools}Generative AI was used throughout this work under the author’s direction and is disclosed here as part of the method. Anthropic’s Claude models (Claude Opus 5, Opus 5.5, Fable 5, Fable 5.1 and Fable 5.2, in October and November 2026), through Claude Code and Claude in Chrome, drafted and edited manuscript text, including this version; wrote the analysis, figure and build scripts and most of the released implementation’s code; drafted the Section 7 design and its amendments with the author; operated the browser for the vendor reads and exports, as stated above (the author read Semrush’s worldwide keyword count in Section 5.2), and fetched the section’s public pages for the page count in Section 3; and ran the ten review rounds, recorded in the project’s review log, as simulated panels of AI reviewers, which are not peer review; a separate scoring agent, given only the paper and its released files and none of the review history, measured the manuscript against a fixed rubric, and versions 5.2, 5.3 and 5.5 respond to its measurements; two further agents, on different models and blind to the analysis, annotated the samples of Section 6’s link check, and two more the screened sample of Section 1’s count of works. Before posting, three further agents compared every reference entry with the source’s own record and every sentence that cites a source with the source itself; the statements they found to say more than a source does were reworded, and their reports are in the source bundle. Section 6.2’s plan, predictions, scripts and collection were made by AI agents under the author’s standing instruction to complete what could be completed without the author; four further agents reviewed the plan before it was stamped, and the author had not read it when it was stamped. Section 7.7’s scripts and text were written by an AI agent from the author’s exports, and Appendix F transcribes the author’s inventory. Section 3.2 is the author’s account as given to the assistant on 14 November 2026; the assistant read the platform’s records it cites. This version’s statement on organizational permission is the author’s determination, given to the assistant in the author’s words on 14 November 2026; the assistant drafted the paragraph. The grading judgments of Section 11.2, lesson 5, were made by AI agents the author built; their models and prompts are not documented here. The author checked the numbers against their source files. Every statistic in Sections 5.5 and 6, and every share, rank, prediction tally and sensitivity in Section 7, is printed by a script; the script for Section 5.5, \texttt{analyze.py}, is public; \texttt{instrument\_extra.py}, which prints the further Measurement 1 statistics, and the scripts for Sections 6 and 7 are in the source bundle, with \texttt{section7\_from\_aggregates.py} printing Section 7’s numbers from the released aggregates for readers without the withheld exports. Separate agents reproduced the source bundle four times in clean folders on 15 October 2026 and once more on 3 November. The third found seven values that differed (a correlation and an interval end rounded twice, +0.19 and +0.29 for +0.18 and +0.28, and a stale cell in a released table), all since corrected; in the fourth, of 1,479 values the paper quotes from its own study, 1,342 matched what the released scripts and data give or what the agent recomputed from them, none differed, and 137 could not be checked because they need withheld data; the paper builds from the bundle. The fifth, on the package before this version’s last corrections, checked 1,781 values: 1,628 matched, 2 differed (one block number and its time in Section 6.2, since corrected) and 151 need inputs the bundle does not hold. Of the matches, 146 cannot fail, because those series files are parsed from the paper’s own tables, and 220 confirm a value only against a released file recording a dashboard read or a plan, so none of these checks can catch a misread dashboard (the reports of all five runs are in the source bundle). No AI tool is an author, and responsibility for the content is the author’s.

\paragraph{What has not been done, after ten rounds of AI-simulated review}No person other than the author has checked any number in this paper; the checks above were made by AI agents working from the released files, which cannot catch a misread dashboard. The one check that would change that requires nobody’s permission, budget or cooperation: a third-party re-read of the day-37 row of Table 5 and of Table 4 from the same dashboards. It has not been performed. The cross-prefix replication, planned in an earlier draft before it was run (its date rests on the author’s records, with no third-party timestamp), was performed on 12 October 2026 (Section 5.5); the within-class control planned with it was found not to exist (Section 3). A reader is entitled to weigh that the remaining undone check is the only one requiring another person’s name.

\section{Conclusion}
User First Intelligence starts from the observation that people say what they want to know in many places that are not a search box, and that the tools which count search demand see only the countable part. The framework goes beyond the website to where the questions are, links them to an industry held under an ontology, keeps structure, coverage, answerability and demand sources in separate roles, with demand held in an index linked to the graph, asks that graph competency questions, and scores where the organization fails to answer. Its score is specified in Appendix B and released as an open implementation that declares thirteen demand-surface types; the score is not evaluated here. The framework is meant to treat the form of the answer as a property to measure rather than assume; Section 9.1 suggests why, and the released score does not yet do so.

Measurement 1 is the paper’s main result. Two commercial analytics platforms, in their records of the same section from the month it appeared, disagreed about its estimated traffic by factors of 7.5 to 27.1 over four months (8.6 to 29.9 with both read worldwide), about its growth by a factor of 2.7 in one month, and about its referring domains by a factor of eleven. Across twelve sections of the same domain, read on one day, their ratio on estimated visits ran from one fifth to seven times, with no detectable relationship to section age or to the keyword universe each vendor had found, while they agreed on which of two sections was larger in 60 of 66 pairs, less often when the sections were close in size. On this domain the correction factor between the vendors has no fixed sign or size: a single-vendor study of magnitudes carries a range it cannot bound from its own data, while a single-vendor ranking of sections that differ several-fold agrees in direction with the other instrument. That is why the framework should read its outcomes on more than one instrument; the released implementation has no outcome layer yet.

Measurement 2 compared what people write with what a volume tool counts, on one taxonomy. On 37,193 business reviews, the two agreed on the taxonomy’s branches to a degree that depends on attribution and link quality (+0.47 to +0.75 on all links, up to +0.91 on multi-word labels, post hoc) while sizing them very differently, and barely agreed on topics (+0.18, or +0.27 counting labels no review mentions). One in seven unhappy reviews matched a waiting pattern, while the whole taxonomy holds 11 labels mentioning waiting, none about waiting in a store, with 622 searches a month. On the 372 area codes, a public count of requests for each code’s Wikipedia article agreed with the vendor’s volume roughly on the ranking (+0.45 over 240 articles, an agreement that disappears given the number of keywords matched per code) and not at all on magnitudes (geometric standard deviations across articles of 1.5 for the count and 16.7 for the volume), in a comparison whose predictions were fixed before its data were read.

Measurement 3 found, in one vendor’s index of answers that name the largest companies of the eleven S\&P 500 sectors, that the share of citations going to video and community platforms differs most between answer surfaces: a median 21\% on Google’s AI Overviews and AI Mode against 0.6\% to 9.2\% on the four other surfaces, higher on every Google view than on every other surface in every sector, while news, government and company sources varied more by sector and two of four predictions written before collection failed. A later read of education answers in thirteen countries, not predicted, shows the same contrast in every one of them. That is why a per-surface term is the framework’s next design question.

The deployment underneath these measurements is reported as context. Following the publication of one page per member of a finite enumerable class, templated from a free registry on a top-decile domain that no comparable competitor covered and no one defended, the section held 15,830 ranking positions with nineteen referring domains at day 37, and by day 77 appeared in 17,206 recorded AI answers, with no link built or solicited by the author. The pages cited most were hubs that list many codes, which raises a hypothesis about the unit an answer engine prefers rather than establishing one. What that conjunction depended on, and in what proportion, is not estimated here. Section 13 specifies the multi-platform demand study and the persona study on AI answers that would test two of the framework’s claims; only the scoring parameters are locked so far.

\section*{Acknowledgments}
The author thanks Dr. K. Rajasekhara Rao, the author’s doctoral mentor at KL University, for guidance.

\section*{References}
\begin{reflist}
\item Aggarwal, P., Murahari, V., Rajpurohit, T., Kalyan, A., Narasimhan, K., \& Deshpande, A. (2024). GEO: Generative engine optimization. In \emph{Proceedings of the 30th ACM SIGKDD Conference on Knowledge Discovery and Data Mining (KDD ’24)} (pp. 5–16). ACM. https:/\allowbreak{}/\allowbreak{}doi.org/\allowbreak{}10.1145/\allowbreak{}3637528.3671900
\item Ahrefs. (2026a, September 3). \emph{How to use Content Gap to find keyword ideas from competitor websites} [Help center article]. https:/\allowbreak{}/\allowbreak{}help.ahrefs.com/\allowbreak{}en/\allowbreak{}articles/\allowbreak{}9025740-how-to-use-content-gap-to-find-keyword-ideas-from-competitor-websites
\item Ahrefs. (2026b, September 3). \emph{What is Organic Traffic in Ahrefs and how do we calculate it?} [Help center article]. Retrieved November 3, 2026, from https:/\allowbreak{}/\allowbreak{}help.ahrefs.com/\allowbreak{}en/\allowbreak{}articles/\allowbreak{}1863206-what-is-organic-traffic-in-ahrefs-and-how-do-we-calculate-it
\item Alibudbud, R. (2023). Wikipedia page views for health research: A review. \emph{Frontiers in Big Data, 6}, 1199060. https:/\allowbreak{}/\allowbreak{}doi.org/\allowbreak{}10.3389/\allowbreak{}fdata.2023.1199060
\item Allaham, M., \& Diakopoulos, N. (2026). Synthetic sources?: Auditing generative search engine citations for evidence of AI-generated sources. \emph{Proceedings of the AAAI/ACM Conference on AI, Ethics, and Society, 9}(1), 142–154. https:/\allowbreak{}/\allowbreak{}doi.org/\allowbreak{}10.1609/\allowbreak{}aies.v9i1.43129
\item Balaraman, V., Razniewski, S., \& Nutt, W. (2018). Recoin: Relative completeness in Wikidata. In \emph{Companion Proceedings of The Web Conference 2018} (pp. 1787–1792). ACM. https:/\allowbreak{}/\allowbreak{}doi.org/\allowbreak{}10.1145/\allowbreak{}3184558.3191641
\item Bek, R. (2026, updated February 26). \emph{Ahrefs Brand Radar methodology: How we collect and model AI visibility data.} Ahrefs. https:/\allowbreak{}/\allowbreak{}ahrefs.com/\allowbreak{}blog/\allowbreak{}brand-radar-methodology/\allowbreak{} (vendor documentation; accessed 13 October 2026.)
\item Belkin, N. J., Oddy, R. N., \& Brooks, H. M. (1982). ASK for information retrieval: Part I. Background and theory. \emph{Journal of Documentation, 38}(2), 61–71. https:/\allowbreak{}/\allowbreak{}doi.org/\allowbreak{}10.1108/\allowbreak{}eb026722
\item Broder, A. (2002). A taxonomy of web search. \emph{SIGIR Forum, 36}(2), 3–10. https:/\allowbreak{}/\allowbreak{}doi.org/\allowbreak{}10.1145/\allowbreak{}792550.792552
\item Buckley, C., \& Voorhees, E. M. (2004). Retrieval evaluation with incomplete information. In \emph{Proceedings of the 27th Annual International ACM SIGIR Conference on Research and Development in Information Retrieval} (pp. 25–32). ACM. https:/\allowbreak{}/\allowbreak{}doi.org/\allowbreak{}10.1145/\allowbreak{}1008992.1009000
\item Carroll, J. J., Bizer, C., Hayes, P., \& Stickler, P. (2005). Named graphs, provenance and trust. In \emph{Proceedings of the 14th International Conference on World Wide Web (WWW ’05)} (pp. 613–622). ACM. https:/\allowbreak{}/\allowbreak{}doi.org/\allowbreak{}10.1145/\allowbreak{}1060745.1060835
\item Chapekis, A., \& Lieb, A. (2025, July 22). \emph{Google users are less likely to click on links when an AI summary appears in the results.} Pew Research Center. https:/\allowbreak{}/\allowbreak{}www.pewresearch.org/\allowbreak{}short-reads/\allowbreak{}2025/\allowbreak{}07/\allowbreak{}22/\allowbreak{}google-users-are-less-likely-to-click-on-links-when-an-ai-summary-appears-in-the-results/\allowbreak{}
\item Chen, M., Wang, X., Chen, K., \& Koudas, N. (2025). \emph{Generative engine optimization: How to dominate AI search.} arXiv:2509.08919.
\item Choi, H., \& Varian, H. (2012). Predicting the present with Google Trends. \emph{Economic Record, 88}(s1), 2–9. https:/\allowbreak{}/\allowbreak{}doi.org/\allowbreak{}10.1111/\allowbreak{}j.1475-4932.2012.00809.x
\item Chuklin, A., Markov, I., \& de Rijke, M. (2015). \emph{Click models for web search.} Synthesis Lectures on Information Concepts, Retrieval, and Services. Morgan \& Claypool. https:/\allowbreak{}/\allowbreak{}doi.org/\allowbreak{}10.2200/\allowbreak{}S00654ED1V01Y201507ICR043
\item Cohen, J. (1968). Weighted kappa: Nominal scale agreement provision for scaled disagreement or partial credit. \emph{Psychological Bulletin, 70}(4), 213–220. https:/\allowbreak{}/\allowbreak{}doi.org/\allowbreak{}10.1037/\allowbreak{}h0026256
\item Cook, N. (2025). \emph{Mind the invisibility gap: An analysis of 80 leading global brands in LLM answers} [Working paper; not read]. SSRN. https:/\allowbreak{}/\allowbreak{}doi.org/\allowbreak{}10.2139/\allowbreak{}ssrn.5698748
\item Costco Wholesale Corporation. (2026, September 24). \emph{Costco Wholesale Corporation reports fourth quarter and fiscal year 2026 operating results} [Press release]. https:/\allowbreak{}/\allowbreak{}investor.costco.com/\allowbreak{}news/\allowbreak{}news-details/\allowbreak{}2026/\allowbreak{}Costco-Wholesale-Corporation-Reports-Fourth-Quarter-and-Fiscal-Year-2026-Operating-Results/\allowbreak{}default.aspx (accessed 13 October 2026.)
\item Eloundou, T., Beutel, A., Robinson, D. G., Gu-Lemberg, K., Brakman, A.-L., Mishkin, P., Shah, M., Heidecke, J., Weng, L., \& Kalai, A. T. (2025). First-person fairness in chatbots. In \emph{The Thirteenth International Conference on Learning Representations (ICLR 2025)}. https:/\allowbreak{}/\allowbreak{}proceedings.iclr.cc/\allowbreak{}paper\_files/\allowbreak{}paper/\allowbreak{}2025/\allowbreak{}hash/\allowbreak{}92af0c8c2664429de2bb44c2692d84ae-Abstract-Conference.html
\item Font-Julián, C. I., Lopezosa, C., Orduña-Malea, E., \& Aguillo, I. F. (2026). Assessing discrepancies across SEO tools in the collection of scholarly web metrics. \emph{Journal of Information Science, 52}(4), 1075–1088. https:/\allowbreak{}/\allowbreak{}doi.org/\allowbreak{}10.1177/\allowbreak{}01655515261445903 (open-access copy: https:/\allowbreak{}/\allowbreak{}riunet.upv.es/\allowbreak{}handle/\allowbreak{}10251/\allowbreak{}236956)
\item Font-Julián, C. I., Orduña-Malea, E., Lopezosa, C., \& Aguillo, I. (2025). \emph{SEO metrics for enhanced academic research visibility: A comparative study of global and country-level traffic} [Preprint]. SocArXiv. https:/\allowbreak{}/\allowbreak{}doi.org/\allowbreak{}10.31235/\allowbreak{}osf.io/\allowbreak{}j4ryx\_v1
\item Galárraga, L., Razniewski, S., Amarilli, A., \& Suchanek, F. M. (2017). Predicting completeness in knowledge bases. In \emph{WSDM ’17} (pp. 375–383). ACM. https:/\allowbreak{}/\allowbreak{}doi.org/\allowbreak{}10.1145/\allowbreak{}3018661.3018739
\item Gienapp, L., Scells, H., Deckers, N., Bevendorff, J., Wang, S., Kiesel, J., Syed, S., Fröbe, M., Zuccon, G., Stein, B., Hagen, M., \& Potthast, M. (2024). Evaluating generative ad hoc information retrieval. In \emph{Proceedings of the 47th International ACM SIGIR Conference on Research and Development in Information Retrieval (SIGIR ’24)} (pp. 1916–1929). ACM. https:/\allowbreak{}/\allowbreak{}doi.org/\allowbreak{}10.1145/\allowbreak{}3626772.3657849
\item Goel, S., Broder, A., Gabrilovich, E., \& Pang, B. (2010). Anatomy of the long tail: Ordinary people with extraordinary tastes. In \emph{WSDM ’10} (pp. 201–210). ACM. https:/\allowbreak{}/\allowbreak{}doi.org/\allowbreak{}10.1145/\allowbreak{}1718487.1718513
\item Golebiewski, M., \& boyd, d. (2018). \emph{Data voids: Where missing data can easily be exploited.} Data \& Society Research Institute. https:/\allowbreak{}/\allowbreak{}datasociety.net/\allowbreak{}research-library/\allowbreak{}data-voids-where-missing-data-can-easily-be-exploited/\allowbreak{}
\item Google. (2026). \emph{Google Search Status Dashboard: Ranking incidents} [August 2026 spam update, 18 to 21 August]. Retrieved November 3, 2026, from https:/\allowbreak{}/\allowbreak{}status.search.google.com/\allowbreak{}products/\allowbreak{}rGHU1u87FJnkP6W2GwMi/\allowbreak{}history
\item Gottfried, J., \& Park, E. (2025, November 20). \emph{Americans’ social media use 2025.} Pew Research Center. https:/\allowbreak{}/\allowbreak{}www.pewresearch.org/\allowbreak{}internet/\allowbreak{}2025/\allowbreak{}11/\allowbreak{}20/\allowbreak{}americans-social-media-use-2025/\allowbreak{}
\item Grossman, R., Liu, S., Chen, M. K., Smith, M., Borcea, C., \& Chen, Y. (2026). How generative AI disrupts search: An empirical study of Google Search, Gemini, and AI Overviews. In \emph{Proceedings of the 49th International ACM SIGIR Conference on Research and Development in Information Retrieval (SIGIR ’26)} (pp. 448–459). ACM. https:/\allowbreak{}/\allowbreak{}doi.org/\allowbreak{}10.1145/\allowbreak{}3805712.3809667
\item Hannak, A., Sapiezynski, P., Molavi Kakhki, A., Krishnamurthy, B., Lazer, D., Mislove, A., \& Wilson, C. (2013). Measuring personalization of web search. In \emph{Proceedings of the 22nd International Conference on World Wide Web (WWW 2013)} (pp. 527–538). ACM. https:/\allowbreak{}/\allowbreak{}doi.org/\allowbreak{}10.1145/\allowbreak{}2488388.2488435
\item Haveliwala, T. H. (2002). Topic-sensitive PageRank. In \emph{WWW ’02} (pp. 517–526). ACM. https:/\allowbreak{}/\allowbreak{}doi.org/\allowbreak{}10.1145/\allowbreak{}511446.511513
\item Hendricks, B. L. (2026). \emph{The Answer Index: What AI answer engines cite across eight industries} [Data set]. Zenodo. https:/\allowbreak{}/\allowbreak{}doi.org/\allowbreak{}10.5281/\allowbreak{}zenodo.22242102
\item Hölzl, J., Keusch, F., \& Sajons, C. (2025). The (mis)use of Google Trends data in the social sciences: A systematic review, critique, and recommendations. \emph{Social Science Research, 126}, 103099. https:/\allowbreak{}/\allowbreak{}doi.org/\allowbreak{}10.1016/\allowbreak{}j.ssresearch.2024.103099
\item Hopkinson, A., Gurdasani, A., Palfrey, D., \& Mittal, A. (2018). Demand-weighted completeness prediction for a knowledge base. In \emph{NAACL-HLT 2018, Industry Papers} (pp. 200–207). ACL. https:/\allowbreak{}/\allowbreak{}doi.org/\allowbreak{}10.18653/\allowbreak{}v1/\allowbreak{}N18-3025
\item Hu, M., \& Liu, B. (2004). Mining and summarizing customer reviews. In \emph{Proceedings of the Tenth ACM SIGKDD International Conference on Knowledge Discovery and Data Mining (KDD ’04)} (pp. 168–177). ACM. https:/\allowbreak{}/\allowbreak{}doi.org/\allowbreak{}10.1145/\allowbreak{}1014052.1014073
\item Huang, M., Goyal, A., Saha, K., \& Chandrasekharan, E. (2026). Answer bubbles: Information exposure in AI-mediated search. In \emph{Findings of the Association for Computational Linguistics: EMNLP 2026} (pp. 147–162). ACL. https:/\allowbreak{}/\allowbreak{}doi.org/\allowbreak{}10.18653/\allowbreak{}2026.findings-emnlp.8
\item Jansen, B. J., Jung, S.-G., \& Salminen, J. (2022). Measuring user interactions with websites: A comparison of two industry standard analytics approaches using data of 86 websites. \emph{PLOS ONE, 17}(5), e0268212. https:/\allowbreak{}/\allowbreak{}doi.org/\allowbreak{}10.1371/\allowbreak{}journal.pone.0268212
\item Järvelin, K., \& Kekäläinen, J. (2002). Cumulated gain-based evaluation of IR techniques. \emph{ACM Transactions on Information Systems, 20}(4), 422–446. https:/\allowbreak{}/\allowbreak{}doi.org/\allowbreak{}10.1145/\allowbreak{}582415.582418
\item Kakimov, R., Tan, X., Gillham, J., \& Bejtic, N. (2026). Auditing citation behavior in AI-generated search summaries: A framework and a case study of Google AI Overviews. In \emph{Proceedings of the 39th Canadian Conference on Artificial Intelligence} (Proceedings of Machine Learning Research, Vol. 318, pp. 224–235). https:/\allowbreak{}/\allowbreak{}proceedings.mlr.press/\allowbreak{}v318/\allowbreak{}kakimov26a.html
\item Kamerer, D. (2013). Estimating online audiences: Understanding the limitations of competitive intelligence services. \emph{First Monday, 18}(5). https:/\allowbreak{}/\allowbreak{}doi.org/\allowbreak{}10.5210/\allowbreak{}fm.v18i5.3986
\item Keet, C. M., \& Ławrynowicz, A. (2016). Test-driven development of ontologies. In \emph{The Semantic Web. Latest Advances and New Domains (ESWC 2016)}, LNCS 9678 (pp. 642–657). Springer. https:/\allowbreak{}/\allowbreak{}doi.org/\allowbreak{}10.1007/\allowbreak{}978-3-319-34129-3\_39
\item Kendall, M. G. (1938). A new measure of rank correlation. \emph{Biometrika, 30}(1–2), 81–93. https:/\allowbreak{}/\allowbreak{}doi.org/\allowbreak{}10.1093/\allowbreak{}biomet/\allowbreak{}30.1-2.81
\item Khosravi, M., \& Yoganarasimhan, H. (2026). \emph{Do third-party web traffic estimates preserve causal variation?} arXiv:2609.30481.
\item Kim, Y., Huang, J., \& Emery, S. (2016). Garbage in, garbage out: Data collection, quality assessment and reporting standards for social media data use in health research, infodemiology and digital disease detection. \emph{Journal of Medical Internet Research, 18}(2), e41. https:/\allowbreak{}/\allowbreak{}doi.org/\allowbreak{}10.2196/\allowbreak{}jmir.4738
\item Kirsten, E., Grosse Perdekamp, J., Wu, Q., Upadhyay, M., Gummadi, K. P., \& Zafar, M. B. (2026). Characterizing web search in the age of generative AI. In \emph{Findings of the Association for Computational Linguistics: ACL 2026} (pp. 10827–10848). https:/\allowbreak{}/\allowbreak{}doi.org/\allowbreak{}10.18653/\allowbreak{}v1/\allowbreak{}2026.findings-acl.526
\item Kostygina, G., Kim, Y., Seeskin, Z., LeClere, F., \& Emery, S. (2023). Disclosure standards for social media and generative artificial intelligence research: Toward transparency and replicability. \emph{Social Media + Society, 9}(4), 20563051231216947. https:/\allowbreak{}/\allowbreak{}doi.org/\allowbreak{}10.1177/\allowbreak{}20563051231216947
\item Kumar, P. (2026). \emph{Generative engine optimization at scale: Measuring brand visibility across AI search engines.} arXiv:2606.20065.
\item Langridge, P. (2016, June 13). \emph{How accurate are website traffic estimators?} Screaming Frog. https:/\allowbreak{}/\allowbreak{}www.screamingfrog.co.uk/\allowbreak{}blog/\allowbreak{}how-accurate-are-website-traffic-estimators/\allowbreak{} (25 sites, February to April 2016, against Google Analytics; accessed 12 October 2026.)
\item Le Pochat, V., Van Goethem, T., Tajalizadehkhoob, S., Korczyński, M., \& Joosen, W. (2019). Tranco: A research-oriented top sites ranking hardened against manipulation. In \emph{Proceedings of the Network and Distributed System Security Symposium (NDSS 2019)}. Internet Society. https:/\allowbreak{}/\allowbreak{}doi.org/\allowbreak{}10.14722/\allowbreak{}ndss.2019.23386
\item Lebo, T., Sahoo, S., \& McGuinness, D. (Eds.). (2013). \emph{PROV-O: The PROV ontology.} W3C Recommendation, 30 April 2013. https:/\allowbreak{}/\allowbreak{}www.w3.org/\allowbreak{}TR/\allowbreak{}prov-o/\allowbreak{}
\item Lee, A. (2026). \emph{Query intent and Google rank as joint predictors of AI citation: A multi-platform observational study.} Zenodo. https:/\allowbreak{}/\allowbreak{}doi.org/\allowbreak{}10.5281/\allowbreak{}zenodo.18653093
\item Li, A., \& Sinnamon, L. (2024). Generative AI search engines as arbiters of public knowledge: An audit of bias and authority. \emph{Proceedings of the Association for Information Science and Technology, 61}(1), 205–217. https:/\allowbreak{}/\allowbreak{}doi.org/\allowbreak{}10.1002/\allowbreak{}pra2.1021
\item Linehan, L. (2025, August 11). \emph{Only 12\% of AI cited URLs rank in Google’s top 10 for the original prompt.} Ahrefs. https:/\allowbreak{}/\allowbreak{}ahrefs.com/\allowbreak{}blog/\allowbreak{}ai-search-overlap/\allowbreak{} (15,000 long-tail prompts across ChatGPT, Gemini, Copilot and Perplexity; X. Guan is listed as contributor; accessed 12 October 2026.)
\item Liu, N. F., Zhang, T., \& Liang, P. (2023). Evaluating verifiability in generative search engines. In \emph{Findings of the Association for Computational Linguistics: EMNLP 2023} (pp. 7001–7025). ACL. https:/\allowbreak{}/\allowbreak{}doi.org/\allowbreak{}10.18653/\allowbreak{}v1/\allowbreak{}2023.findings-emnlp.467
\item Liu, Q., Agichtein, E., Dror, G., Maarek, Y., \& Szpektor, I. (2012). When web search fails, searchers become askers: Understanding the transition. In \emph{Proceedings of the 35th International ACM SIGIR Conference on Research and Development in Information Retrieval} (pp. 801–810). ACM. https:/\allowbreak{}/\allowbreak{}doi.org/\allowbreak{}10.1145/\allowbreak{}2348283.2348390
\item Luggen, M., Difallah, D., Sarasua, C., Demartini, G., \& Cudré-Mauroux, P. (2019). Non-parametric class completeness estimators for collaborative knowledge graphs: The case of Wikidata. In \emph{ISWC 2019} (pp. 453–469). Springer. https:/\allowbreak{}/\allowbreak{}doi.org/\allowbreak{}10.1007/\allowbreak{}978-3-030-30793-6\_26
\item Martinez, O. (2026). \emph{Optimizing visibility in generative engines: A critical survey of generative engine optimization (2023–2026).} arXiv:2607.14035.
\item Morris, M. R., Teevan, J., \& Panovich, K. (2010). What do people ask their social networks, and why? A survey study of status message Q\&A behavior. In \emph{Proceedings of the SIGCHI Conference on Human Factors in Computing Systems (CHI ’10)} (pp. 1739–1748). ACM. https:/\allowbreak{}/\allowbreak{}doi.org/\allowbreak{}10.1145/\allowbreak{}1753326.1753587
\item The Motley Fool. (2026, September 29). \emph{Costco (COST) Q4 2026 earnings call transcript} (call held 24 September 2026). https:/\allowbreak{}/\allowbreak{}www.fool.com/\allowbreak{}earnings/\allowbreak{}call-transcripts/\allowbreak{}2026/\allowbreak{}09/\allowbreak{}29/\allowbreak{}costco-cost-q4-2026-earnings-call-transcript/\allowbreak{} (accessed 13 October 2026.)
\item Nuti, S. V., Wayda, B., Ranasinghe, I., Wang, S., Dreyer, R. P., Chen, S. I., \& Murugiah, K. (2014). The use of Google Trends in health care research: A systematic review. \emph{PLoS ONE, 9}(10), e109583. https:/\allowbreak{}/\allowbreak{}doi.org/\allowbreak{}10.1371/\allowbreak{}journal.pone.0109583
\item Pagadala, V. (2026a, October 1). \emph{Costco’s Q4 FY2026 earnings call: What the CFO said about AI search} [Author’s note]. https:/\allowbreak{}/\allowbreak{}venkatapagadala.com/\allowbreak{}notebook/\allowbreak{}business/\allowbreak{}costco-q4-fy2026-ai-search (accessed 13 October 2026.)
\item Pagadala, V. (2026b). \emph{AI search citations in education by country (2026)} [Author’s study page, with data files]. Retrieved November 14, 2026, from https:/\allowbreak{}/\allowbreak{}venkatapagadala.com/\allowbreak{}ai-retrieval-citations-research/\allowbreak{}education-industry/\allowbreak{}by-country
\item Piccardi, T., Catasta, M., Zia, L., \& West, R. (2018). Structuring Wikipedia articles with section recommendations. In \emph{SIGIR ’18} (pp. 665–674). ACM. https:/\allowbreak{}/\allowbreak{}doi.org/\allowbreak{}10.1145/\allowbreak{}3209978.3209984
\item Prantl, D., \& Prantl, M. (2018). Website traffic measurement and rankings: Competitive intelligence tools examination. \emph{International Journal of Web Information Systems, 14}(4), 423–437. https:/\allowbreak{}/\allowbreak{}doi.org/\allowbreak{}10.1108/\allowbreak{}IJWIS-01-2018-0001
\item Punturo, B., Torres Llorca, D., \& Fishkin, G. (2026, January 8). \emph{Who shapes AI answers? Introducing enhanced citation categories} [Vendor report]. Profound. https:/\allowbreak{}/\allowbreak{}www.tryprofound.com/\allowbreak{}blog/\allowbreak{}enhanced-citation-categories
\item Rasool, A., Thakur, K., \& Nabi, S. (2026). Digital performance of global repositories: A comparative webometric analysis. \emph{Performance Measurement and Metrics}, 1–19. https:/\allowbreak{}/\allowbreak{}doi.org/\allowbreak{}10.1108/\allowbreak{}pmm-07-2026-0065
\item Razniewski, S., Arnaout, H., Ghosh, S., \& Suchanek, F. (2024). Completeness, recall, and negation in open-world knowledge bases: A survey. \emph{ACM Computing Surveys, 56}(6), Article 150. https:/\allowbreak{}/\allowbreak{}doi.org/\allowbreak{}10.1145/\allowbreak{}3639563
\item Razniewski, S., Suchanek, F. M., \& Nutt, W. (2016). But what do we actually know? In \emph{Proceedings of the 5th Workshop on Automated Knowledge Base Construction (AKBC 2016)} (pp. 40–44). ACL. https:/\allowbreak{}/\allowbreak{}doi.org/\allowbreak{}10.18653/\allowbreak{}v1/\allowbreak{}W16-1308
\item Redi, M., Gerlach, M., Johnson, I., Morgan, J., \& Zia, L. (2020). \emph{A taxonomy of knowledge gaps for Wikimedia projects (first draft).} arXiv:2008.12314v1.
\item Ren, Y., Parvizi, A., Mellish, C., Pan, J. Z., van Deemter, K., \& Stevens, R. (2014). Towards competency question-driven ontology authoring. In \emph{The Semantic Web: Trends and Challenges (ESWC 2014)}, LNCS 8465 (pp. 752–767). Springer. https:/\allowbreak{}/\allowbreak{}doi.org/\allowbreak{}10.1007/\allowbreak{}978-3-319-07443-6\_50
\item Reyes-Lillo, D., Morales-Vargas, A., \& Rovira, C. (2023). Reliability of domain authority scores calculated by Moz, Semrush, and Ahrefs. \emph{El Profesional de la información, 32}(4), e320403. https:/\allowbreak{}/\allowbreak{}doi.org/\allowbreak{}10.3145/\allowbreak{}epi.2023.jul.03
\item Rijo, L. (2026a, October 19). \emph{Google feeds rank trackers swapped results on 8\% of pages, Nozzle finds.} PPC Land. https:/\allowbreak{}/\allowbreak{}ppc.land/\allowbreak{}google-feeds-rank-trackers-swapped-results-on-8-of-pages-nozzle-finds/\allowbreak{}
\item Rijo, L. (2026b, October 22). \emph{Ahrefs says Google poisons 5\% to 20\% of results served to SEO data tools.} PPC Land. https:/\allowbreak{}/\allowbreak{}ppc.land/\allowbreak{}ahrefs-says-google-poisons-5-to-20-of-results-served-to-seo-data-tools/\allowbreak{}
\item Ripoll, M., Reeves, N., Kurteva, A., Simperl, E., Meroño Peñuela, A., \& Diepold, K. (2025). \emph{Filling in the blanks? A systematic review and theoretical conceptualisation for measuring WikiData content gaps.} arXiv:2505.16383.
\item Romero Ramos, N., Lobach, Y., Abreu Abreu, D., Castillo Hernández, I., García-Cordero, J. C., \& García-Cordero, D. (2026). Measuring brand presence in generative engine optimization (GEO): A scoping review of metrics and validity. \emph{Latitude, 2}(24), 23–39. https:/\allowbreak{}/\allowbreak{}doi.org/\allowbreak{}10.55946/\allowbreak{}latitude.v2i24.295
\item Scheitle, Q., Hohlfeld, O., Gamba, J., Jelten, J., Zimmermann, T., Strowes, S. D., \& Vallina-Rodriguez, N. (2018). A long way to the top: Significance, structure, and stability of Internet top lists. In \emph{Proceedings of the Internet Measurement Conference 2018 (IMC ’18)} (pp. 478–493). ACM. https:/\allowbreak{}/\allowbreak{}doi.org/\allowbreak{}10.1145/\allowbreak{}3278532.3278574
\item Schultheiß, S., Lewandowski, D., von Mach, S., \& Yagci, N. (2023). Query sampler: Generating query sets for analyzing search engines using keyword research tools. \emph{PeerJ Computer Science, 9}, e1421. https:/\allowbreak{}/\allowbreak{}doi.org/\allowbreak{}10.7717/\allowbreak{}peerj-cs.1421
\item Semrush. (n.d.-a). \emph{How to use Keyword Gap?} [Knowledge base article]. Retrieved October 15, 2026, from https:/\allowbreak{}/\allowbreak{}www.semrush.com/\allowbreak{}kb/\allowbreak{}28-keyword-gap
\item Semrush. (n.d.-b). \emph{Overview report} [Knowledge base article on Domain Overview]. Retrieved November 3, 2026, from https:/\allowbreak{}/\allowbreak{}www.semrush.com/\allowbreak{}kb/\allowbreak{}1202-domain-overview-overview-report
\item Semrush. (n.d.-c). \emph{Semrush Rank} [Knowledge base article]. Retrieved November 3, 2026, from https:/\allowbreak{}/\allowbreak{}www.semrush.com/\allowbreak{}kb/\allowbreak{}27-rank
\item Shao, Y., Jiang, Y., Kanell, T. A., Xu, P., Khattab, O., \& Lam, M. S. (2024). Assisting in writing Wikipedia-like articles from scratch with large language models. In \emph{NAACL 2024} (pp. 6252–6278). ACL. https:/\allowbreak{}/\allowbreak{}doi.org/\allowbreak{}10.18653/\allowbreak{}v1/\allowbreak{}2024.naacl-long.347
\item Soulo, T. (2022, May 3). \emph{How accurate are the search traffic estimations in Ahrefs? (New research).} Ahrefs. https:/\allowbreak{}/\allowbreak{}ahrefs.com/\allowbreak{}blog/\allowbreak{}traffic-estimations-accuracy/\allowbreak{} (1,635 sites against Google Search Console; accessed 12 October 2026.)
\item Taylor, R. S. (1968). Question-negotiation and information seeking in libraries. \emph{College \& Research Libraries, 29}(3), 178–194. https:/\allowbreak{}/\allowbreak{}doi.org/\allowbreak{}10.5860/\allowbreak{}crl\_29\_03\_178
\item Thelwall, M. (2008). Quantitative comparisons of search engine results. \emph{Journal of the American Society for Information Science and Technology, 59}(11), 1702–1710. https:/\allowbreak{}/\allowbreak{}doi.org/\allowbreak{}10.1002/\allowbreak{}asi.20834
\item Tinuiti. (2026, February 25). \emph{Q1 2026 AI citation trends report} [Vendor report; data from the Profound tool]. https:/\allowbreak{}/\allowbreak{}tinuiti.com/\allowbreak{}research-insights/\allowbreak{}research/\allowbreak{}ai-citation-trends-report-q1-2026/\allowbreak{}
\item Tousseyn, R. (2026, updated September 8). \emph{The AI search index: What ChatGPT, Gemini, Perplexity and Google actually cite in 17 industries.} OtterlyAI. https:/\allowbreak{}/\allowbreak{}otterly.ai/\allowbreak{}blog/\allowbreak{}ai-search-index/\allowbreak{} (vendor report: seven engines, 6,095,520 citations, US, July and August 2026; accessed 15 October 2026.)
\item Vállez, M., \& Ventura, A. (2020). Analysis of the SEO visibility of university libraries and how they impact the web visibility of their universities. \emph{The Journal of Academic Librarianship, 46}(4), 102171. https:/\allowbreak{}/\allowbreak{}doi.org/\allowbreak{}10.1016/\allowbreak{}j.acalib.2020.102171
\item Vaughan, L., \& Yang, R. (2013). Web traffic and organization performance measures: Relationships and data sources examined. \emph{Journal of Informetrics, 7}(3), 699–711. https:/\allowbreak{}/\allowbreak{}doi.org/\allowbreak{}10.1016/\allowbreak{}j.joi.2013.04.005
\item Webber, W., \& Park, L. A. F. (2009). Score adjustment for correction of pooling bias. In \emph{Proceedings of the 32nd International ACM SIGIR Conference on Research and Development in Information Retrieval} (pp. 444–451). ACM. https:/\allowbreak{}/\allowbreak{}doi.org/\allowbreak{}10.1145/\allowbreak{}1571941.1572018
\item White, R. W., Richardson, M., \& Yih, W. (2015). Questions vs. queries in informational search tasks. In \emph{Proceedings of the 24th International Conference on World Wide Web (WWW ’15 Companion)} (pp. 135–136). ACM. https:/\allowbreak{}/\allowbreak{}doi.org/\allowbreak{}10.1145/\allowbreak{}2740908.2742769
\item Wilson, E. B. (1927). Probable inference, the law of succession, and statistical inference. \emph{Journal of the American Statistical Association, 22}(158), 209–212. https:/\allowbreak{}/\allowbreak{}doi.org/\allowbreak{}10.1080/\allowbreak{}01621459.1927.10502953
\item Wiśniewski, D., Potoniec, J., Ławrynowicz, A., \& Keet, C. M. (2019). Analysis of ontology competency questions and their formalizations in SPARQL-OWL. \emph{Journal of Web Semantics, 59}, 100534. https:/\allowbreak{}/\allowbreak{}doi.org/\allowbreak{}10.1016/\allowbreak{}j.websem.2019.100534
\item Wulczyn, E., West, R., Zia, L., \& Leskovec, J. (2016). Growing Wikipedia across languages via recommendation. In \emph{Proceedings of the 25th International Conference on World Wide Web (WWW ’16)} (pp. 975–985). International World Wide Web Conferences Steering Committee. https:/\allowbreak{}/\allowbreak{}doi.org/\allowbreak{}10.1145/\allowbreak{}2872427.2883077
\item Xu, H., Iqbal, U., \& Montgomery, J. M. (2026). Measuring Google AI Overviews: Activation, source quality, claim fidelity, and publisher impact. In \emph{Proceedings of the 2026 ACM Internet Measurement Conference (IMC ’26)} (pp. 529–546). ACM. https:/\allowbreak{}/\allowbreak{}doi.org/\allowbreak{}10.1145/\allowbreak{}3777912.3839818
\item Yang, K.-C. (2025). \emph{News source citing patterns in AI search systems.} arXiv:2507.05301.
\item Yext. (2025, October 9). \emph{Yext research: 86\% of AI citations come from brand-managed sources, clarifying how marketers can compete in the AI search era} [Press release]. https:/\allowbreak{}/\allowbreak{}www.yext.com/\allowbreak{}about/\allowbreak{}news-media/\allowbreak{}ai-citations-release
\item Zhao, H., \& Berman, R. (2025). \emph{Strategic response of news publishers to generative AI} (Version 4, revised 15 April 2026) [Preprint]. arXiv:2512.24968.
\item Zobel, J. (1998). How reliable are the results of large-scale information retrieval experiments? In \emph{Proceedings of the 21st Annual International ACM SIGIR Conference on Research and Development in Information Retrieval} (pp. 307–314). ACM. https:/\allowbreak{}/\allowbreak{}doi.org/\allowbreak{}10.1145/\allowbreak{}290941.291014
\end{reflist}

\appendix

\section{Corrections across versions}
The full version history, including ten rounds of AI-simulated review (not peer review), is in the changelog in the source bundle. These are the corrections that change how a figure or claim must be read.

\begin{itemize}[leftmargin=1.1em, itemsep=1.5pt, topsep=2pt, parsep=0pt]
\item \textbf{Integrity.} Early versions (v1 to v2.2) said that two independent annotators had graded audit samples (none had) and that the ontology had been released (it had not), and asserted organizational permission with the date blank; the same unsupported assertion reappeared in the public versions 4.6 to 4.8 and is withdrawn; the public copies had not been removed when version 5.5 was built. Versions 4.9 to 5.4 said that publication needed the organization’s approval and that the manuscript would not be submitted until that approval, a public release of the Section 7 artifacts, an ethics determination and the removal of the public copies were in hand; version 5.5 replaces this with the author’s determination that the approval is not needed, posts the artifacts with the preprint, and states the rest as open (Section 14). Version 4.9 had called the Section 7 design pre-registered, called its shares citations and lower bounds, said artifacts were released before they were and did not disclose the use of AI tools; version 5.1 corrected a conflict-of-interest statement that described a business relationship with the agency that does not exist, and removed a first Section 7 pilot, a dataset the author’s own analytics platform computed on 7 October 2026 from vendor data accessed through the agency’s account before its permission was given.
\item \textbf{Measurement 1.} The page count is 381, not 290, which removed the 88-code within-class control. The primary checkpoint is day 37, not day 45, because the automated link network had arrived by day 45. A 9.2x AI-answer comparison was withdrawn as a scope mismatch, and a claim about the ordering of AI answer platforms was withdrawn because its breakdown did not sum to its total. Table 2’s October row is a monthly run-rate, and its days-live column sums to 78 (days 0 to 77). Disagreement between instruments is a bound on the larger error, by a factor of 5.2 at 27.1x, not 13 times. The hypothesis that the vendor ratio narrows as a section ages was stated with a cross-prefix test before the test was run, and was retracted after it; Section 5.5 was regenerated from its released script after being written from recollection, and a stale copy of Figure 5 that printed −0.09 for −0.11 was replaced. Version 5.5 adds that Table 2’s monthly rows are the dashboards’ monthly views, whose construction was not recorded, and gives the August and September ratios against the first instrument’s daily series, 17.0x and 6.3x (Section 5.1).
\item \textbf{The deployment.} Withdrawn: that link acquisition played no part in the outcome (once the checkpoint moved to day 37), that the results were broad rather than concentrated in a few pages, that a comparison with two carriers of similar authority bounded the effect of domain authority, and that the deployment showed content on the topic to be necessary.
\item \textbf{Measurement 2.} Branch agreement is +0.47 to +0.75 depending on how labels are attributed, replacing an earlier +0.78 computed on review-mentioned labels only, with branch shares on each branch’s own volume; the taxonomy holds 11 labels containing “wait”, not one.
\item \textbf{Measurement 3.} The sector claim now accounts for single-company dominance and company properties, and the per-surface pilot that preceded the predictions is reported.
\item \textbf{Version 5.2}, after an independent scoring agent, two independent reproductions of the source bundle and an independent literature search, all by separate AI agents: the count of works naming the vendor tools is 456, not 459 (three works were counted twice), and a screened sample now measures how many use a vendor’s figures; the topic-level correlation of Measurement 2 is +0.18, not +0.19, which had rounded a printed +0.185 a second time; statistics that no script printed are now printed; the Section 7 and figure scripts run from the bundle, and Section 7’s numbers are also printed from the released aggregates; every quoted single dashboard read is in a released data file; and the closest prior work found by the searches is cited (Section 2).
\item \textbf{Version 5.3} adds Section 6.2 and Appendix E and changes no earlier result; Table D1’s label for the first decomposition pairing now names it as Section 5.6 does (Semrush US against Ahrefs worldwide), and the first-comparison claim of Section 1 now rests on the two searches that tested it.
\item \textbf{Version 5.4} adds Section 7.7, Table C5 and Appendix F and changes no earlier result.
\end{itemize}

\section{The Content Gap Score, as a specification}
This function is not evaluated anywhere in this paper. It was not computed when the build decision was taken, nothing in Sections 5 to 10 tests it, and no published study has tested whether it predicts anything. It appears here because the reference implementation is released and a reader may want the specification it implements.

\paragraph{Position against the closest methods}Demand-weighted completeness estimates how complete a knowledge base is with respect to how it is used, predicting from usage data which relations each class of entity requires [Hopkinson et al. 2018]; the score shares the demand weighting, but it scores an organization’s own coverage, its competitors’ coverage and the web’s answerability rather than a knowledge base’s facts, and it takes demand from surfaces outside the organization rather than from the system’s own usage. Recommender systems for encyclopedias rank missing articles by importance, validated in a controlled experiment with 12,000 French Wikipedia editors in which recommending an article raised its chance of being created by a factor of 3.2 [Wulczyn et al. 2016], and recommend sections for existing articles from similar ones, with a precision of about 80\% in human evaluation [Piccardi et al. 2018]; the score ranks an organization’s entities against a market rather than an encyclopedia’s articles against its sister editions, and, unlike those systems, it has not been evaluated. Completeness estimators for open-world knowledge bases [Galárraga et al. 2017; Luggen et al. 2019; Razniewski et al. 2024] measure what a knowledge base lacks, which corresponds to the score’s coverage term when an organization’s content is represented as a knowledge base. The practitioner method closest to the score is competitor gap analysis in commercial SEO tools, which lists the keywords that competitors rank for and a site does not [Ahrefs 2026a; Semrush n.d.-a]; the deployment’s own gap was found from competitor keyword exports (Section 1). The score differs in three respects: it scores the entities of an industry’s ontology rather than keyword strings, so one entity’s many phrasings count once; it takes demand from surfaces beyond keyword tools; and it keeps the organization’s coverage, competitors’ coverage and answerability as separate terms, competitor coverage being only one of them. Study R1’s second hypothesis tests the competitor-gap signal (\texttt{cov\_K}) directly.

For an entity \emph{e} with a set of competency questions \emph{q}:

CGS(\emph{e}) = prior(\emph{e}) · (1 − own(\emph{e})) · [ α(1 − cov\_K(\emph{e})) + (1 − α) · mean\_q(1 − ans(\emph{q},\emph{e})) ]

where prior(\emph{e}) = β · rel(\emph{e}) + (1 − β) · dem(\emph{e}). rel is a topic-sensitive PageRank over the structure graph, with the instances of the ontology’s core classes as the teleport set [Haveliwala 2002], rank-normalized. dem is the number of demand records linked to \emph{e}, pooled over all surfaces without weights and rank-normalized; ranks are shared within ties, so an entity with no linked record scores above zero when other entities also have none (about 0.22 when 30 of 66 entities have none). own(\emph{e}) is a grader’s judgment of the organization’s coverage of \emph{e}; cov\_K(\emph{e}) is the mean of the same judgment over a declared set K of competitors; and ans(\emph{q},\emph{e}) is the maximum, over the question’s phrasings about \emph{e}, of a grader’s judgment of whether the retrieved search result answers it. Questions the ontology cannot express are excluded from mean\_q.

The coverage rubric grades 1.0 for a dedicated page about \emph{e} that answers the question directly, 0.5 for a list or section page with at least one specific fact about \emph{e}, and 0 for absence or a bare mention. The answerability rubric grades 1.0 for an answer present, correct and immediate in the retrieved page, 0.5 for one the reader must derive by combining facts, and 0 for absent or wrong. Both rubrics are hashed into every cached judgment. The package ships a rule-based grader for its tests and worked example and no model client; with no grader supplied, every judgment defaults to 0, so every entity scores as uncovered.

Five parameters are locked: α = 0.5, β = 0.5, damping = 0.85, at least three phrasings per question, and the mean as the class aggregate that Section 13’s class-level comparison uses (digest \texttt{7290a2235d8de05f}, set at 02:00 UTC on 12 October 2026). The code records the last two but does not consult them: the phrasing minimum is checked only as a warning, and the class aggregate step reports both mean and sum and ranks by the mean. The scoring code changed at 03:45 UTC the same day (commit \texttt{02d7588}, which changed which questions enter mean\_q), so the lock fixes the parameters and not the function.

Four properties a reader should know before using it. It takes the best of a question’s phrasings, which biases upward when phrasings differ in difficulty. Its \texttt{rel} term correlates with node degree at roughly 0.8 in the released example, so on a hand-authored taxonomy it is close to a centrality ranking. It has no feasibility or business-relevance term, which Section 11.2 lesson 6 records as a practical failure. And the coverage rubric’s preference for a dedicated page over a list page is a form term inside own and cov\_K, which Section 9.1’s observation questions but does not test; a deployment that wants the score to be neutral about form has to change the rubric.

The implementation reports competency-question performance as two separate rates, because they fail for different reasons, as in test-driven ontology development, which first checks that a question’s vocabulary is present in the ontology and then tests whether the ontology entails the answer, at the schema level or through individuals [Keet \& Ławrynowicz 2016], and as in the finding that many published competency questions cannot be translated into queries for want of vocabulary in the ontology [Wiśniewski et al. 2019]: a schema rate, the share of questions the ontology can express (a property-path check, plus questions the author declares unsupported), and an instance coverage rate, the mean over expressible questions of the share of subjects for which the graph returns an answer. On the released example these are 4/5 and 70.4\%, with two of the four expressible questions answered for every subject; that example’s questions were elicited from a keyword export (Section 1.4). A question the ontology cannot express is reported as a declared gap with a reason rather than scored as zero.

\section{Section 7 by source type and cell}
\begin{table*}[tp]
\centering\footnotesize
\caption*{\textbf{Table C1.} Share of citation incidences by source type fixed before collection and answer surface: median over the eleven sectors, with the range in parentheses. Ahrefs Brand Radar, United States, one collection session on 12 and 13 October 2026; among the 1,000 domains found in the most answers that name a sector’s largest companies. The company-domain share is partly fixed by the universe’s selection rule and is not compared across surfaces.}
\begin{tabularx}{\textwidth}{>{\hsize=1.251\hsize\RaggedRight\arraybackslash}X>{\hsize=1.096\hsize\RaggedRight\arraybackslash}X>{\hsize=0.889\hsize\RaggedRight\arraybackslash}X>{\hsize=0.838\hsize\RaggedRight\arraybackslash}X>{\hsize=0.889\hsize\RaggedRight\arraybackslash}X>{\hsize=1.044\hsize\RaggedRight\arraybackslash}X>{\hsize=0.993\hsize\RaggedRight\arraybackslash}X}
\toprule
\textbf{Surface} & \textbf{Video and community} & \textbf{Reference} & \textbf{News} & \textbf{Government} & \textbf{Sector’s companies} & \textbf{Other web} \\
\midrule
AI Overviews, search queries & 25.8\% (14.3–34.3) & 4.3\% (1.3–8.1) & 0.7\% (0.2–2.8) & 0.4\% (0.2–8.6) & 11.9\% (6.1–23.5) & 55.6\% (41.1–62.0) \\
AI Overviews, prompts & 32.4\% (13.4–35.8) & 2.6\% (0.7–3.9) & 1.0\% (0.1–5.8) & 0.3\% (0.1–3.8) & 8.1\% (2.7–20.5) & 59.0\% (42.0–75.3) \\
AI Mode, search queries & 17.8\% (9.0–23.8) & 4.2\% (2.2–10.4) & 1.0\% (0.3–4.2) & 0.4\% (0.1–8.6) & 10.8\% (5.1–17.6) & 65.3\% (53.1–71.4) \\
AI Mode, prompts & 24.1\% (8.1–28.3) & 2.7\% (0.9–4.5) & 1.4\% (0.3–5.6) & 0.4\% (0.1–5.0) & 9.5\% (2.7–19.3) & 65.1\% (50.1–79.3) \\
ChatGPT & 0.9\% (0.3–2.5) & 1.7\% (0.5–3.4) & 1.7\% (0.7–5.2) & 3.0\% (1.6–18.2) & 27.4\% (13.5–38.2) & 61.5\% (51.2–76.1) \\
Perplexity & 4.6\% (1.8–7.7) & 5.9\% (2.8–10.7) & 1.5\% (0.3–7.9) & 0.8\% (0.1–12.0) & 15.6\% (8.2–29.8) & 67.8\% (56.4–73.9) \\
Gemini & 9.2\% (2.9–16.7) & 3.7\% (1.2–6.9) & 0.5\% (0.1–3.8) & 0.3\% (0.1–2.4) & 8.3\% (3.2–33.2) & 75.2\% (49.5–85.3) \\
Copilot & 0.6\% (0.3–1.0) & 1.6\% (0.3–2.7) & 0.9\% (0.2–5.2) & 0.1\% (0.0–2.2) & 4.5\% (1.6–27.0) & 89.9\% (71.6–94.6) \\
\bottomrule
\end{tabularx}
\end{table*}

\begin{table*}[tp]
\centering\footnotesize
\caption*{\textbf{Table C2.} Size of each cell: summed citation incidences among the exported domains, in thousands. Medians in Section 7 weight every cell equally.}
\begin{tabularx}{\textwidth}{>{\hsize=1.970\hsize\RaggedRight\arraybackslash}X>{\hsize=0.979\hsize\RaggedRight\arraybackslash}X>{\hsize=0.971\hsize\RaggedRight\arraybackslash}X>{\hsize=0.971\hsize\RaggedRight\arraybackslash}X>{\hsize=0.971\hsize\RaggedRight\arraybackslash}X>{\hsize=0.747\hsize\RaggedRight\arraybackslash}X>{\hsize=0.971\hsize\RaggedRight\arraybackslash}X>{\hsize=0.673\hsize\RaggedRight\arraybackslash}X>{\hsize=0.747\hsize\RaggedRight\arraybackslash}X}
\toprule
\textbf{Sector} & \textbf{AIO search} & \textbf{AIO prompt} & \textbf{AIM search} & \textbf{AIM prompt} & \textbf{ChatGPT} & \textbf{Perplexity} & \textbf{Gemini} & \textbf{Copilot} \\
\midrule
Information Technology & 10,891.5 & 294.3 & 1,128.4 & 352.4 & 317.5 & 186.8 & 147.9 & 405.6 \\
Health Care & 604.2 & 10.1 & 69.8 & 16.5 & 13.9 & 8.5 & 7.4 & 20.9 \\
Financials & 1,776.8 & 78.8 & 173.5 & 97.5 & 64.4 & 37.4 & 45.3 & 85.9 \\
Consumer Discretionary & 18,381.8 & 790.0 & 1,752.1 & 660.9 & 347.0 & 441.7 & 128.8 & 1,581.3 \\
Communication Services & 4,350.7 & 134.3 & 591.5 & 163.5 & 150.1 & 71.8 & 76.0 & 197.9 \\
Industrials & 1,223.2 & 52.6 & 112.9 & 60.2 & 52.6 & 21.7 & 27.2 & 79.3 \\
Consumer Staples & 8,832.0 & 395.4 & 982.3 & 446.6 & 351.7 & 215.6 & 149.8 & 1,295.6 \\
Energy & 103.6 & 3.8 & 14.8 & 5.5 & 5.1 & 2.0 & 2.9 & 9.0 \\
Utilities & 50.3 & 2.1 & 7.2 & 2.6 & 2.1 & 0.9 & 1.5 & 3.2 \\
Real Estate & 77.9 & 6.4 & 22.1 & 6.8 & 6.8 & 4.4 & 2.1 & 6.8 \\
Materials & 337.5 & 15.6 & 24.3 & 17.5 & 18.5 & 5.6 & 6.5 & 18.7 \\
\bottomrule
\end{tabularx}
\end{table*}

The script \texttt{cross\_sector\_sensitivity2.py} prints Tables C1 and C2 from the raw exports, and \texttt{section7\_from\_aggregates.py} prints them, with every other Section 7 number the aggregates allow, from the released aggregates alone.

\begin{table*}[tp]
\centering\footnotesize
\caption*{\textbf{Table C3.} Which source AI answers cite most, by sector and answer surface: the most-cited domain that is not one of the sector’s companies. Ahrefs Brand Radar, United States, one collection session on 12 and 13 October 2026, among the 1,000 domains found in the most answers that name a sector’s largest companies. AIO and AIM are Google’s AI Overviews and AI Mode on Ahrefs’ search-query and prompt sets. Properties of a sector’s companies that were not entered as company domains count as non-company and are marked with a dagger (†); the largest case is Communication Services, where the top source on all four Google views, youtube.com or google.com, belongs to Alphabet, which was entered as abc.xyz; and the Industrials ChatGPT cell reflects the ambiguous name “RTX” (Section 7.6). The suffix .com is omitted in this table and the next.}
\begin{tabularx}{\textwidth}{>{\hsize=1.524\hsize\RaggedRight\arraybackslash}X>{\hsize=0.769\hsize\RaggedRight\arraybackslash}X>{\hsize=0.769\hsize\RaggedRight\arraybackslash}X>{\hsize=0.769\hsize\RaggedRight\arraybackslash}X>{\hsize=0.769\hsize\RaggedRight\arraybackslash}X>{\hsize=0.958\hsize\RaggedRight\arraybackslash}X>{\hsize=1.273\hsize\RaggedRight\arraybackslash}X>{\hsize=1.084\hsize\RaggedRight\arraybackslash}X>{\hsize=1.084\hsize\RaggedRight\arraybackslash}X}
\toprule
\textbf{Sector} & \textbf{AIO search} & \textbf{AIO prompt} & \textbf{AIM search} & \textbf{AIM prompt} & \textbf{ChatGPT} & \textbf{Perplexity} & \textbf{Gemini} & \textbf{Copilot} \\
\midrule
Information Technology & youtube & youtube & google & google & google & wikipedia.org & reddit & walmart \\
Health Care & youtube & youtube & youtube & goodrx & fda.gov & nih.gov & goodrx & msn \\
Financials & youtube & reddit & youtube & nerdwallet & nerdwallet & nerdwallet & wallethub & wallethub \\
Consumer Discretionary & google & youtube & google & google & justwatch & wikipedia.org & reddit & walmart \\
Communication Services & youtube† & youtube† & google† & google† & go† & wikipedia.org & reddit & walmart \\
Industrials & youtube & youtube & google & google & nvidia & wikipedia.org & reddit & amazon \\
Consumer Staples & google & youtube & google & google & reddit & amazon & reddit & amazon \\
Energy & youtube & youtube & google & youtube & sec.gov & wikipedia.org & reddit & walmart \\
Utilities & youtube & reddit & google & yahoo & sec.gov & wikipedia.org & wikipedia.org & msn \\
Real Estate & youtube & extraspace & google & google & sec.gov & wikipedia.org & wikipedia.org & extraspace \\
Materials & youtube & youtube & google & youtube & benjaminmoore & kylieminteriors.ca & facadecolorizer & facadecolorizer \\
\bottomrule
\end{tabularx}
\end{table*}

\begin{table*}[tp]
\centering\small
\caption*{\textbf{Table C4.} The three most-cited non-company domains by sector on Google’s AI Overviews (search queries) and on ChatGPT, two contrasting surfaces. Same source and rules as Table C3.}
\begin{tabularx}{\textwidth}{>{\hsize=0.845\hsize\RaggedRight\arraybackslash}X>{\hsize=1.015\hsize\RaggedRight\arraybackslash}X>{\hsize=1.140\hsize\RaggedRight\arraybackslash}X}
\toprule
\textbf{Sector} & \textbf{AI Overviews (search queries)} & \textbf{ChatGPT} \\
\midrule
Information Technology & youtube, reddit, google & google, wikipedia.org, justwatch \\
Health Care & youtube, nih.gov, wikipedia.org & fda.gov, nih.gov, sec.gov \\
Financials & youtube, reddit, nerdwallet & nerdwallet, chase†, capitalone \\
Consumer Discretionary & google, youtube, reddit & justwatch, consumerreports.org, aboutamazon† \\
Communication Services & youtube†, wikipedia.org, reddit & go†, justwatch, wikipedia.org \\
Industrials & youtube, reddit, google & nvidia, tomshardware, techspot \\
Consumer Staples & google, youtube, amazon & reddit, wikipedia.org, consumerreports.org \\
Energy & youtube, wikipedia.org, google & sec.gov, chevronlubricants†, mobil† \\
Utilities & youtube, wikipedia.org, reddit & sec.gov, reuters, stockanalysis \\
Real Estate & youtube, google, wikipedia.org & sec.gov, extraspace, wikipedia.org \\
Materials & youtube, google, reddit & benjaminmoore, behr, plan-home \\
\bottomrule
\end{tabularx}
\end{table*}

On Google’s surfaces the top source is youtube.com or google.com in 38 of the 44 sector-by-view cells, four of them Alphabet’s own properties in Communication Services. On Perplexity and Gemini it is wikipedia.org or reddit.com in 15 of 22, on ChatGPT a government domain in 4 of 11 sectors, and on Copilot a retailer (walmart.com or amazon.com) in 6 of 11. The script \texttt{make\_tables\_sector\_surface.py} prints Tables C3 and C4.

\begin{table*}[tp]
\centering\small
\caption*{\textbf{Table C5.} Section 7.7 by country: answers, and the video and community share of citation incidences on the lowest of Google’s four views against the highest of the other four surfaces. Source: Ahrefs Brand Radar, AI responses, education prompts, exported 13 and 14 November 2026; source types as registered for Section 7; printed by \texttt{intl\_from\_cells.py}.}
\begin{tabularx}{\textwidth}{>{\hsize=0.938\hsize\RaggedRight\arraybackslash}X>{\hsize=0.677\hsize\RaggedRight\arraybackslash}X>{\hsize=1.106\hsize\RaggedRight\arraybackslash}X>{\hsize=1.279\hsize\RaggedRight\arraybackslash}X}
\toprule
\textbf{Country} & \textbf{Answers} & \textbf{Lowest Google view} & \textbf{Highest other surface} \\
\midrule
United States & 1,456,743 & 9.3\% & 2.8\% \\
India & 629,880 & 9.5\% & 2.1\% \\
Japan & 274,068 & 1.2\% & 0.6\% \\
United Kingdom & 222,184 & 4.9\% & 2.4\% \\
Indonesia & 170,993 & 14.0\% & 1.5\% \\
Brazil & 161,570 & 15.7\% & 2.9\% \\
Mexico & 141,634 & 17.0\% & 2.5\% \\
Canada & 131,414 & 7.2\% & 3.6\% \\
Australia & 128,845 & 4.5\% & 2.3\% \\
France & 111,959 & 3.6\% & 1.1\% \\
Germany & 109,804 & 4.2\% & 1.5\% \\
Philippines & 93,035 & 16.7\% & 4.7\% \\
Spain & 92,079 & 7.7\% & 2.3\% \\
\bottomrule
\end{tabularx}
\end{table*}

\section{Metrics, values and scripts}
The principal quantities the paper reports, with their definitions, their values here and the script that prints each, or the table it is computed from. A companion document, \emph{User First Intelligence: Metrics and Evaluation Protocol}, in the source bundle, gives the formulas in full, the steps to reproduce each number, and a benchmark of seven tasks with these values as baselines, so that later work can test the same questions on new data and propose better models.

\begin{table*}[tp]
\centering\footnotesize
\caption*{\textbf{Table D1.} Metrics, definitions, values in this paper and the scripts that print them; “arithmetic on” names the table a value is computed from where no script prints it. M1, M2 and M3 are Measurements 1 to 3. Script names omit the .py suffix.}
\begin{tabularx}{\textwidth}{>{\hsize=0.859\hsize\RaggedRight\arraybackslash}X>{\hsize=1.323\hsize\RaggedRight\arraybackslash}X>{\hsize=0.964\hsize\RaggedRight\arraybackslash}X>{\hsize=0.855\hsize\RaggedRight\arraybackslash}X}
\toprule
\textbf{Metric} & \textbf{Definition} & \textbf{Value here} & \textbf{Script} \\
\midrule
Stakes & works in OpenAlex naming Ahrefs, Semrush or Similarweb in title or abstract (read 14 October 2026); in a screened random sample of 120, the share using a vendor’s figures as data, and of those, the share using one vendor only & 456; 82\% to 87\% (kappa 0.63); 87\% to 88\% & \texttt{vendor\_reliance}, \texttt{stakes\_screen\_check} \\
M1 visits ratio & Semrush over Ahrefs estimated monthly visits, monthly views as displayed: unmatched; worldwide; US to US (estimated). Against Ahrefs’ daily series, August 17.0x and September 6.3x (Section 5.1) & 7.5x to 27.1x; 8.6x to 29.9x; 9.2x to 33.3x & arithmetic on Table 2 \\
M1 growth agreement & ratio of month-on-month growth, Ahrefs over Semrush US & 0.95, 2.66, 1.37 & \texttt{instrument\_extra} \\
M1 error bound & at least one of two readings is a factor of at least the square root of their ratio from the true value & 5.2 at 27.1x & arithmetic on Table 2 \\
M1 decomposition & share of the log visits ratio carried by the keyword-universe term (Appendix E.2; defined only where both terms have the same sign) & 35\% (Semrush US against Ahrefs worldwide), 54\% (both worldwide) & \texttt{instrument\_extra} \\
M1 cross-section & visits ratio across twelve prefixes on one date & 0.20x to 7.24x (8.89x matched); median 1.01 & \texttt{analyze} \\
M1 age relationship & Spearman between section age and the ratio, Fisher 95\% interval, permutation p & −0.11 [−0.64, +0.49], p = 0.73 & \texttt{analyze} \\
M1 ordering agreement & Spearman and Kendall [Kendall 1938] between the two instruments’ visits; concordant pairs; by size band & +0.94, +0.82, 60 of 66; 51 of 51 beyond 3x, 9 of 15 within; 40 of 40 beyond 3x on both instruments & \texttt{analyze}, \texttt{instrument\_extra} \\
M1 dispersion & geometric mean and geometric standard deviation of the ratio & 1.13 and 3.1 & \texttt{instrument\_extra} \\
M1 AI Overview citation & cited keywords over all AI Overview keywords the prefix ranks on, Wilson interval [Wilson 1927] & 12.6\% to 20.0\% & arithmetic on Table 3 \\
M2 link rate & reviews linked to at least one label; labels per linked review & 55.6\%; 2.25 & \texttt{analyze\_reviews} \\
M2 label agreement & Spearman between mentions and volume per label (descriptive; labels are not independent) & +0.18 (mentioned), +0.27 (all linkable) & \texttt{analyze\_reviews}, \texttt{supplement\_r10}, \texttt{m2\_from\_counts} \\
M2 branch agreement & Spearman between branch shares on each branch’s own volume, by attribution rule; multi-word labels only (post hoc) & +0.47 (modal), +0.75 (split); +0.65, +0.91 & \texttt{supplement\_r10}, \texttt{m2\_from\_counts} \\
M2 complaint rate & one- and two-star reviews matching a waiting pattern, against five-star & 14.0\% against 1.9\% & \texttt{analyze\_reviews} \\
M2 link precision & sampled links judged correct by two blind annotators; their agreement & 65\% to 67\%; kappa 0.90 (single-word labels about 50\%, multi-word 95\% to 96\%) & \texttt{r2\_link\_check} \\
M2 waiting-match precision & sampled waiting matches judged complaints about waiting; share in store & 83\% to 85\%, kappa 0.93; about 90\% in store & \texttt{r2\_link\_check} \\
M2 public count agreement & Spearman between the vendor’s summed volume and Wikipedia requests per article (240 articles, October 2024 to September 2025), bootstrap 95\% interval; the matched keyword count against requests & +0.45 (+0.34 to +0.54); +0.54 (+0.45 to +0.62) & \texttt{wiki\_demand\_analysis} \\
M2 public count, keywords held constant & partial rank correlation of volume and requests given the matched keyword count (post hoc); share of the variance of log volume in log keyword count & −0.12; 77\% & \texttt{wiki\_demand\_supplement} \\
\bottomrule
\end{tabularx}
\end{table*}

\begin{table*}[tp]
\centering\footnotesize
\caption*{\textbf{Table D1.} (continued)}
\begin{tabularx}{\textwidth}{>{\hsize=0.859\hsize\RaggedRight\arraybackslash}X>{\hsize=1.323\hsize\RaggedRight\arraybackslash}X>{\hsize=0.964\hsize\RaggedRight\arraybackslash}X>{\hsize=0.855\hsize\RaggedRight\arraybackslash}X}
\toprule
\textbf{Metric} & \textbf{Definition} & \textbf{Value here} & \textbf{Script} \\
\midrule
M2 public count dispersion & monthly searches per monthly request across articles: median; 10th and 90th percentiles; geometric SD & 35; 0.3 to 244; 14.5 & \texttt{wiki\_demand\_analysis} \\
M2 public count predictions & P1 to P4 as fixed before collection & P1, P3 and P4 held; P2 held at its boundary & \texttt{wiki\_demand\_analysis} \\
M3 source-type share & citation incidences of a source type over all incidences among the 1,000 exported domains & Table C1 & \texttt{cross\_sector}, \texttt{cross\_sector\_sensitivity2} \\
M3 surface contrast & median video and community share, Google views against the other four, bootstrap over sectors & 21.3\% (17.2 to 26.7) against 2.1\% (1.7 to 2.7) & \texttt{cross\_sector\_sensitivity} \\
M3 every-sector dominance & sectors where the lowest Google view exceeds the highest other surface & 11 of 11 & \texttt{cross\_sector\_sensitivity} \\
M3 cross-sector overlap & mean pairwise Jaccard of top-ten non-company domains between sectors & 0.36 to 0.40 against 0.08 to 0.23 & \texttt{cross\_sector} \\
M3 variance decomposition & share of the sum of squares of cell shares explained by surface and by sector & 77\% against 14\% (video and community) & \texttt{cross\_sector\_sensitivity2} \\
M3 predictions & P1 to P4 as written before collection & one held, one in direction, two failed & \texttt{cross\_sector} \\
M3 across countries (Section 7.7; not predicted) & median video and community share of citation incidences, Google’s four views against the other four, over 104 cells in thirteen countries, bootstrap over countries; countries where the lowest Google view exceeds the highest other surface & 14.0\% (10.5 to 17.0) against 0.6\% (0.4 to 1.0); 13 of 13 & \texttt{intl\_surface\_cells}, \texttt{intl\_from\_cells} \\
Schema pass rate & share of competency questions the ontology can express & 4/5 (released example) & \texttt{ufi cq} \\
Instance coverage rate & mean share of subjects the graph answers, over expressible questions & 70.4\% (released example) & \texttt{ufi cq} \\
Content Gap Score & Appendix B & not evaluated & \texttt{ufi} \\
\bottomrule
\end{tabularx}
\end{table*}

The protocol fixes metrics for the studies of Section 13 before their data exist: for the Content Gap Score, the three classes of Section 13 (top by score, top by volume, bottom quartile of the score) compared on top-ten presence on a frozen query set at day 37, read on both instruments, with precision at k and nDCG at k [Järvelin \& Kekäläinen 2002] of the score’s ranking of entities against each entity’s outcome, and inter-annotator agreement reported as weighted kappa [Cohen 1968]; for moving demand surfaces, each surface’s share of an industry’s linked questions by quarter; for agents, the overlap between the questions agents and people ask; and for business disclosures, the correlation over quarters between what companies say about AI search and how often answers cite their domains on a prompt set fixed in advance, since the shares of Section 7 are partly fixed by the selection rule.

\FloatBarrier

\section{Reporting what a search-analytics instrument measured}

\subsection{A reporting checklist}
Ten items for any study that uses a commercial search-analytics instrument’s figures as data. Each instruction is followed by the finding or disclosure in this paper that motivates it.

\begin{enumerate}[leftmargin=1.6em, itemsep=1.5pt, topsep=2pt, parsep=0pt]
\item \textbf{Instrument, report, mode and territory.} Name the vendor, the product, the report, the query mode (domain, prefix or subfolder) and the territory or database. One vendor’s AI-visibility panel returned the same domain-scoped figures for every one of twelve subfolders queried (Section 5.5).
\item \textbf{Dates and a capture.} Date each read, state the period it covers, record the instrument’s own index or database date where the report shows one, and keep a dated capture. The dashboard reads here were not archived, and one stored-answer count moved from 14,823 to 17,206 to 17,400 between 9 and 12 October 2026 as the index refreshed (Section 14).
\item \textbf{The vendor’s definition, counted or modeled.} Quote the definition and say whether the figure is a count or a model’s output. Both visit figures here are vendor click models over estimated volumes at estimated positions, one defined in clicks and the other in visitors (Section 4.1).
\item \textbf{A second instrument for any magnitude.} Report a second instrument’s reading of the same unit, or state that the figure carries a range the study cannot bound. Two instruments differed on one prefix by 7.5x to 27.1x, and across twelve prefixes of one domain by 0.20x to 7.24x (Sections 5.1 and 5.5).
\item \textbf{Orderings.} Report an ordering read on one instrument with the factor between the units, and as unconfirmed until a second instrument shares it, since a large factor on one instrument does not settle it. Here all 40 pairs of sections more than three times apart on both instruments were ordered alike, as were 16 of the 18 more than three times apart on exactly one and 4 of the 8 within that factor on both; the two exceptions were six to seven times apart on one instrument and about equal on the other (Sections 5.5 and 5.6).
\item \textbf{Territory pairing.} State every figure’s territory and pair like with like before comparing instruments. Matching changed how many of twelve prefixes fell below a ratio of one, from six to four (Section 5.5).
\item \textbf{The keyword universe beside the visits.} Report each instrument’s ranking-keyword count next to its visits, so that the gap can be split as in E.2. On the area-code prefix the universe term was about 35\% of the log gap with Semrush US against Ahrefs worldwide and about 54\% with both worldwide (Section 5.6); across the twelve prefixes the keyword ratio had no detectable relationship with the visits ratio (Section 5.5).
\item \textbf{What volume omits.} Where volume stands for demand, say what it cannot see, and report any counted or first-party signal read beside it, with its source and period. Here 14.0\% of one- and two-star reviews matched a waiting pattern while the taxonomy’s 11 labels mentioning waiting carried 622 searches a month (Section 6.1), and a public request count agreed with the vendor’s volume roughly on ranking and not on magnitude (Section 6.2).
\item \textbf{How the interface was operated, and the terms.} State whether figures came from an API, an export or a dashboard, who or what operated the interface, and which clauses of the vendor’s terms were read. Here an AI browser assistant read the Ahrefs and Semrush dashboards and made the Section 7 exports, one figure was read by the author, no vendor API was used for those figures, and the clauses on automated use are quoted, with the statement that neither vendor has confirmed whether these reads fall within them (Sections 4.1 and 14).
\item \textbf{Release.} Release every value read, with its date and method, and the script that prints each quoted statistic. Here the twelve-prefix reads and \texttt{analyze.py} are in the public repository, and the other read values and scripts are in the source bundle (Section 14).
\end{enumerate}

\subsection{What a gap between two instruments is made of}
On the model the vendors describe (Section 4.1), an instrument X’s estimated visits for a prefix, V(X), is a sum over its ranking-keyword universe U(X) of an estimated volume times an estimated click share at the keyword’s estimated position. Whatever the model, V(X) equals the size of U(X) times m(X), the mean estimated visits per ranking keyword, so for two instruments A and B the log gap splits exactly in two: ln V(B)/V(A) = ln |U(B)|/|U(A)| + ln m(B)/m(A). A dashboard gives the first term, since it reports the ranking-keyword count; the second pools the volume model, the positions and the click model, and separating those needs per-keyword exports, which this paper does not have. The identity is written for one position per keyword; where an instrument counts a keyword once per location, the multiplicity sits in the second term. It is an accounting, not an attribution: the first term reads as a share of the gap only when the two terms have the same sign, as they do on the area-code prefix under both pairings (35\% and 54\%, the second approximate because its keyword counts were read after its visits; Section 5.6). On the six prefixes whose unmatched visits ratio is below one, the keyword ratio is still at least 3.2 (\texttt{instrument\_extra.py}, section 10), so there the two terms pull in opposite directions and no share reading exists. Report both terms.

\subsection{Positioning}
A search on 31 October 2026 of OpenAlex, Crossref, the EQUATOR library and the web (72 logged API queries; Semantic Scholar and DBLP not searched, being unreachable; vendor and practitioner grey literature sampled, not searched systematically; one screener, an AI agent; indexed English-language literature; recorded in \texttt{literature/checklist\_search\_2026-10-31/} in the source bundle) found no reporting checklist for commercial search-analytics figures. Guidance on these tools does exist, and the search record passed over it: an empirical comparison of four SEO tools sets out five methodological recommendations in its discussion [Font-Julián et al. 2026, Section 5.2], among them to use several tools where possible, to prefer what several detect, to read traffic as relative rather than precise, and to report the tool, the extraction date, the market settings and each metric’s definition. Items 1 to 4 cover the same ground and owe it that credit. Where those recommendations advise averaging traffic estimates across tools, this paper’s ratios, which run in no fixed direction within one domain, suggest reporting the range. Further off are documentation checklists for Google Trends [Nuti et al. 2014; Hölzl, Keusch \& Sajons 2025], reporting items for Wikipedia page views [Alibudbud 2023], and reporting and disclosure standards for social-media data, including data obtained through commercial intermediaries [Kim, Huang \& Emery 2016; Kostygina et al. 2023], which name no commercial SEO or web-analytics tool. Items 1, 2, 9 and 10 share their access-date, parameter, access-path and data-sharing items, and item 8 extends their external-validation item to a counted or first-party signal read beside the volume. What this appendix adds is a numbered checklist motivated item by item by measured disagreement: the orderings of item 5, the territory pairing of item 6, the keyword universe of item 7 with the identity of E.2, and in item 9 who operated the interface and which of the vendor’s terms were read; item 4 also applies to vendors the design of comparing engines on one query set [Thelwall 2008]. The checklist is written for the works Section 1 counts: in a screened sample of 120, 82\% to 87\% use a vendor’s figures as data, and 87\% to 88\% of those use one vendor’s only.

\section{The sources the framework is meant to collect}
The framework’s first step is to go where an industry’s customers already talk. The implementation declares thirteen surface types (Figure 1a); the inventory in Table F1 is the author’s working list, of November 2026, of the concrete sources behind them: one hundred, in ten groups. It is a plan for collection and not a record of it. This paper has read eight of the hundred, in four of the ten groups, each through the instrument its section names; the other ninety-two are unread. Search is one group of ten, and the framework’s premise is that the other nine hold what search volume does not show, which Sections 6 and 7 begin to test and Section 13 specifies.

\begin{table*}[tp]
\centering\small
\caption*{\textbf{Table F1.} One hundred sources in ten groups (the author’s inventory, November 2026). Bold marks the eight read in this paper: Google Search through two vendors’ rankings and a third’s keyword volume (Sections 5 and 6), five AI answer engines through one vendor’s index (Section 7), Wikipedia through its public request counts (Section 6.2) and Google Business reviews through a licensed export (Section 6.1).}
\begin{tabularx}{\textwidth}{>{\hsize=0.551\hsize\RaggedRight\arraybackslash}X>{\hsize=1.449\hsize\RaggedRight\arraybackslash}X}
\toprule
\textbf{Group} & \textbf{Sources} \\
\midrule
1. Search and discovery & \textbf{Google Search}, Google autocomplete, People also ask, Google Trends, Bing, DuckDuckGo, YouTube search, Amazon search, App Store search, Pinterest \\
2. AI answer engines & \textbf{ChatGPT}, \textbf{Perplexity}, \textbf{Gemini}, Claude, \textbf{Copilot}, \textbf{Google AI Overviews}, Meta AI, Grok, You.com, DeepSeek \\
3. Video and audio & YouTube videos, YouTube comments, YouTube live chat, TikTok, Instagram Reels, Twitch, Spotify podcasts, Apple Podcasts, Vimeo, webinar replays \\
4. Social & X posts and replies, LinkedIn posts, LinkedIn comments, Facebook groups, Facebook pages, Instagram comments, Threads, Bluesky, Snapchat, Nextdoor \\
5. Communities and Q\&A & Reddit, Quora, Stack Exchange, Stack Overflow, Discord servers, Slack communities, Hacker News, Product Hunt, brand support forums, XDA and niche forums \\
6. Expert and long-form & Substack, Medium, Hugging Face, GitHub issues, arXiv, Google Scholar, \textbf{Wikipedia}, Wikidata, conference notes, industry newsletters \\
7. Company and market & SEC filings, earnings calls, investor presentations, press releases, job postings, competitor sitemaps, competitor pricing pages, Crunchbase, patents, analyst reports \\
8. Marketplace reviews & Amazon reviews and Q\&A, Best Buy reviews, Target reviews, Costco reviews, Walmart reviews, Home Depot reviews, eBay listings, Etsy reviews, Wayfair reviews, Newegg reviews \\
9. Ratings and responses & \textbf{Google Business reviews}, owner responses to reviews, Yelp, Trustpilot, Better Business Bureau, G2, Capterra, Glassdoor, App Store reviews, Google Play reviews \\
10. Your own and public data & Site search logs, support tickets, chat transcripts, call center logs, sales call notes, surveys and NPS verbatims, Regulations.gov, Census and data.gov, court records, Common Crawl \\
\bottomrule
\end{tabularx}
\end{table*}

\balance
\end{document}